\documentclass[preprint]{elsarticle}
\usepackage{breqn}
\usepackage{amsmath,amssymb,amsfonts,gensymb}
\usepackage[ruled,vlined]{algorithm2e}
\usepackage{graphicx}
\usepackage{subcaption}
\usepackage{cite}
\graphicspath{ {./images/} }
\SetKwComment{Comment}{/* }{ */}

\author[1]{Adedapo Alabi\corref{cor1}}\ead{alabiaa@mail.uc.edu}
\author[2]{Dieter Vanderelst}\ead{vanderdt@ucmail.uc.edu}
\author[3]{Ali Minai}\ead{ali.minai@uc.edu}
\cortext[cor1]{Corresponding author}
\fntext[fn1]{The authors are with the Department of Electrical Engineering \& Computer Science, University of Cincinnati, Cincinnati, OH, 45221 USA}

\begin{document}

\title{Rapid Learning of Spatial Representations for Goal-Directed Navigation Based on a Novel Model of Hippocampal Place Fields}

\begin{abstract}
    The discovery of place cells and other spatially modulated neurons in the hippocampal complex of rodents has been crucial to elucidating the neural basis of spatial cognition. More recently, the replay of neural sequences encoding previously experienced trajectories has been observed during consummatory behaviour -- potentially with implications for rapid learning, quick memory consolidation, and behavioral planning. Several promising models for robotic navigation and reinforcement learning have been proposed based on these and previous findings. Most of these models, however, use carefully ingineered neural networks, and sometimes require long learning periods. In this paper, we present a self-organizing model incorporating place cells and replay, and demonstrate its utility for rapid one-shot learning in non-trivial environments with obstacles. 
\end{abstract}

\maketitle

\section{Introduction}
\label{sec:Introduction}

Animals have over time evolved to possess, among other things, advanced spatial cognition. In a bid to understand the computational processes and neural mechanisms underpinning this ability, the rodent in particular has been studied extensively. These studies have lead to the current consensus that rodent spatial cognition is supported by a context dependent topological map encoded in the hippocampus with mechanisms for quick one-shot memory consolidation and recall~\citep{moser2008place,o1978hippocampus,ulanovsky2007hippocampal,redish2016vicarious, papale2016interplay}. Computational models of this system have the potential to improve robot spatial cognition.

Bio-inspired approaches to navigation are particularly interesting because they naturally lend themselves to neuromorphic implementation with its inherent advantages of energy efficiency and parallelizability, which is better suited to maintaining multiple pose estimates under uncertainty than the traditional von Neumann architecture. More ambitiously, it has been hypothesized that the same basic circuitry that is used for spatial cognition in the hippocampal complex evolved into the cortical columns of the cerebral cortex that enable more general cognition or navigation in other abstract (non physical) spaces~\citep{hawkins2019framework}. Thus, understanding how this system works and building computational models of it can serve as a precursor to realizing the elusive dream of artificial general intelligence. The use of biologically-inspired features in our model is also driven by the strong conviction that the solutions to many of the deep, unsolved problems in learning systems, e.g., rapid reinforcement learning, can be found by moving closer to the biological systems that already solve these problems. 

Over the years, many research groups have developed mapping and navigation methods inspired by the hippocampal system \citep{burgess1996neuronal,skaggs1998spatial,samsonovich1997path,erdem2012goal}. The model presented in this paper builds on some of the ideas in these models, and adds others from recent discoveries in neuroscience such as replay and preplay. The model shows how a self-organized dynamical system can learn to encode actionable place representations using a biologically plausible reinforcement learning mechanism, and exploit it to learn and generalize goal-oriented spatial representations from very limited experience.

\section{Motivation}
The work in this paper had two primary motivations:

\begin{enumerate}

    \item Animals are able to learn quickly and generalize from limited experience. This is in sharp contrast to most of today's artificial intelligence systems which require extensive training. Rodents appear to augment their real-time exploration with offline replay -- a discovery that has inspired a number of successful reinforcement learning algorithms such as AlphaZero and Deep Q-learning~\citep{silver2017mastering, mnih2015human, mnih2013playing}. This is putatively coupled with the use of mental preplay to ''imagine" the outcome of possible actions at decision points~\citep{redish2016vicarious, papale2016interplay, schmidt2013conflict}. The current model integrates both mechanisms in a model that can learn to demonstrate efficient goal-directed navigation from limited experience.
   The model shows that replay can be used to assign reward values to previously explored states rapidly and preplay can be used to exploit these reward values during subsequent navigation.
    
    \item Place cell models typically include carefully tuned connections between the place cells and/or with their afferent input to generate convex place fields~\citep{samsonovich1997path,grieves2018boundary}. We show how a competitive self-organized place cell network requiring minimal pre-configurations can learn convex place fields with implicit directional information, resulting in a spatial representation that supports efficient goal-directed navigation in complex environments with multiple obstacles.
    
\end{enumerate}

\section{Background}
\label{sec:Background}
It is now generally accepted that the hippocampal complex plays a central role in spatial cognition in mammals~\citep{moser2008place,o1978hippocampus,ulanovsky2007hippocampal}. Place cells in the hippocampus \citep{o1971hippocampus, o1978hippocampus, muller1987spatial, muller1989firing}, grid cells in the entorhinal cortex \citep{hafting2005microstructure,fyhn2008grid,moser2008place, mcnaughton2006path, bush2015using}, and head direction cells in the postsubiculum \citep{taube2007head} and entorhinal cortex are some of the spatially modulated cells found in this region. 
In a new environment, place cells exhibit spatially localized activity called {\it place fields} \citep{o1978hippocampus,muller1987spatial}. Grid cells effectively code for displacement, firing in multiple locations across an environment in a regular hexagonal lattice~\citep{hafting2005microstructure,fyhn2008grid} and are believed to perform path integration to support localization in place cells~\citep{mcnaughton2006path,solstad2006grid,bush2014grid}. Both systems work together to allow animals to form spatial cognitive maps \citep{o1978hippocampal,redish1997cognitive} supporting localization and navigation by integrating sensory and ideothetic information. As the name suggests, the head direction cells of the entorhinal cortex \citep{sargolini2006conjunctive} and postsubiculum \citep{taube1990head} encode the allocentric heading of the animal, firing maximally when the heading matches their preferred directions.

Along with mapping, intelligent agents must also learn important locations within the map and how to navigate efficiently to them. Animals are known to exhibit quick one-shot learning of key locations in new environments~\citep{morris1984developments, vorhees2006morris}. In both biological and artificial agents, this problem is solved through reinforcement learning (RL)~\citep{sutton2018reinforcement}, where the agent learns through rewards and penalties elicited by exploratory behavior. RL is particularly useful when complete knowledge of the environment is unavailable, rendering supervised training impossible 

A major difference between RL in animals and computational systems is that the former can learn very rapidly from just a few trials while the latter requires extensive training. The reasons for this difference are not well-understood. The availability of good pre-configured priors, transfer learning, and accurate extrapolation may play a role, but an important contributor may be the ability to learn from mental rehearsal. The discovery that previously experienced neural sequences are replayed in the hippocampus during sleep~\citep{maingret2016hippocampo, klinzing2019mechanisms, diekelmann2010memory} and consummatory behaviour~\citep{buzsaki1989two,carr2011hippocampal,epsztein2022mental} suggests the possibility that animals augment real-time learning with off-line replay. The replay that occurs upon reward receipt is particularly interesting as it theoretically provides a mechanism for the quick assignment of expected reward values to previously traversed locations if accompanied by spike time-dependent potentiation (STDP) between place cells and neurons that encode rewards~\citep{alabi2020one,michon2019post,olafsdottir2018role}.

Beyond replay, rodents navigating towards known goals have also been observed to pause and look down the available paths at decision points in real-time navigation. This is accompanied by a mental \textit{preplay} of the place cell sequences encoding the paths in a process termed vicarious trial-and-error (VTE)~\citep{redish2016vicarious, papale2016interplay, schmidt2013conflict}. Assuming that expected reward values have previously been assigned to the place cells - either by replay or a different means; this could hypothetically allow the rodent evaluate the outcomes of its available actions so it can decide which one to take.  This approach has also been used in some hippocampally-inspired navigation models~\citep{erdem2012goal,Erdem2014}. However, the success of both replay-based RL and preplay-based decision-making depends on the agent's ability to generalize by extrapolation as well as interpolation. The model presented in this paper shows such generalization. 

A number of research groups have developed robotic navigation algorithms incorporating findings in neuroscience. RatSLAM~\citep{milford2004ratslam,wyeth2009spatial,milford2008mapping} is one such example and has been shown to navigate office spaces \citep{milford2010persistent} and map large outdoor environments \citep{milford2008mapping}. Extensions of RatSLAM have also been shown to be sensor-agnostic and to support sensor fusion \citep{Berkvens2015,Jacobson2015,Steckel2013}. Some algorithms also incorporate the experimentally observed replay~\citep{JOHNSON20051163, foster2012sequence} for memory consolidation. We have recently proposed and developed a model~\citep{alabi2020one} to enable quick one-shot learning and generalization in simple open mazes using replay. In this work, we refine that model by defining it as a self-organizing dynamical system and demonstrate it's ability to learn quickly in more complex mazes with obstacles.

\section{Model Description}
\label{sec:Model}
\begin{figure*}[!h]
    \centering
    \includegraphics[width=.8\textwidth]{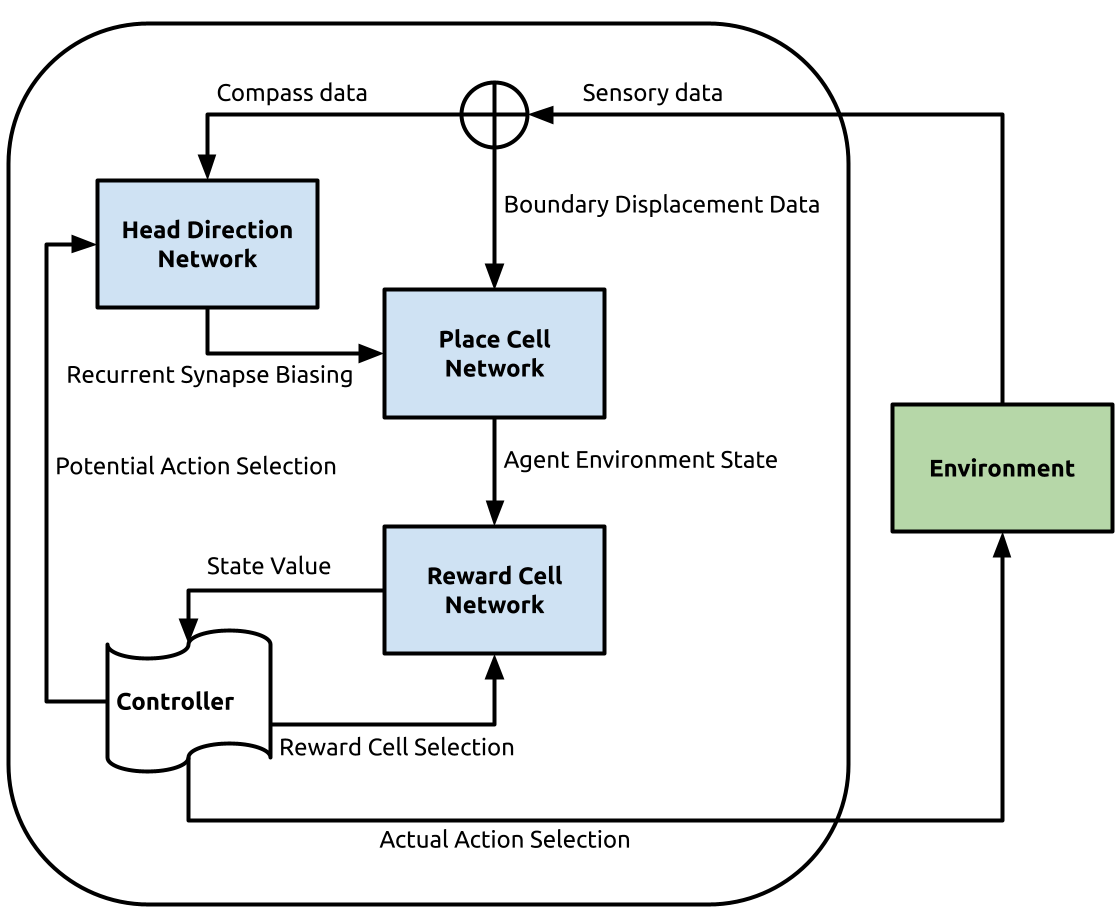}
    \caption{An overview of our model architecture}
    \label{fig:model_architecture}
\end{figure*}

The model consists of four types of cells ~\citep{alabi2020one}: the head-direction cells; boundary vector cells; place cells, and reward cells. These four types of cells are denoted by the symbols $h$ (head-direction cells), $b$ (boundary vector cells), $p$ (place cells), and $r$ (reward cells). The time-varying firing rate of the $i^{th}$ cell from cell type $k$ is denoted as $ {v_i^k}(t)$ or $ {v_i^k}$ for brevity. The time-varying synapse from the $j^{th}$ cell of type $l$ to the $i^{th}$ cell of type $k$ as $ {W_{ij}^{kl}}(t)$ (or $ {W_{ij}^{kl}}$). The current model does not include grid cells, which will be added in the future.

The networks operate in three states: regular, replay and preplay. In the regular state, the networks directly encode the sensory input and the environment is learnt. In preplay, the agent uses the learnt map to \textit{``imagine''} the expected network states if it were to move a step from its location in a specific direction. Replay is a transient spreading activation state that occurs upon reward receipt and serves to learn the value map of the environment. 
\subsection{Head Direction Network}

The head-direction cells comprise a single layer of neurons, each of which has a preferred direction in allocentric coordinates. It fires maximally when the animal's heading is in its preferred direction, with a symmetric dropoff on both sides. The model by Erdem \& Hasselmo~\citep{erdem2014biologically} is used to represent the head-direction cells. In particular, the firing rate of head-direction cell $i$ is given by:

\begin{equation}
     {v^h_i} = \sum_{j=0}^2  {x'_j} \left[\begin{array}{c}
        \cos( {\theta_i^h} + \theta_0) \\[1ex] 
        \sin( {\theta_i^h} + \theta_0) \\
    \end{array}
    \right]
\end{equation}
where $\theta_0$ is the heading angle of the anchor cue, $ {\theta_i^h}$ is the preferred direction of head direction cell, and $ {x'}$ is the instantaneous velocity  of the agent in Cartesian coordinates.

\subsection{Place Cell Network}

\begin{figure*}
    \centering
    \includegraphics[clip,trim={2cm 1cm 1.5cm 0 },width=52mm]{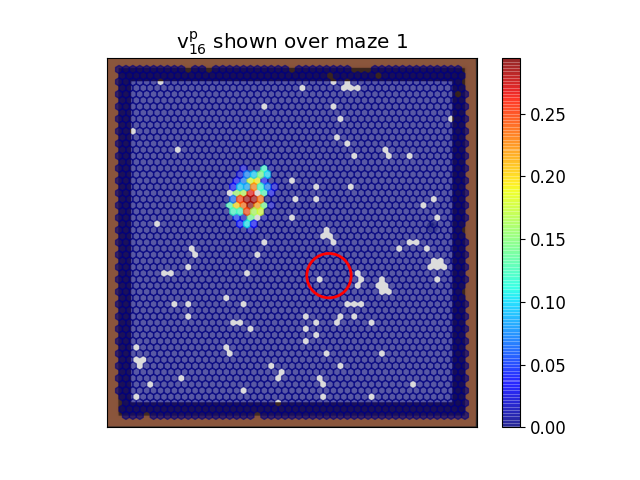}
    \includegraphics[clip,trim={2cm 1cm 1.5cm 0 },width=52mm]{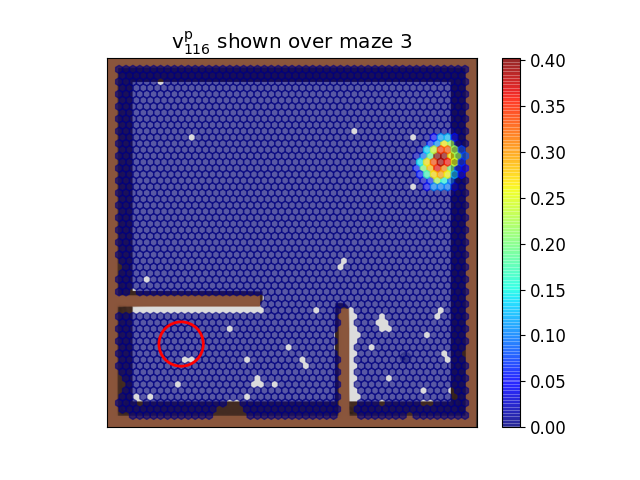}
    \includegraphics[clip,trim={2cm 1cm 1.5cm 0 },width=52mm]{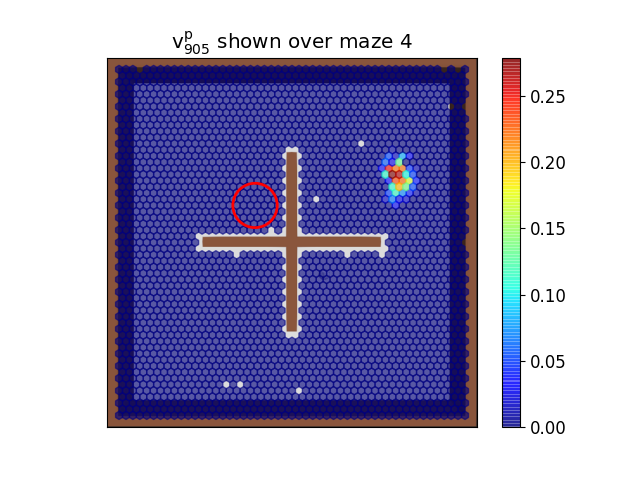}
    \includegraphics[clip,trim={2cm 1cm 1.5cm 0 },width=52mm]{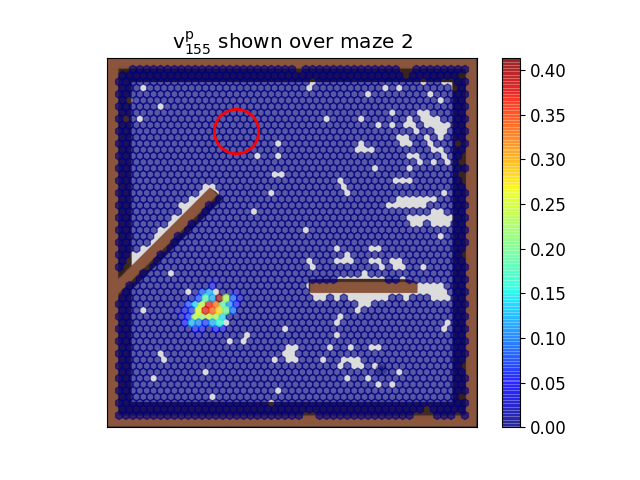}
    \includegraphics[clip,trim={2cm 1cm 1.5cm 0 },width=52mm]{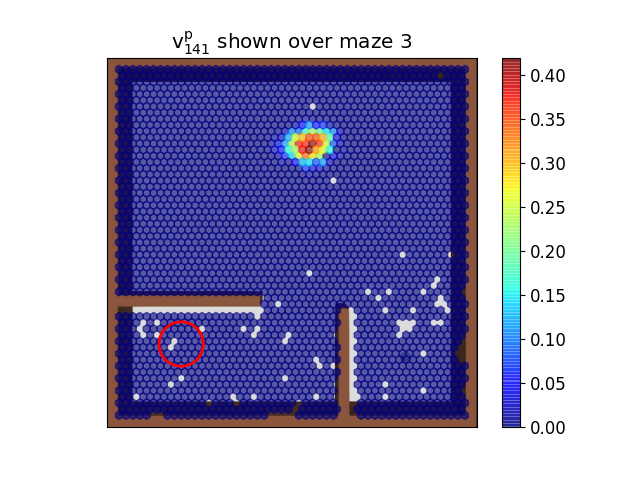}
    \includegraphics[clip,trim={2cm 1cm 1.5cm 0 },width=52mm]{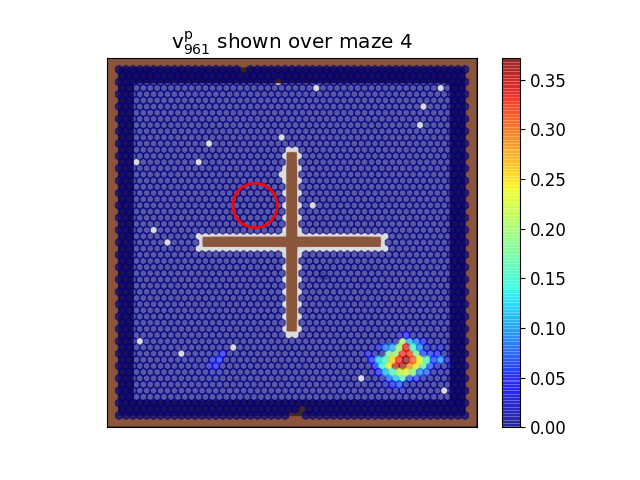}
    \caption{Examples of place fields developed in the mazes. These plots were generated using a color coded hexagonal bin plot of the firing rate of the probed place cell during a random run after exploration. It should be noted that here and in other plots, the environment is continuous and the binning is only done for plotting. Unvisited bins have no values and are left unfilled, with the grey arena underneath visible. The red circles indicate the goal locations in the mazes.}
    \label{fig:place_fields}
\end{figure*}

Place cells have localized, convex activity fields (\emph{place fields}) in specific locations in a given environment. While their activity is known to be based on several input sources, two of the primary sources are boundary vector cells~\citep{o1996geometric,lever2009boundary} and other place cells via recurrent connections~\citep{miles2014recurrent,solstad2014place,bains1999reciprocal}, which are the sources considered in the current model. 

The regular state activity of place cells in the model is based on input from boundary vector cells (BVCs). BVCs are neurons that fire when the animal is at specific distance and allocentric directions from obstacles or boundaries. These cells are found in the subiculum~\citep{lever2009boundary}. Following the model by Barry et al ~\citep{barry2006boundary}, for BVC $i$ tuned to obstacles at displacement $(d_i , \; \phi_i)$ in polar coordinates with tuning widths $(\sigma_r ,\; \sigma_\theta)$, the firing rate in response to an obstacle at displacement $(r ,\; \theta)$, subtending an angle $\delta \theta$ is given by:

\begin{dmath}
    \label{eq:bvc}
    \delta  {v_i^b} = \frac{\exp \left[-\left( r-d_i\right) ^2/2\sigma^2_{r}\right]}{\sqrt{2\pi\sigma^2_{r}}} \times \frac{\exp \left[-\left( \theta-\phi_i\right) ^2/2\sigma^2_{\theta}\right]}{\sqrt{2\pi\sigma^2_{\theta}}} \delta \theta
\end{dmath}
The equations for the firing rate of the $i^{th}$ place cell are then:
\begin{subequations}
\label{eq:place_cells}
\begin{equation}
    \tau_{p} \frac{d {s_i^{p}}}{dt} = - {s_i^{p}} + \sum_{j=0}^{n_b}  {W_{ij}^{pb} v_j^b} - \Gamma^{pb} \sum_{j=0}^{n_b}  {v_j^b} - \Gamma^{pp} \sum_{j=0}^{n_p}  {v_j^p} 
    \label{eq:s_pb}
\end{equation}

\begin{equation}
     {v_i^p} =\tanh{\left(\left[ \psi {s_i^{p}}\right]_+\right)}
    \label{eq:v_p}
\end{equation}
\end{subequations}

Equation \eqref{eq:s_pb} models the dynamics of $ {s_i^{p}}$, the membrane potential of the place cell as a function of its BVC input. This response is subject to inhibition from both feedforward and feedback sources to ensure spatial specificity. The feedforward inhibition term $\sum_{j=0}^{n_b}  {v_j^b}$ scaled by gain $\Gamma^{pb}$ sets a threshold on the amount of feedforward input to a place cell required for depolarization while the recurrent inhibition term $\sum_{j=0}^{n_p}  {v_j^p}$ scaled by gain $\Gamma^{pp}$ serves to keep the total network activity from diverging. Equation \eqref{eq:v_p} determines the firing rate $ {v_i^p}$ as the hyperbolic tangent of its rectified membrane potential scaled by a parameter $\psi$. 

It is important to ensure that place cells cover the entire environment. To guarantee uniform coverage over environments, competitive learning is used in the place cell network. Place cells can develop localized place fields by competing for boundary vector cell inputs. The following rule proposed by Oja~\citep{oja1982simplified} captures competitive learning. Specifically, the synaptic strength from BVC cell $j$ to place cell $i$ evolves as:
\begin{equation}
    \tau_{w^{pb}} \frac{d {W^{pb}_{ij}}}{dt} =  {v^p_i} \left(  {v^b_j} - \frac{1}{\alpha_{pb}}  {v^p_i W^{pb}_{ij}} \right) \label{eq:w_pb}
\end{equation}
where $\tau_{w^{pb}}$ parameterizes the speed of learning, $ {W^{pb}_{ij}}(0) = 1$ with a probability $p_{pb}$ and $\alpha_{pb}$ is a normalizing factor.

We propose that place cells switch to being driven by recurrent input during preplay state. This distinction between the contributions of the recurrent and boundary vector cell input is inspired by both the morphology and physiology of the hippocampus. Afferent input to place cells such as that from boundary vector cells targets distal apical dendritic tufts while excitatory recurrent synapses are typically at basal dendrites \citep{klausberger2008neuronal, takahashi2009pathway, miles2014recurrent}. Negative recurrent feedback -- mediated by oriens lacunosum-moleculare (OLM) cells -- also target the apical dendritic tufts \citep{bloss2016structured}, justifying the negative feedback term in equation \ref{eq:s_pb}. 

Beyond the physical separation of the input sources, it has been proposed that, similar to what has been observed in the piriform cortex \citep{hasselmo1991selective, hasselmo1992cholinergic, tang1994selective}, selective suppression of recurrent but not feedforward synaptic transmission by acetylcholine and other neurotransmitters in the hippocampus can allow for the learning of new information over recall\citep{hasselmo1994laminar, hasselmo1993acetylcholine, wallenstein1997gabaergic, booker2020presynaptic}. We extend this hypothesis in Section \ref{subsec:MO}. 

During preplay, the place cells are driven by recurrent input from other place cells along the direction $a \in  {\theta^h}$.  
\begin{equation}
\label{eq:v_p_p}
     {\Bar{v}_i^p}\, |\, a = \tanh{\left( \left[ \sum_{j=0}^{n_p} {W^{pp}_{aij}} {v^p_j} -  {v^p_i} \right]_+ \right)}
\end{equation}
$ {\Bar{v}_i^p}$ is the resultant value of $ {v_i^p}$ when the agent imagines taking one step in direction $a$. The synapse $ {W^{pp}_{aij}}$ between place cell $i$ and place cell $j$ encodes the proximity of their place fields along direction $a$. 

While synapses have conventionally been represented as dyadic objects, newer evidence is beginning to make clear that their effect can depend on signals from sources other than just the activity of pre-synaptic and post-synaptic neurons. For example, external modulation and local computations occurring within dendritic branches creates complex dependencies between synapses and with modulating inputs~\citep{yang2016dendritic, chiu2013compartmentalization, takahashi2009pathway,poirazi2003pyramidal, losonczy2006integrative}. We extend a previous proposal that co-active pre-synaptic neurons that respond to related sensory input preferentially innervate the same dendritic branches~\citep{takahashi2012locally, kleindienst2011activity, druckmann2014structured,alabi2022one} to suggest that the dendritic branches of place cell recurrent collaterals can be modulated in direction-specific ways, thus resulting in triadic synapses. We represent the direction encoding branches as the first dimension of a three-dimensional weight tensor, $[ {W^{pp}_{aij}}]$. Potentation occurs on individual branches in a direction-specific manner so that $  {W^{pp}_{aij}}$ -- the synapse from place cell $j$ to place cell $i$ along the branch responsive to direction $a$ encodes the proximity of their place fields going in the direction $a$ from the place field of $j$ to the place field of $i$.

Along with the directionality in the spatial relationship between two place fields, the temporal order is also crucial, i.e., which of the place fields leads to the other along the specified direction. To capture this causality or temporal order, a static rule focusing only on the instantaneous firing rate would be insufficient. We formulated a learning rule that integrates the pre-synaptic firing rate of $  {v^{p}_j}$ as well as the head direction cell activity $ {v^h_k}$ over the preceding time period $\tau_{pre}$ to determine temporal order. 

To update the synaptic strengths in the weight tensor, $W^{pp}_{kij}$, in a directionality-specific way, we compute variables $\Upsilon^p_j$, $\Upsilon^p_i$, and $\Upsilon^p_k$, each representing the integrated recent activation of pre-synaptic place cell $j$, post-synaptic place cell $i$, and head direction cell $k$, respectively, along direction $\theta_k^h$:

\begin{subequations}
    \begin{equation}
        \tau_{h} \frac{d   {\Upsilon^p_j}}{dt} = -  {\Upsilon^p_j} +  {v^p_j} 
    \end{equation}
    \begin{equation}
        \tau_{h} \frac{d   {\Upsilon^p_i}}{dt} = -  {\Upsilon^p_i} +  {v^p_i} 
    \end{equation}
    \begin{equation}
        \tau_{h} \frac{d   {\Upsilon^h_k}}{dt} =  -  {\Upsilon^h_k} +  {v^h_k}
    \end{equation}
\end{subequations}
\noindent
where $\tau_h$ is a time constant parameterizing the length of the history taken into account.

The weights in the tensor then change as:

\begin{equation}
   \tau_{w^{pp}} \frac{d   {W^{pp}_{kij}}}{dt} =  {\Upsilon^h_k} \left( {v^p_i}  {\Upsilon^p_j} -  {v^p_j} {\Upsilon^p_i} \right) \label{eq:w_pp}
\end{equation}
where $\tau_{w^{pp}}$ parameterizes the speed of learning, and , $ {W^{pp}_{kij}}(0) = 0$. The product $ {\Upsilon^p_i}  {v^p_j}$ has a high value if post-synaptic place cell $i$ has strong activation following strong recent activation of pre-synaptic place cell $j$, indicating movement \emph{towards} the place field of $i$ from the place field of $j$. Similarly, $ {\Upsilon^p_j}  {v^p_i}$ has a high value if pre-synaptic place cell $i$ has strong activation following strong recent activation of post-synaptic place cell $i$, indicating movement \emph{away from} the place field of $i$ towards the place field of $j$. The sign of the difference term in equation (6) thus indicates the direction of movement relative to the place fields of $i$ and $j$, and the $ {\Upsilon^h_k}$ term gates the difference term by the directionality. As a result, weight $  {W^{pp}_{kij}}$ increases when the path in direction $ {\theta_k^h}$ activates the place field of $j$ before the place field of $i$, and decreases if the order is reversed.

\subsection{Reward Cell Network}
This network comprises reward cells that exhibit a binary response to the receipt of their preferred rewards. Reward cells receive synapses from place cells, which are modified to learn the proximity of each place cell's place field to the  preferred reward locations for reward cells. These weights are learnt during \textit{replay} states as described further in Section \ref{subsec:replay}. Once learnt, the weights enable the evaluation of actions as explained in Section \ref{subsec:exploitation}.

\subsection{Modes of Operation}
\label{subsec:MO}
The simulated scenario is that of an environment with visible boundaries, a variety of possible obstacles, and one or more goal locations that elicit rewards when the agent reaches them. The environment is initially unfamiliar to the agent, and the goal locations are not known. The agent operates in two modes:

\begin{figure}
    \centering
    \includegraphics[clip,trim={0 1cm 0 1.45cm},width=.8\columnwidth]{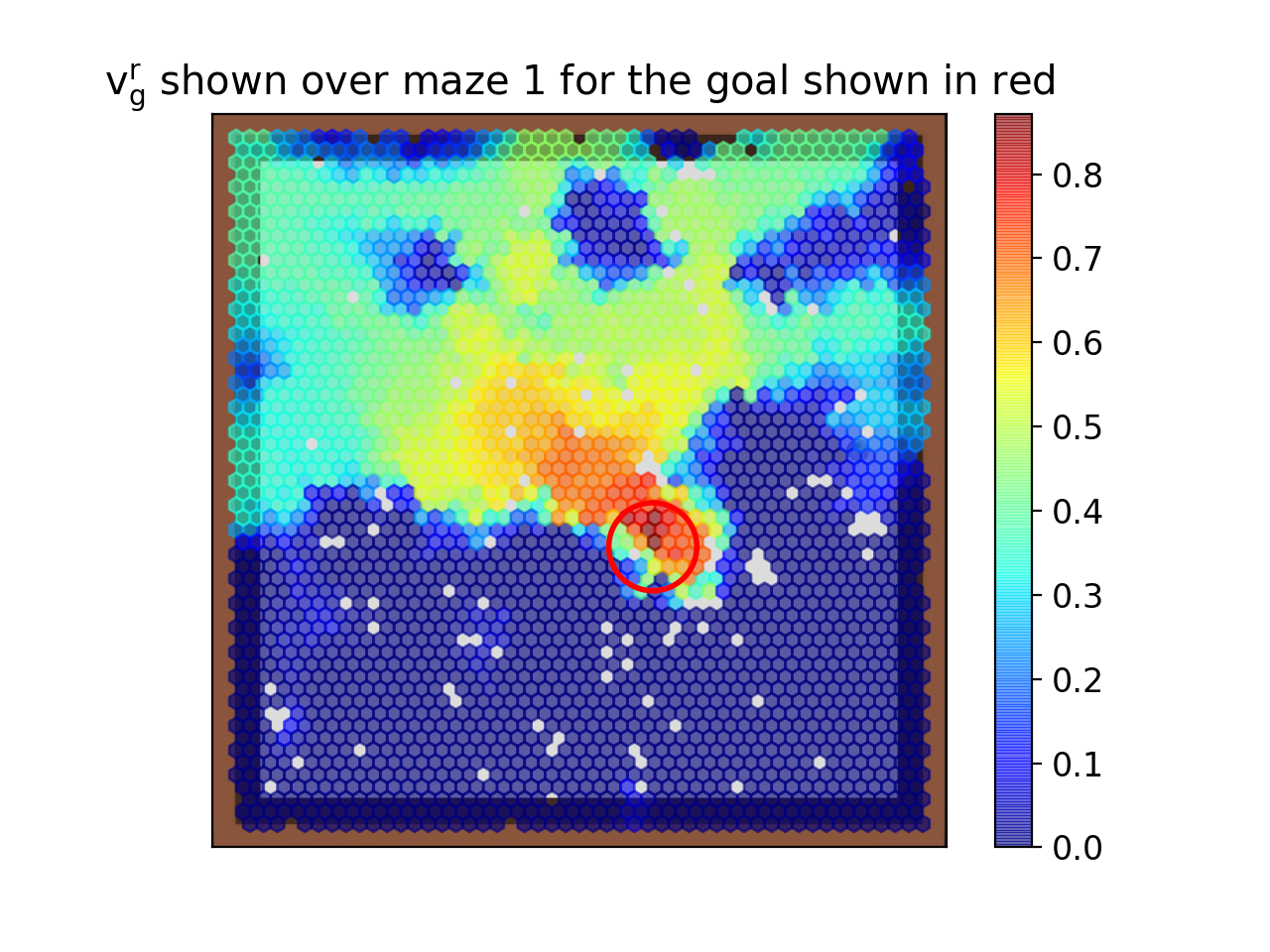}
    \includegraphics[clip,trim={0 1cm 0 1.45cm},width=.8\columnwidth]{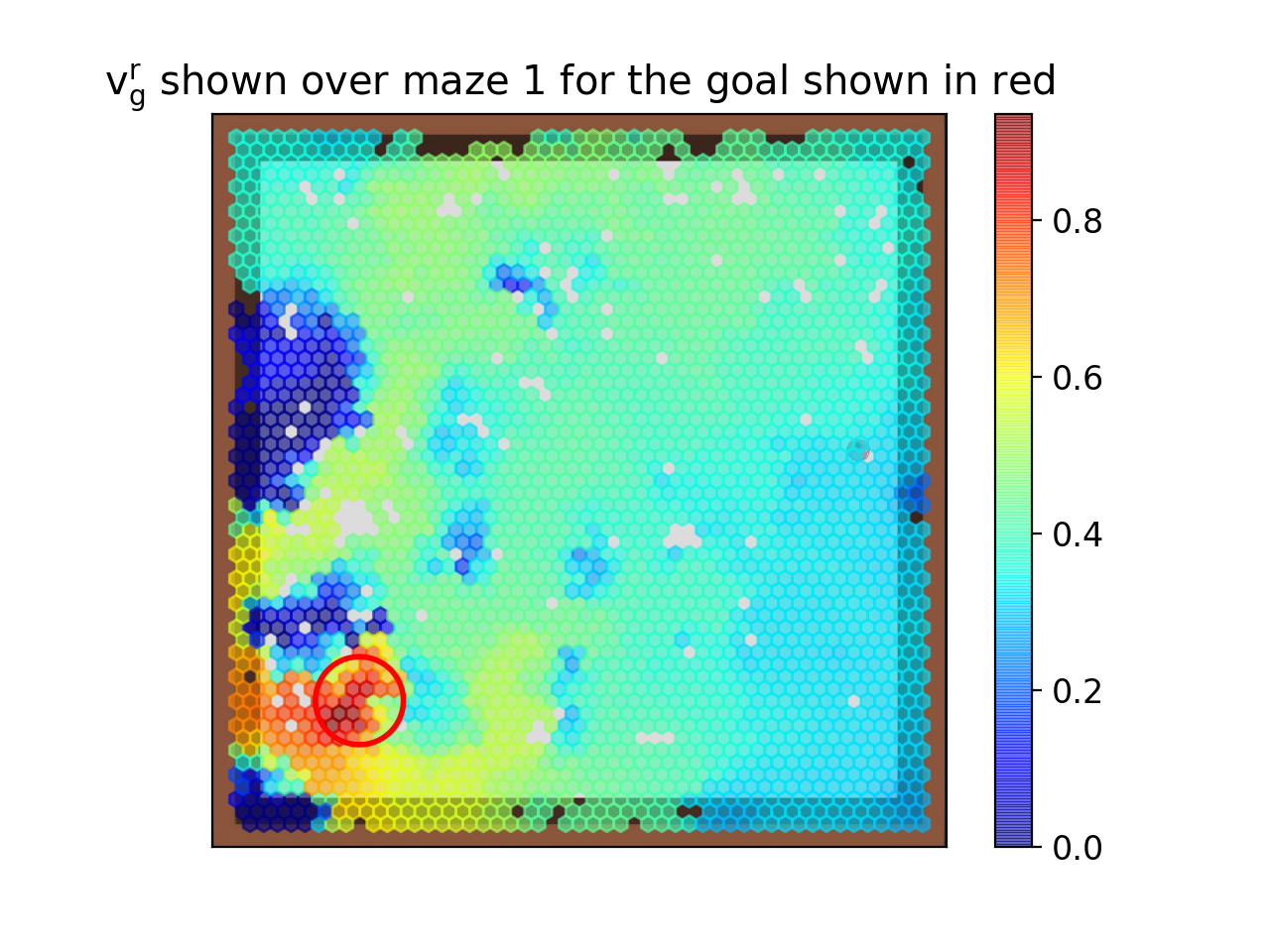}
    \includegraphics[clip,trim={0 1cm 0 1.45cm},width=.8\columnwidth]{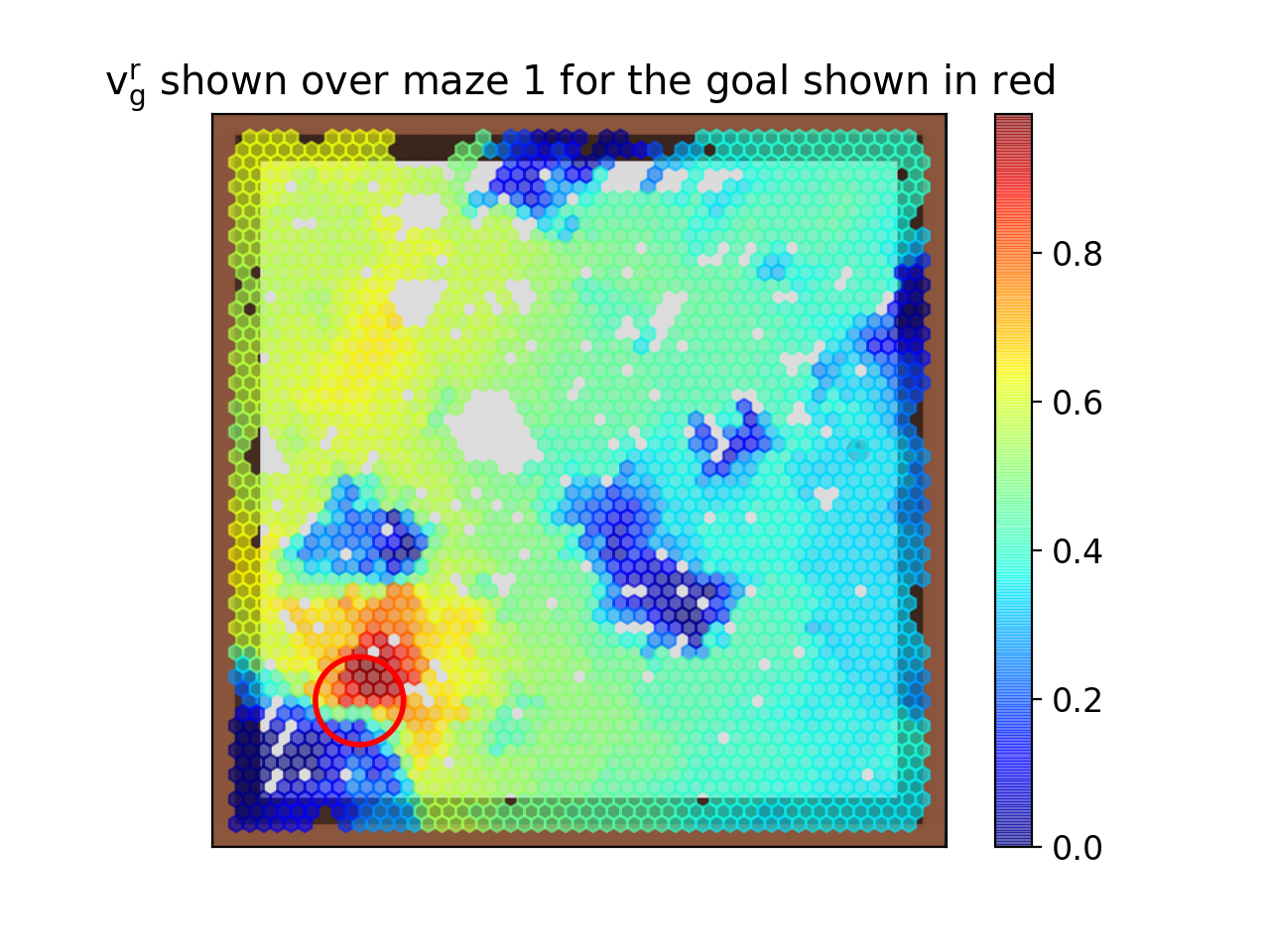}
    \caption{Implied values maps developed in maze 1 after the first runs depicted in Fig. \ref{fig:same_start_open_maze}. The value map was derived by allowing the robot to roam aimlessly around the maze after the reward had been found on the first run without any synaptic modulation. At each location the firing rate of the reward cell as a result of place cell input was recorded. This was then plot as a color coded hexagonal bin plot.}
    \label{fig:reward_maps}
\end{figure}
\begin{figure*}
    \begin{subfigure}[b]{.3\textwidth}
        \includegraphics[clip,trim={2.5cm 1cm 2.5cm 0 },width=40mm]{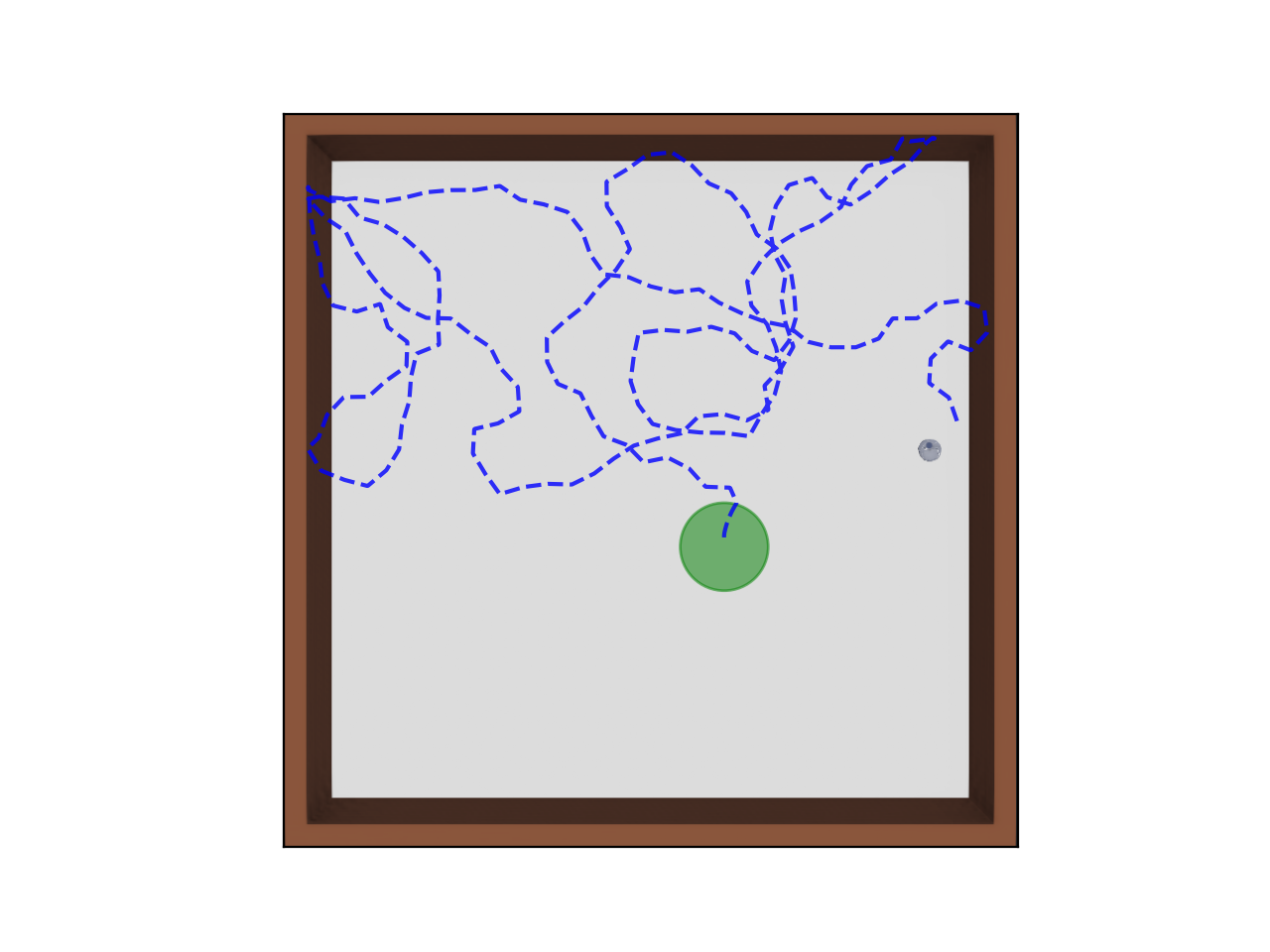}
        \includegraphics[clip,trim={2.5cm 1cm 2.5cm 0 },width=40mm ]{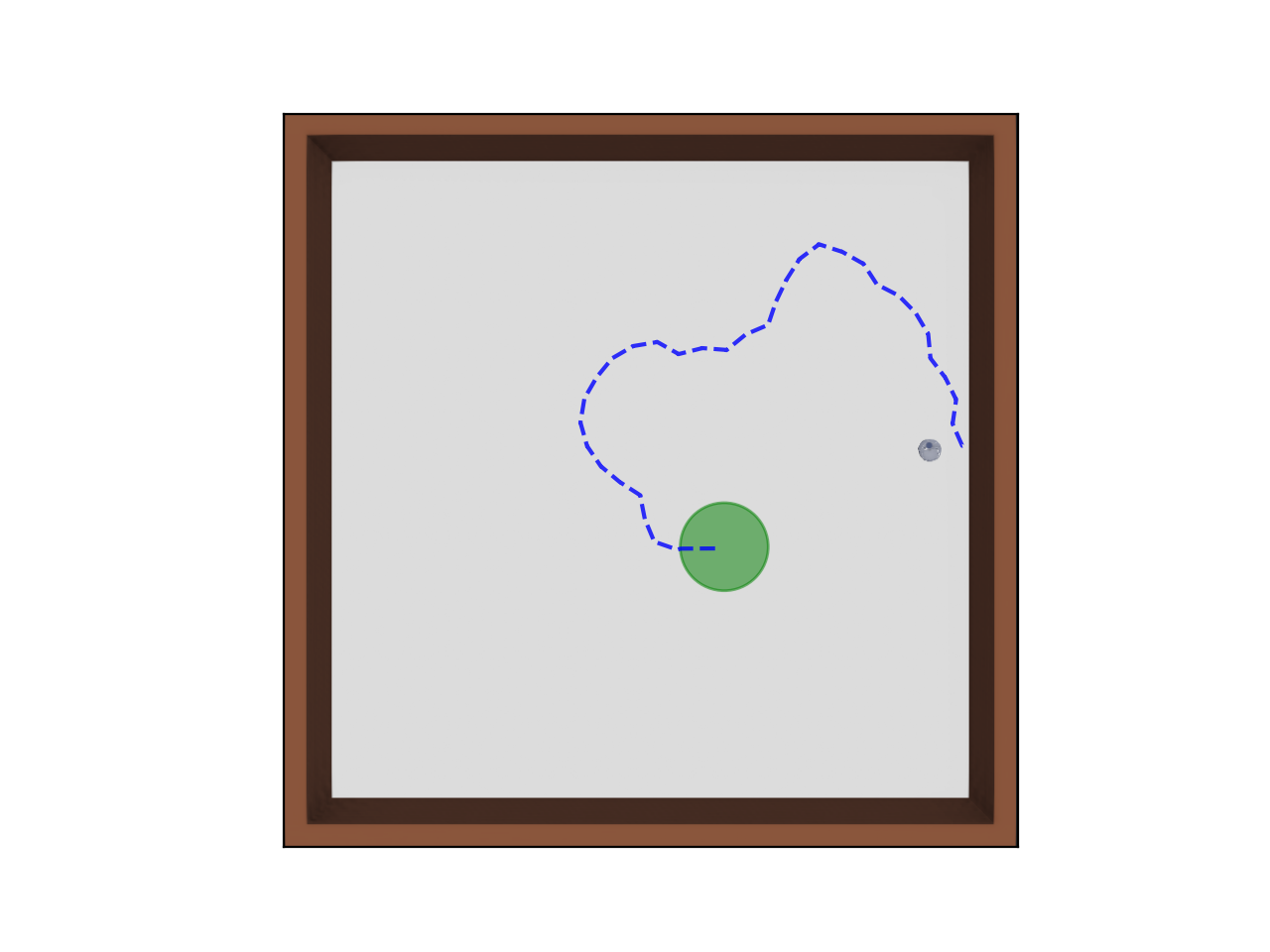}
        \includegraphics[clip,trim={2.5cm 1cm 2.5cm 0 },width=40mm ]{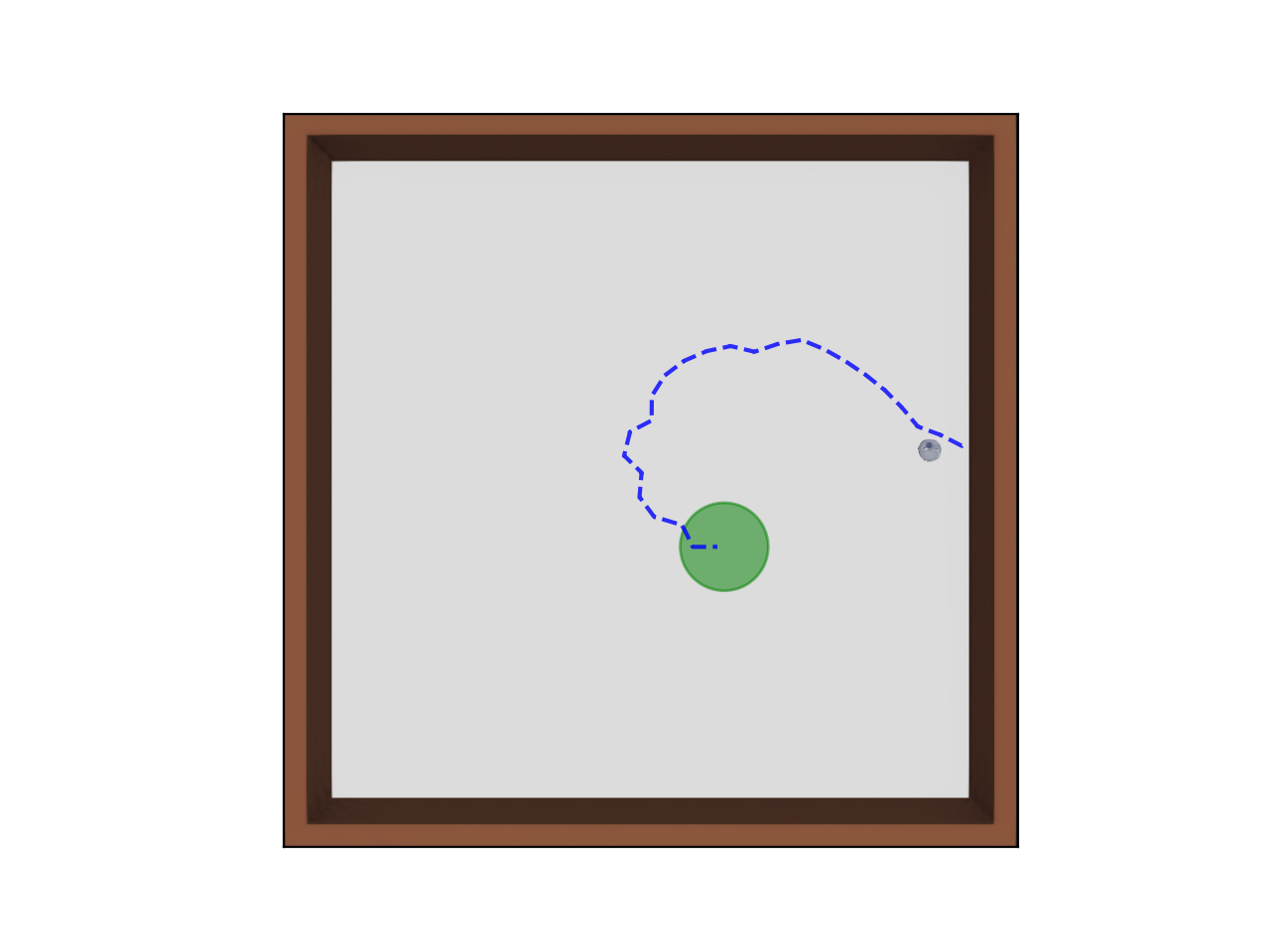}
        \includegraphics[clip,trim={2.5cm 1cm 2.5cm 0 },width=40mm ]{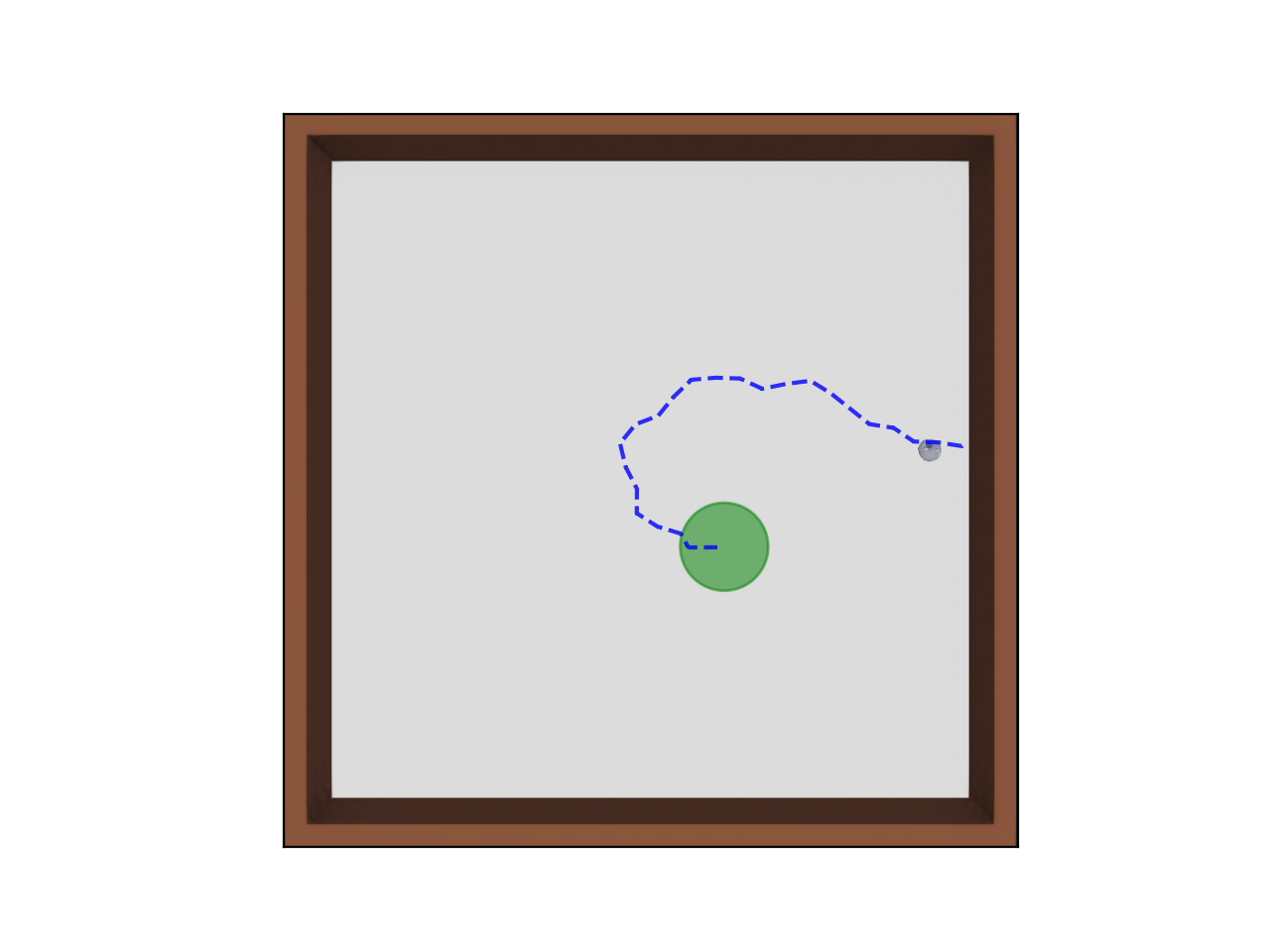}
        \caption{Trial 1}
    \end{subfigure}
        \begin{subfigure}[b]{.3\textwidth}
        \includegraphics[clip,trim={2.5cm 1cm 2.5cm 0 },width=40mm ]{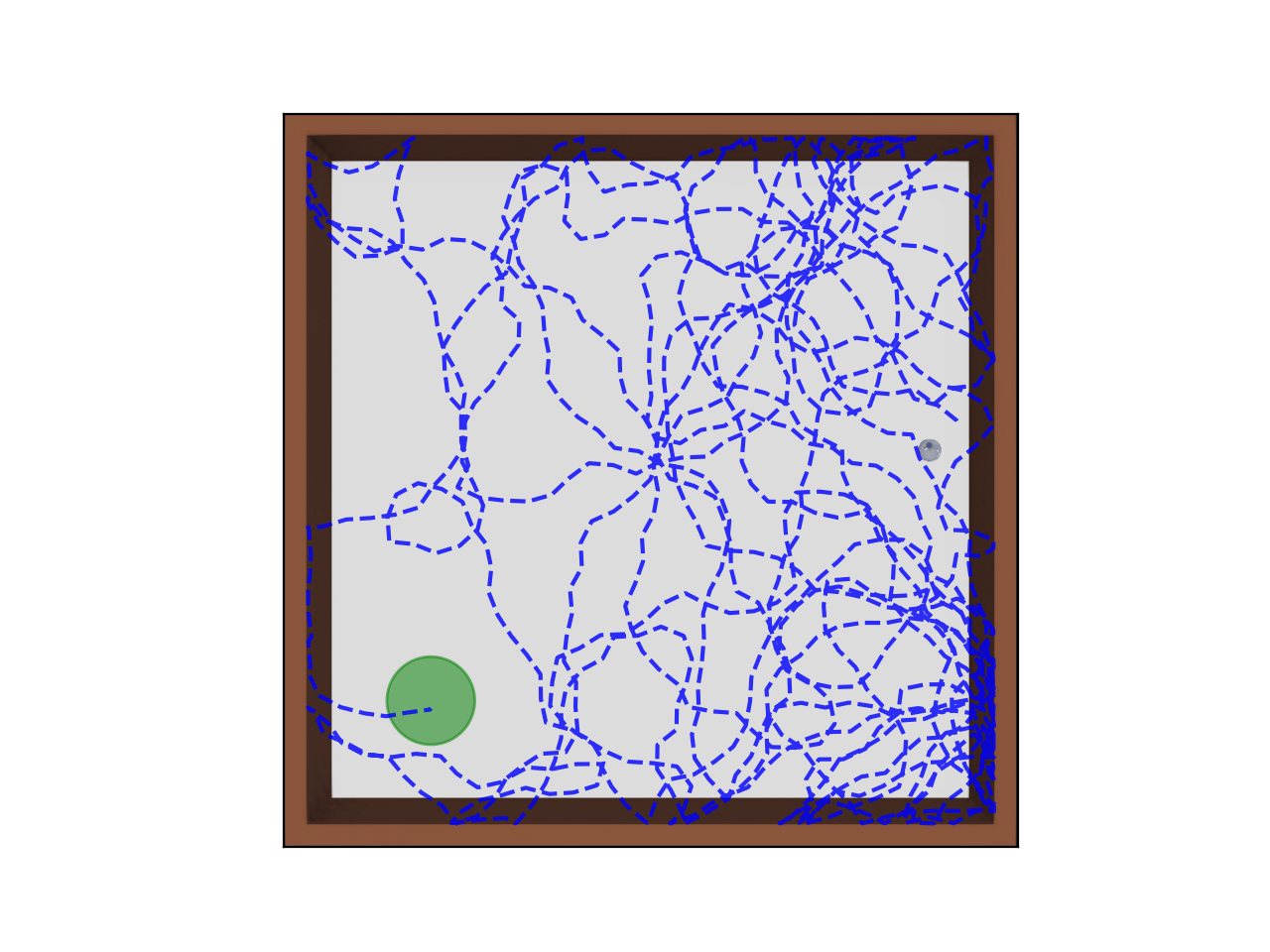}
        \includegraphics[clip,trim={2.5cm 1cm 2.5cm 0 },width=40mm ]{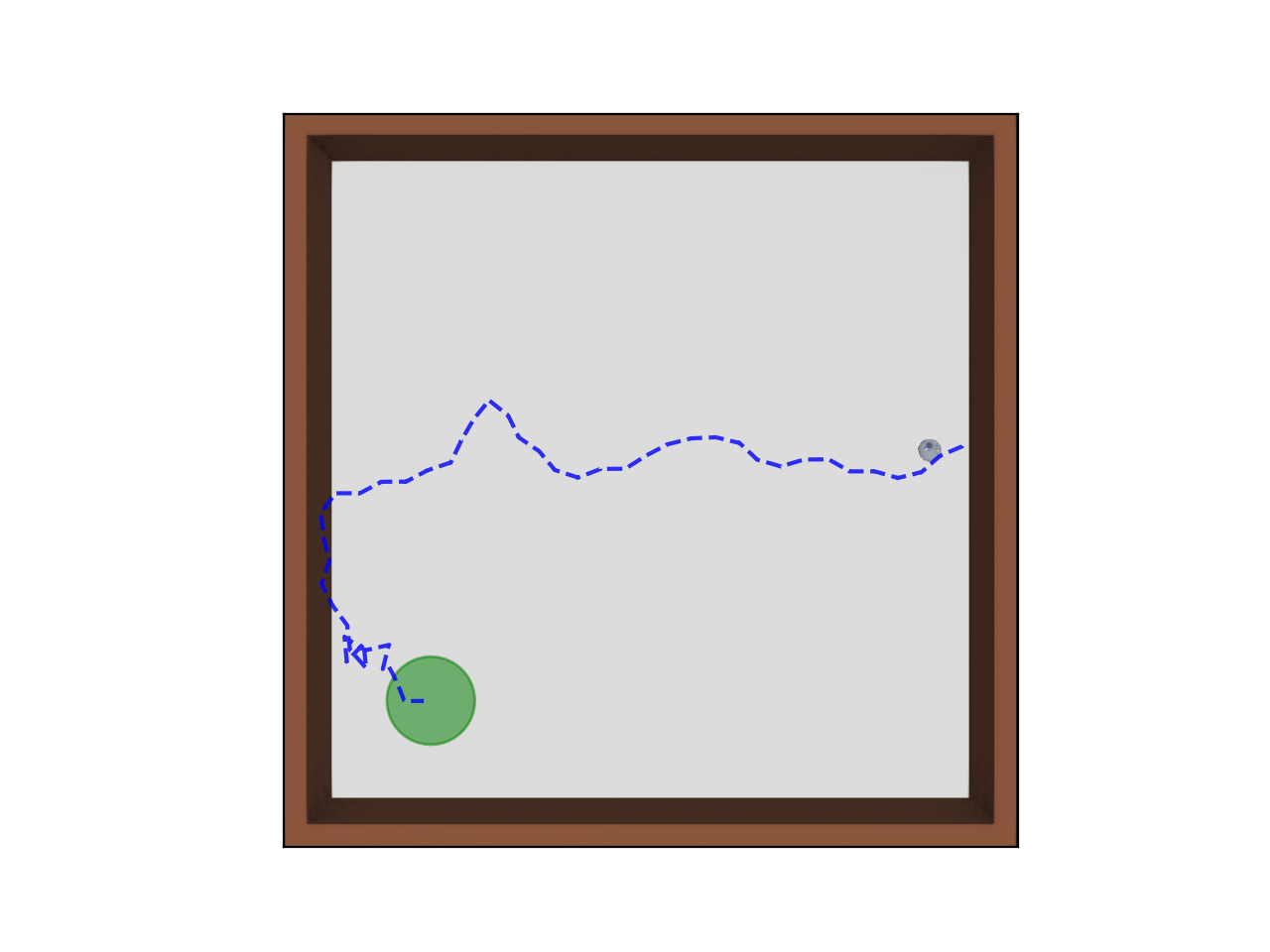}
        \includegraphics[clip,trim={2.5cm 1cm 2.5cm 0 },width=40mm ]{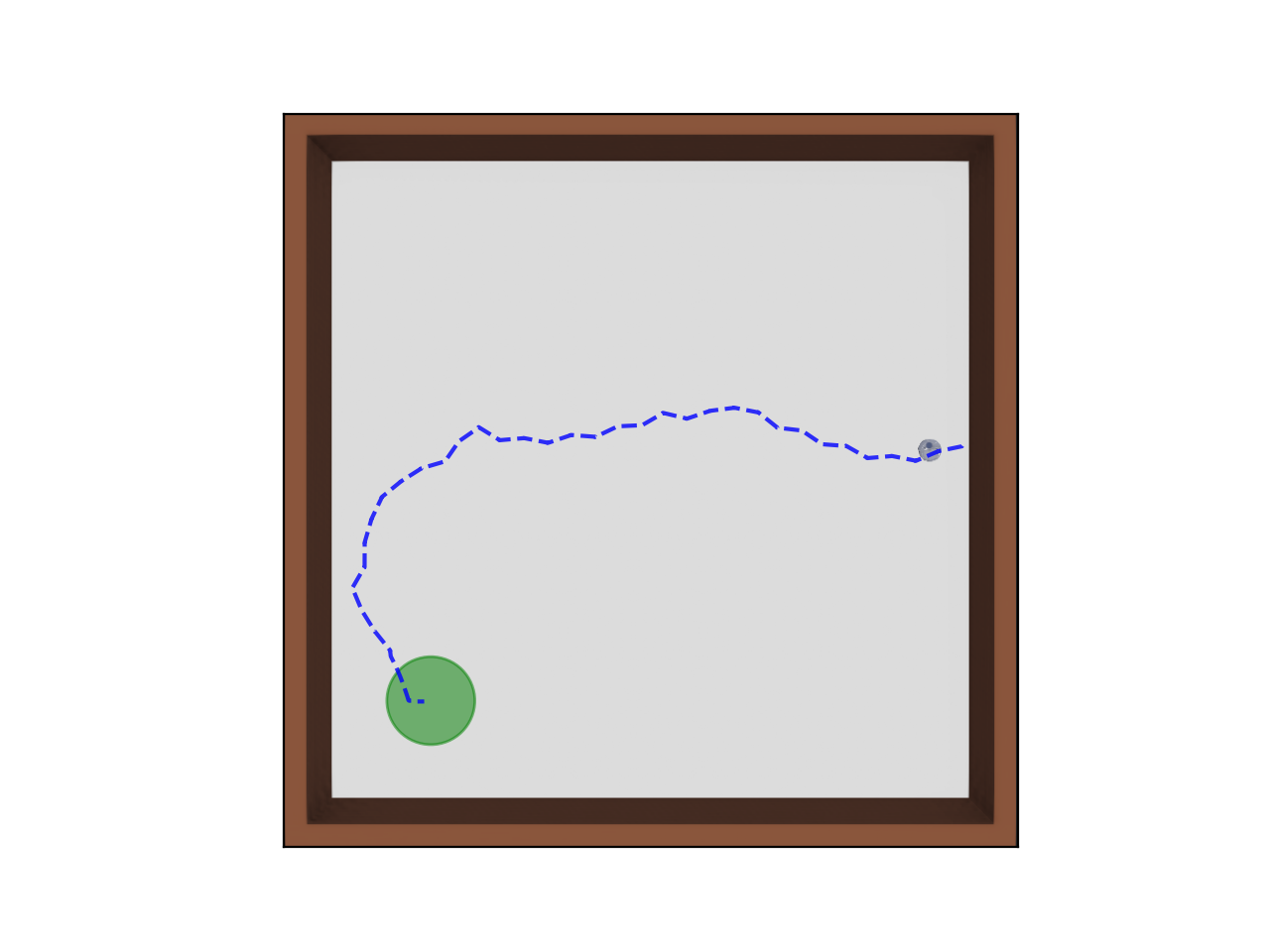}
        \includegraphics[clip,trim={2.5cm 1cm 2.5cm 0 },width=40mm ]{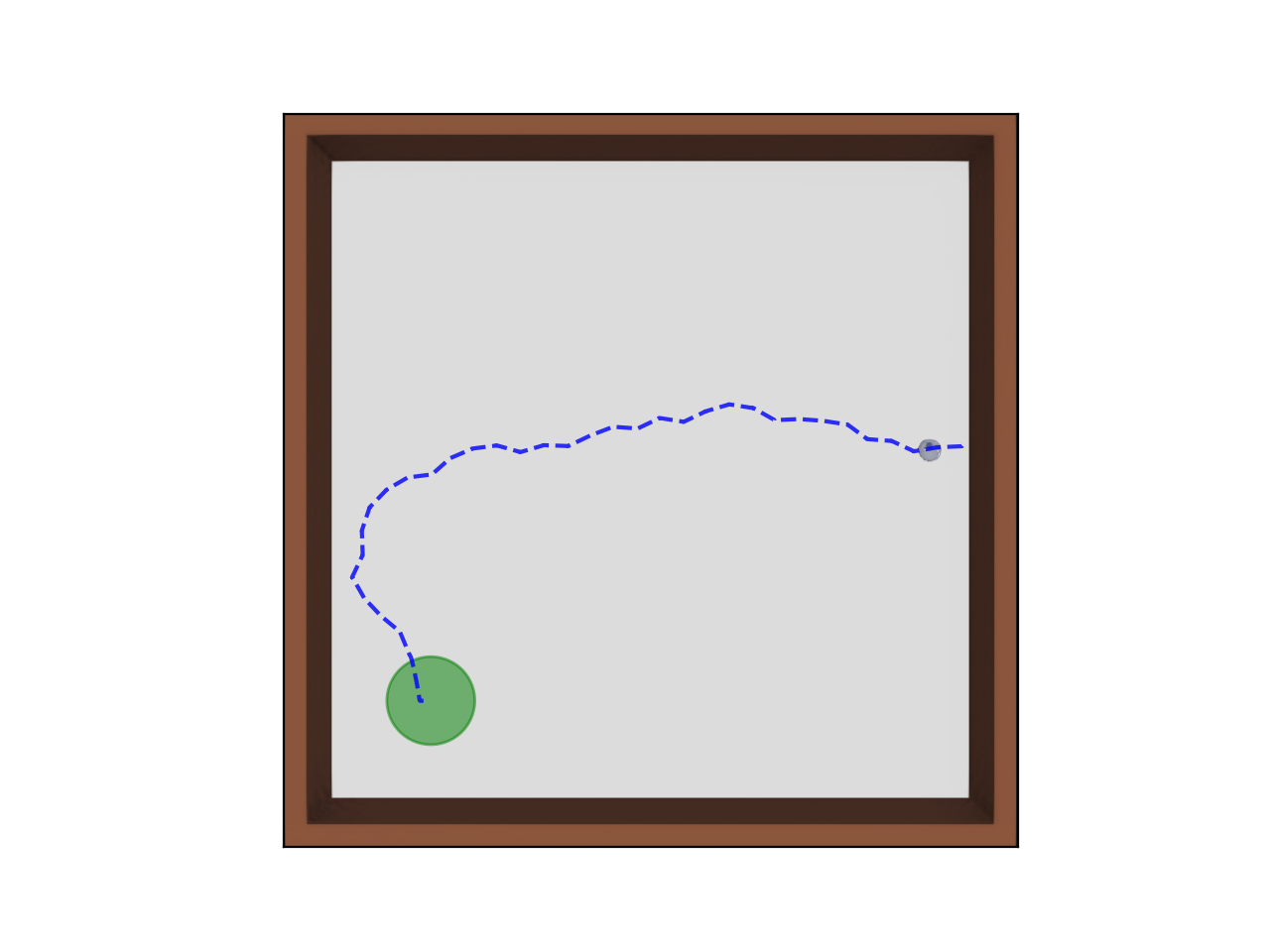}
        \caption{Trial 2}
    \end{subfigure}
    \begin{subfigure}[b]{.3\textwidth}
        \includegraphics[clip,trim={2.5cm 1cm 2.5cm 0 },width=40mm ]{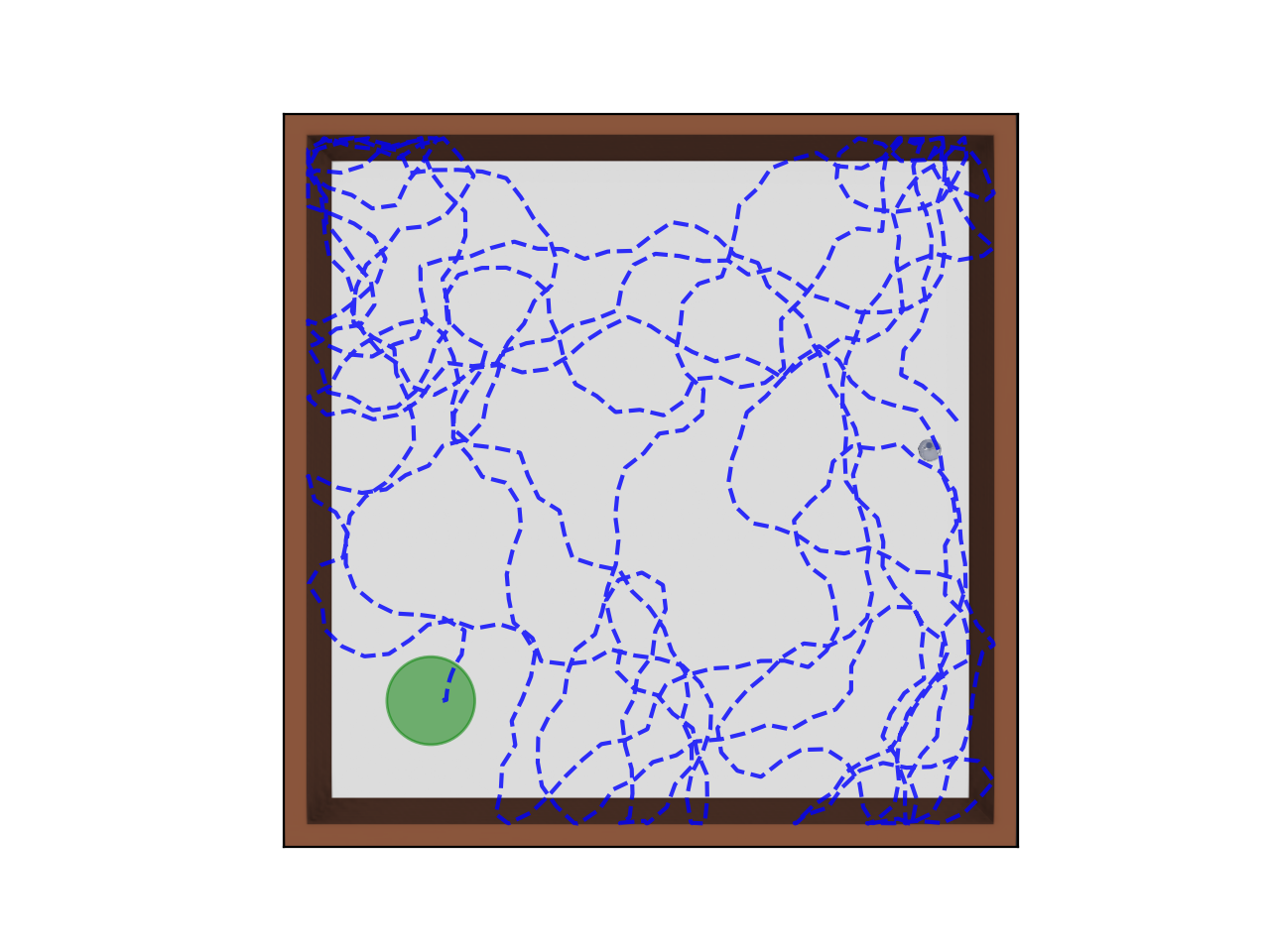}
        \includegraphics[clip,trim={2.5cm 1cm 2.5cm 0 },width=40mm ]{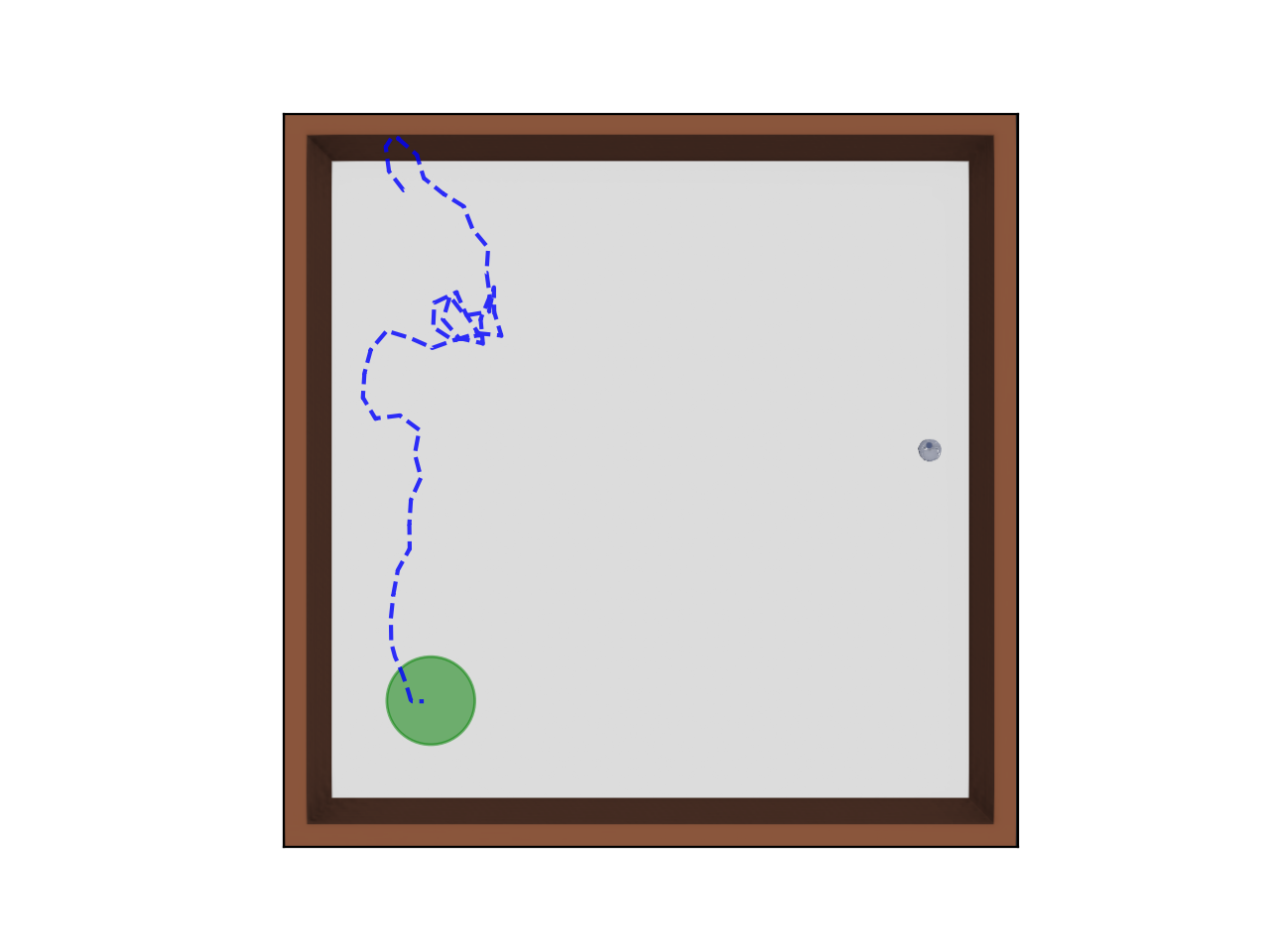}
        \includegraphics[clip,trim={2.5cm 1cm 2.5cm 0 },width=40mm ]{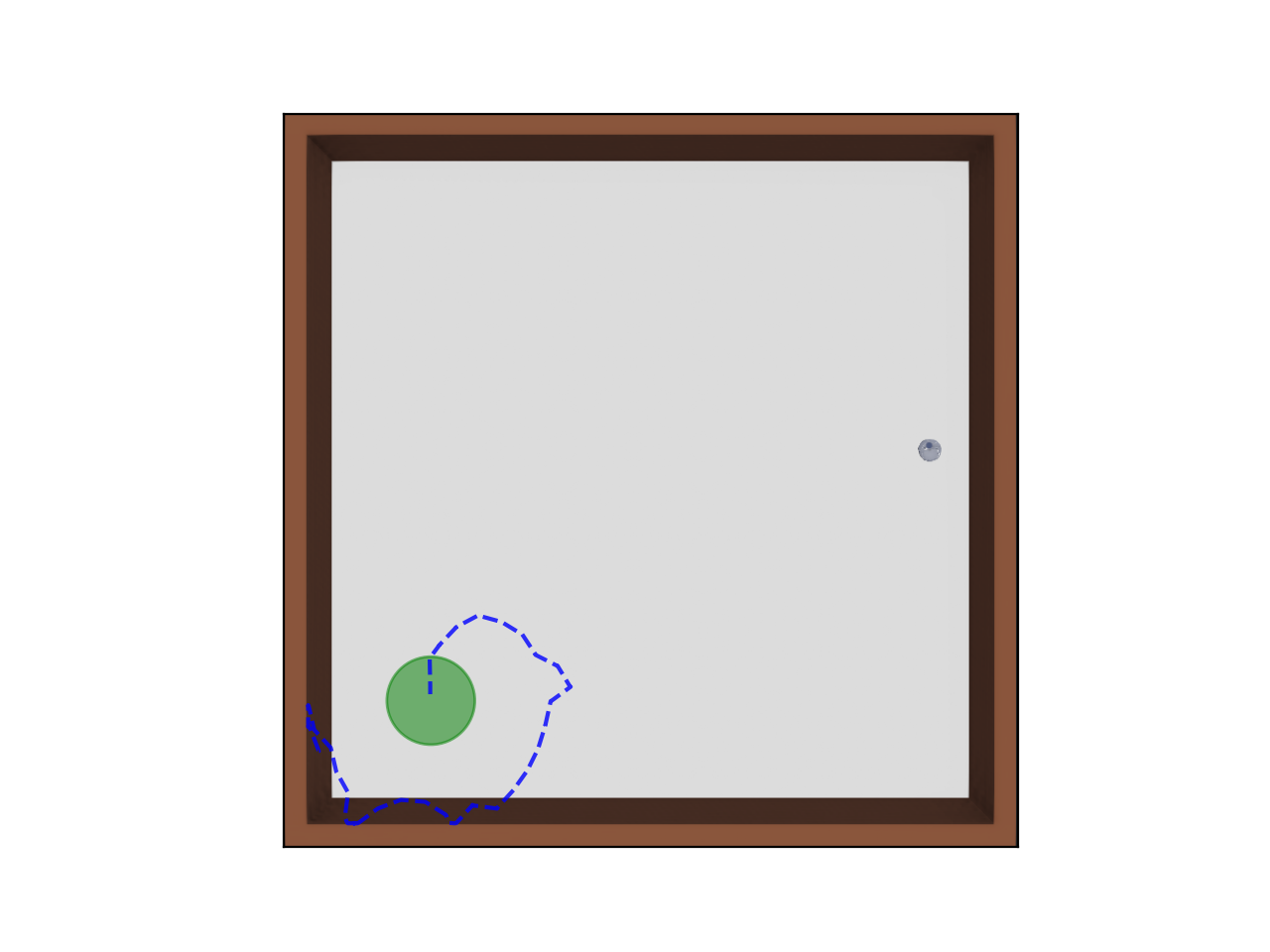}
        \includegraphics[clip,trim={2.5cm 1cm 2.5cm 0 },width=40mm ]{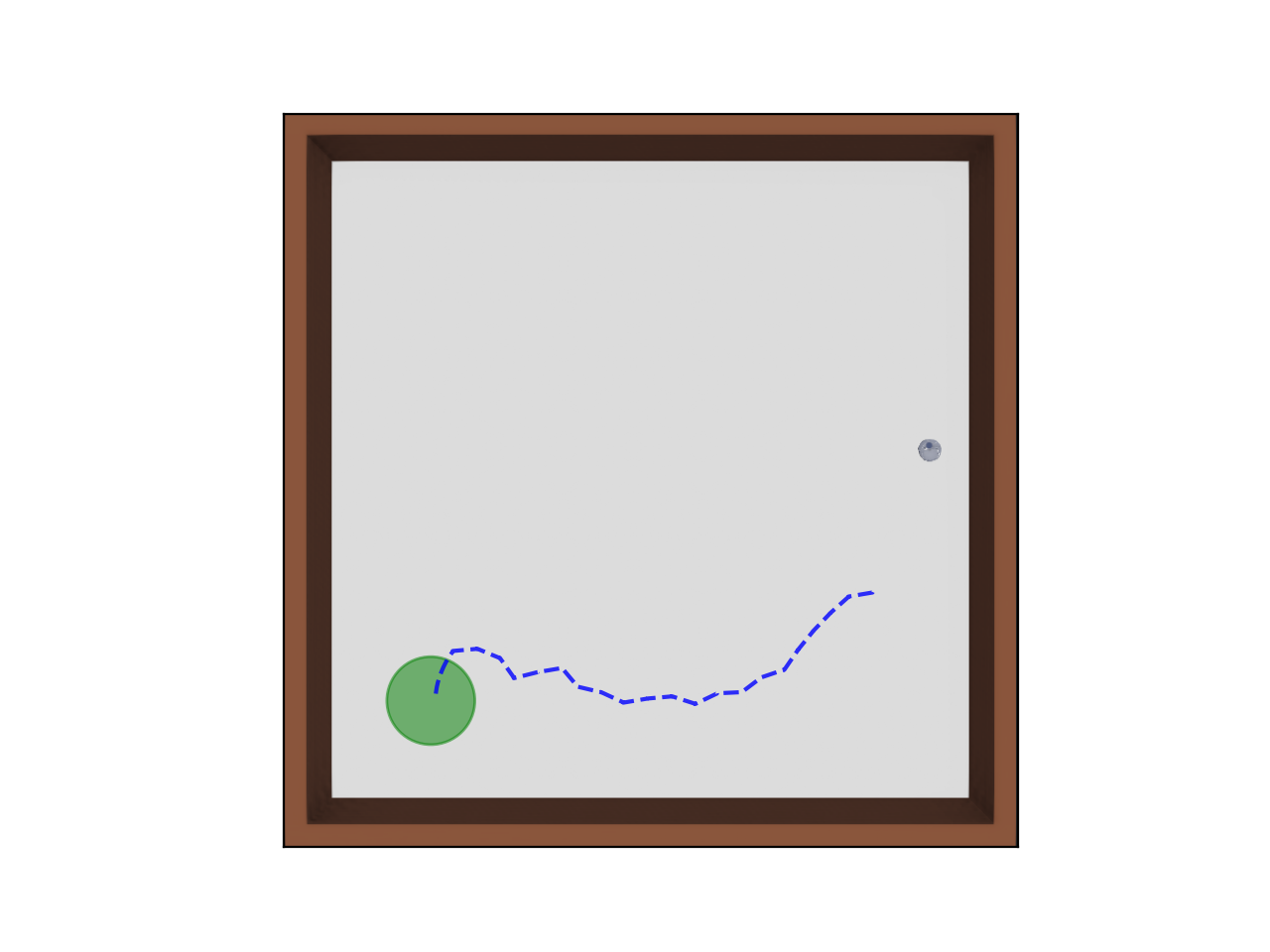}
        \caption{Trial 3}
    \end{subfigure}
    \caption{Three trials in the open maze showing the ability of the model to quickly learn and exploit a useful value map of the environment. In each trial, the robot is tasked with navigating to the goal shown in green (which is invisible to the robot) four times from the same starting point. On the first runs, it has no prior knowledge of the environment, however, subsequent runs exploit and build on the previously learnt maps.}
    \label{fig:same_start_open_maze}
\end{figure*}
\subsubsection{Exploration}
\label{subsec:exploration}
\label{subsec:replay}

In this mode, the agent explores randomly to map the environment. Place cells develop place fields using equation \eqref{eq:w_pb} and the environment's topology is learnt using equation \eqref{eq:w_pp}. On encountering a goal, the agent receives a reward, and initiates the backward replay state. Such time-compressed replay of experienced trajectories by rodents during consummatory behaviour has been observed experimentally \citep{buzsaki1989two,carr2011hippocampal}. The result of this process is to associate expected reward values with the recently experienced states by modifying the place-to-reward cell synapses as described next.

Upon receiving a reward, a specific reward cell is activated. On an initial encounter with a goal, this cell is chosen randomly, but is thereafter associated with that goal. The synapses from the currently active place cells to this reward cell are then potentiated. The currently active place cells then activate the place cells with adjacent place fields via the recurrent synapses that have been potentiated previously according to equation \ref{eq:w_pp}, and the place cell to reward cell synapses for these cells are also potentiated. Activity then spreads out through the recurrent place cell synapses generating a backward replay of place activity and associating it with the active reward cell. This continues for a certain time, causing the reward to become associated with the path taken to reach the goal. At each iteration, the total amount of potentiation in the place-to-reward cell synapses is subject to normalization by the total place cell network activity to enforce competition as well as a temporal decay from the time when the replay started.  This process is summarized in Algorithm \ref{algo:replay}.

\begin{algorithm}
\SetAlgoLined
$\Delta W^{rp} = 0$\;
\For{$t_r \gets 0$ to $n_{rs}$}{
    $\Delta W_{ij}^{rp} \mathrel{{+}{=}}\frac{\bar{v}_j^p}{||\bar{v}^p||_\infty}e^{\frac{-t_r}{\tau_r}}$\; 
    $\bar{v}_j^p = \tanh{\left(\left[\bar{v}_j^p + \sum_{l=0}^{n_p} (\max_k{W_{kjl}^{pp}})  \bar{v}_l^p\right]_+\right)}$\;
}
$W_{ij}^{rp} \mathrel{{+}{=}} \frac{\Delta W_{ij}^{rp}}{||\Delta W^{rp}||_\infty}$\;
\caption{Replay Generation Method}
\label{algo:replay}
\end{algorithm}

In Algorithm \ref{algo:replay}, the synaptic modulation term $\Delta  {W^{rp}}$ is initialized to $0$ for all synapses. For a preset number of time steps $n_{rs}$;
\begin{enumerate}
    \item Each synapse's modulation term is increased by its pre-synaptic place cell's firing rate max normalized and scaled by a temporal decay. This captures how active the place cell is at this point in the replay and the temporal decay serves to discount activity further from the reward.
    \item Place cells activate connected place cells along their most strongly connected direction - indicating the most efficient path between them.
\end{enumerate} 
After this two-step process is completed $n_{rs}$ times, the resulting synaptic modulation term $\Delta  {W^{rp}}$ is normalized by the maximum and added to $ {W^{rp}}$. Here (and in the preplay described below), the activation of place cell $j$ is denoted by $ {\bar{v}_j^p}$ to indicate that this activation is because of the agent's imagination and not because it is actually located in the place field of neuron $j$.

As the agent encounters a goal along more paths, more locations in the environment become associated with the reward at that goal, creating a \emph{reward map} of the environment with respect to that goal. Three aspects of this process are worth noting: 1) The normalization serves to ensure that place cells that are activated later in the replay and are thus presumably further away have their connections to the reward cell potentiated less, thus causing the reward map to encode the distance to the reward; 2) Given that activity spreads out along experienced paths, the learnt reward map is a topological rather than a metric map enabling it to take obstacles into account to the extent they are experienced; and 3) Because place fields have significant width, the map covers not only the exact paths traversed but also a swath of locations around them, resulting in significant generalization to unvisited locations. As a result of this {\it field effect}, a reasonably accurate and actionable reward map is built for a given goal even with a small amount of exploration though, of course, the reward map only covers the general region that has been explored.

\subsubsection{Exploitation}
\label{subsec:exploitation}
In previously explored environments, the agent exploits the learnt place field connectivity and reward map values to navigate to the reward locations. It achieves this by using a 1-step preplay to recall the outcome of taking the actions available in $ {\theta^h}$ and evaluating the expected reward value of the resulting states using the reward map. 

The agent {``imagines''} the outcomes of taking actions by letting activity spread in the place cell network via the recurrent synapses that have been potentiated through exploration. The place cells encoding its current location activate the place cells encoding locations that have been learnt to be adjacent along the direction of the action being evaluated. This imagined place cell activity then drives the reward cell network giving a measure of the value of the imagined states, allowing the robot to pick the action that maximizes the expected value state. 

The imagined activation of place cell $i$ denoted by $  {\Bar{v}_i^p}\, |\, a$ is defined in \eqref{eq:v_p_p}. The expected reward value, $ {\Bar{v}_g^r}\, |\, a$, of the imagined state is computed by allowing the imagined place cell network activity, $ {\Bar{v}}^p$ to activate the reward cell for the current goal $g$ via the place to reward cell synapses and normalizing this by the sum of $ {\Bar{v}_i^p}$. The normalization is done so that the potential reward evaluation can be made relative to other options without any bias from the total place cell network activity. We represent the effective weight matrix as $ {\Bar{W}_{gi}^{rp}}$ for the duration of the exploitation episode. It is initialized as $ {{W}_{gi}^{rp}}$. Thus for all $a \in  {\theta^h}$:

\begin{equation}
\label{eq:vr_bar}
     {\Bar{v}_g^r}\, |\, a = \frac{\sum_{i=0}^{n_p}  {\Bar{W}_{gi}^{rp}}  {\Bar{v}_i^p}}{\sum_{i=0}^{n_p}  {\Bar{v}_i^p}}
\end{equation}

With the expected reward values of each action estimated, the optimal action $a^*$ is then computed as the circular mean direction of the expected reward values as shown in \eqref{eq:a*} below. The optimal action is taken when $\max_a \left( {\Bar{v}_g^r} | a \in  {\theta^h}\right) \geq \xi$ otherwise the robot explores. Here, $\xi$ is a (possibly stochastic) threshold that determines the explore/exploit trade-off.
\begin{equation}
    a^* = \arctan( \sum_{j=1}^n  {\Bar{v}_g^r} |  {a_j}\sin{ {a_j}}, \sum_{j=1}^n   {\Bar{v}_g^r} |  {a_j}\cos{ {a_j}})
    \label{eq:a*}
\end{equation}

where $n$ is the number of possible step directions, and $a_j \in  {\theta^h}$. Synapses from active place cells are temporarily depressed for the rest of the episode by a constant factor $\lambda$ to disincentivize the agent from revisiting locations.
\begin{equation}
    \label{eq:eff_rew_value}
     {\Bar{W}_{gi}^{rp}} =  {\Bar{W}_{gi}^{rp}} - \lambda  {\Bar{W}_{gi}^{rp}} {v_i^p} 
\end{equation}

After each action $ {v}_i^p$ is reset from the imagined state to actual state according to \eqref{eq:place_cells}. 

\subsection{Parallels with Traditional Reinforcement Learning}
% post edit
The components of the model for neural reinforcement learning described above have many similarities with the traditional Markov Decision Process (MDP) formalism of RL, where an MDP is a 4-tuple $\left(\mathcal{S}, \mathcal{A}, \mathcal{T}_a, \mathcal{V}\right)$. The model in this paper is specialized to a place cell-based spatial representation and navigation task, using specific features such as network connectivity and replay of neural activity, and is, thus, not intended to be a general-purpose state-of-the-art RL model. Nevertheless, it might be useful for the reader to view the model through the lens of the RL framework. The MDP variables and their parallels in the model are described below:

\subsubsection{$\mathcal{S}$ - \textit{state space}}
This is the set of agent-environment states. In the model, the place cells encode the possible locations or states of the agent. Thus $\mathcal{S}$ can be taken as $ {v^p}$ the vector of place cell firing rates..

\subsubsection{$\mathcal{A}$ - \textit{action space}}
This is the set of actions that are available to the agent. While we left this unconstrained during exploration, we restrict this to $ {\theta^h}$ - the preferred head directions of the head direction cells i.e the eight allocentric cardinal directions during exploitation. This gives a discrete action space as is the case in most RL models.

\subsubsection{$\mathcal{T}_a$ - \textit{state transition function}}
This gives the expected state if an action is taken from the current state. During exploration, the place cell recurrent synapses are potentiated in an action specific manner to learn the expected network activity if an action is taken from an initial network state. As we've already stated that the place cell activity can be seen as the state, the place cell recurrent synapses $ {W^{pp}}$ can be taken as the state transition function.

\subsubsection{$\mathcal{V}$ - \textit{value function}}
This gives a measure of the expected reward starting from the current state. This is typically a value that is discounted by the number of steps required to reach a rewarded state from the current state. Through the replay process, the place to reward cell synapses encode a proxy of this value. Consequently, the firing rate of the reward cell associated with the currently sought reward due to place cell input is a direct analog of the value function. It is assumed that a higher level brain structure that is not explicitly modeled here assigns reward cells to the rewards discovered, and selects those reward cells for evaluation when seeking known goals.

\begin{table}
\centering
 \begin{tabular}{c c  c}  
 \hline
  \hline
 Values & MDP & Proposed Model\\[.5ex]
  \hline

 State Space & $\mathcal{S}$ & $ {v^p}$ \\ 

 Action Space & $\mathcal{A}$ & $ {\theta^h}$ \\

 Transition Probability Function & $\mathcal{T}_a$  & $ {W^{pp}}$\\ 

 Reward Function & $\mathcal{V}$ & $ {v^r}$ \\
 \hline
 \hline
\end{tabular}
\caption{Interpretation of the proposed model as an MDP}
\label{table:MDP_table}
\end{table}

\section{Experimental Results}
\label{sec:result}
\subsection{Experimental Setup}
\label{subsec:experiment}
The model was deployed on a simulated iRobot Create 2 robot, equipped with an omnidirectional rangefinder in Webots~\citep{Webots} -- an open-source robot simulation platform. The standard delayed-match-to-place (DMTP) procedure of the Morris water maze task~\citep{vorhees2006morris, d2001applications, brandeis1989use} was followed, with the agent seeking to find a disk-shaped goal that becomes visible only when reached. Unlike the original water maze experiment, which used a cylindrical environment, the simulations here is a square one. In addition to simulating the classic water maze test with an obstacle-free open environment, simulations were also carried out with various obstacle configurations that required the agent to learn topological rather than metric maps. 

Ideally, the model agent would develop place cells in is initial exploration of the environment. However, the small size of the simulated model meant that the number of place cells competing for activity at any one time was small, resulting in place fields developing \textit{``tails"} along the receptive fields of their BVC~\citep{barry2006boundary}. To overcome this, an initial \textit{``dry run exploration''} without the goal present was done where only the BVC-to-place cell synapses were plastic to \textit{``clean''} the place fields. This step would not be needed with enough place cells enforcing competition for place fields. The robot was then reintroduced into the same environment and allowed to explore without any restrictions on the network processes previously described. Upon finding the goal, the robot was reinitialized in the environment to exploit its learnt (partial) value map. This was done three times, making a total of four runs -- the first being naive.

\subsection{Open Maze Performance}
\begin{figure}
    \centering
    \includegraphics[width=.8\columnwidth]{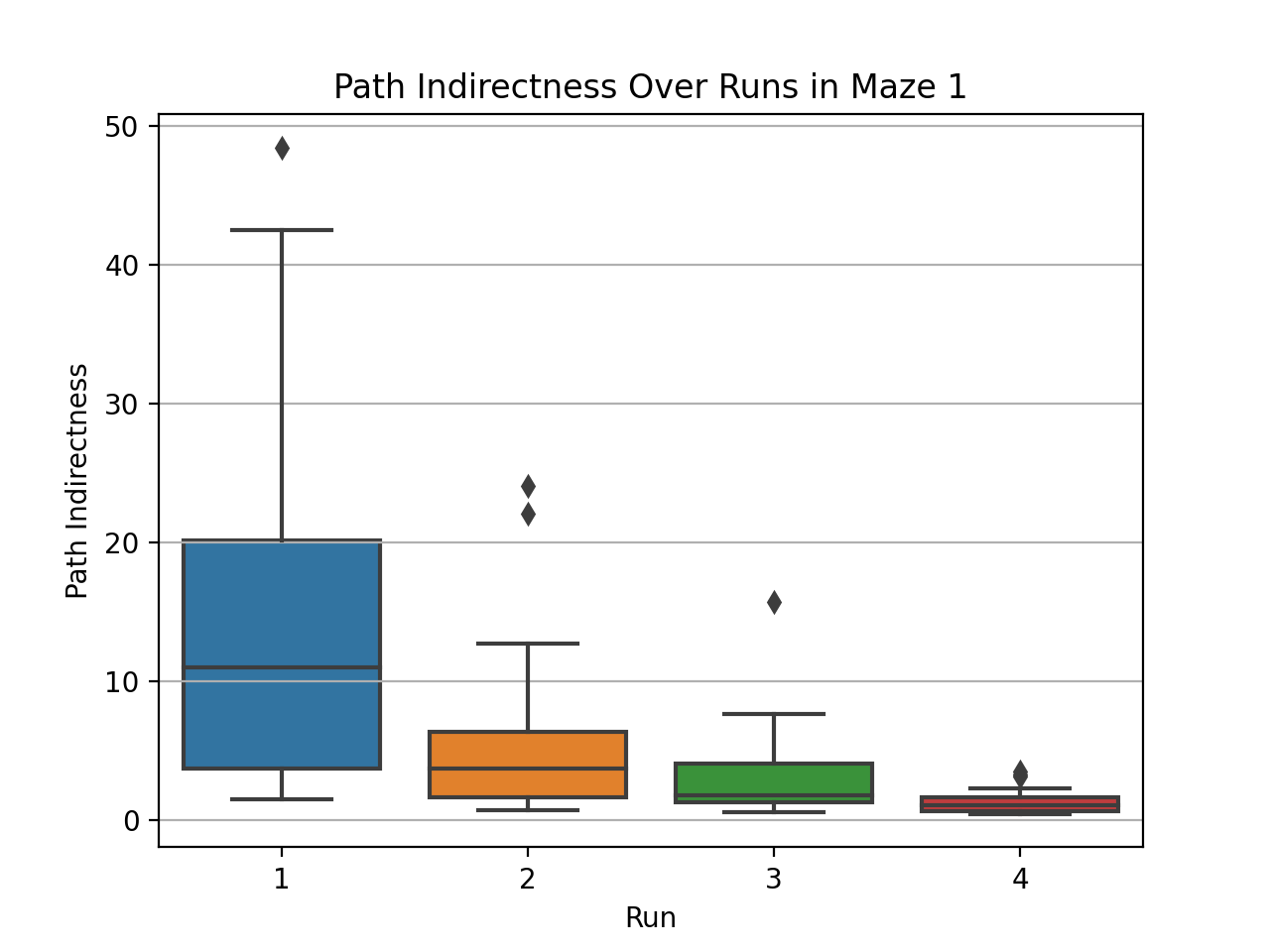}
    \caption{The path indirectness over the four runs is depicted here with a box plot. It shows a negative trend - indicating the ability of the robot to learn a reward map from experience and exploit it to compute more direct paths.}
    \label{fig:path_efficiency}
\end{figure}

\begin{figure}
    \centering
    \includegraphics[width=.8\columnwidth]{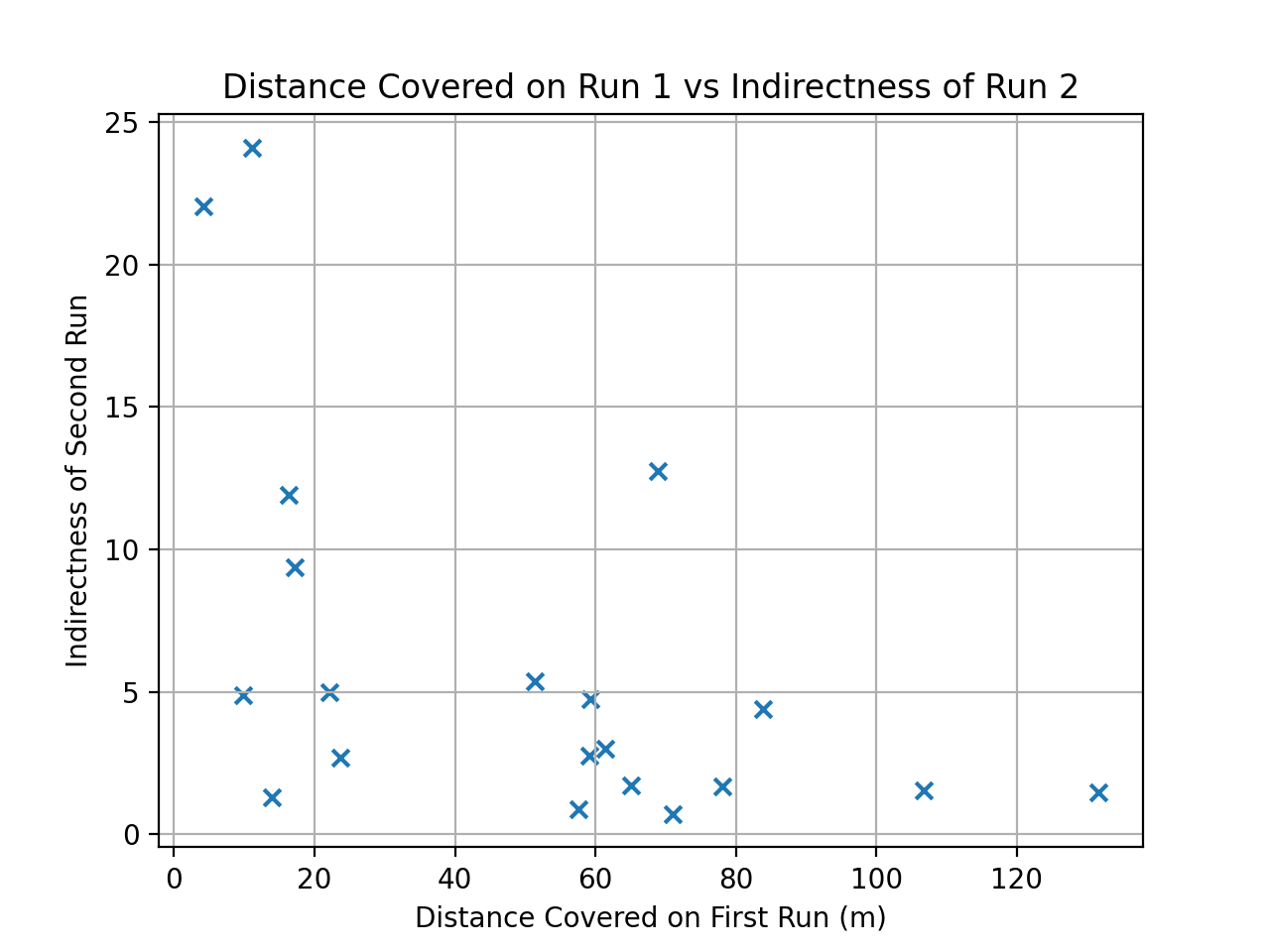}
    \caption{The path indirectness on the second run shows an inverse correlation to the distance covered on the first run. This indicates that covering more ground on the initial run makes it more likely that the starting location of the second run has previously been encountered and thus incorporated into the reward map.}
    \label{fig:dist_v_direct}
\end{figure}

The model was first deployed and evaluated in an open maze environment with no obstacles. Starting from the same location, the robot was able to improve its path over the four runs. Having no prior experience of the maze, the first run showed random undirected searching. Once the goal had been found the first time, the replay process enabled the creation of an implicit reward value map which is visualized in Fig. \ref{fig:reward_maps}. This value map was then exploited on subsequent runs so that the robot follows the previously traversed locations that connect the starting location to the goal. The efficiency of the computed path is constrained by the previously learnt place field connectivity - in turn constrained by the size of the place fields themselves.

Fig. \ref{fig:same_start_open_maze} illustrates three sample runs. On the first two runs, the robot starts from the same location towards the same goal location in each trial. After finding the goal by random exploration on the first run, the robot takes progressively more direct paths towards the goal by exploiting its implicit reward map shown in \ref{fig:reward_maps}. In trial 3, the robot begins each run from a different location with the same goal location. Since the first run involved wide sampling of the maze before finding the goal, an extensive reward map is built and the subsequent runs from random spots were directly towards the goal.

To further quantify the performance of the model, 20 trials were run where the robot started from a different random location on each of the four runs. On each run, a \textit{path indirectness} metric was computed to measure the efficiency of the path taken. This is calculated as the \emph{relative excess length} of the path taken compared to the length straight line path from the starting point to the closest point on the edge of the goal disk, which is always the shortest path in an open environment:
\begin{subequations}
    \begin{equation}
        d = \sum_{t=0}^{T-1} || x(t+1) - x(t) ||_2 \,dt
    \end{equation} 
        \begin{equation}
        d^* =  || {x}(T) -  {g^x}||_2 - \rho
    \end{equation} 
    \begin{equation}
        indirectness = \frac{d - d^*}{d^*}
    \end{equation} 
\end{subequations}
where $d$ is the length of the path taken, $d^*$ is the length of the shortest path to the edge of the goal, $T$ is the duration of the run, $ {x}$ is the robot's trajectory in Cartesian coordinates, $ {g^x_i}$ is the center of the goal location, and $\rho$ is the goal radius.

The results obtained are shown as box plots in Fig. \ref{fig:path_efficiency}. As expected, the path indirectness is seen to decrease over the four runs, given that more of the environment has been learnt on previous runs and a more direct path can be computed. The greatest improvement. however. occurs from the first run to the second as most information is typically learnt in the first run. First runs that result in the robot finding the goal quickly with little exploration are more likely to cause subsequent runs to have higher indirectness values because the initial run did not enable the robot to build a full map. Fig. \ref{fig:dist_v_direct} shows the indirectness on the second run plotted against the total length of the path taken on Run 1. A negative Pearson correlation of $-0.53$ was observed, in line with expectations.

\begin{figure*}
    \centering
    \begin{subfigure}[b]{.3\textwidth}
        \includegraphics[clip,trim={2.5cm 1cm 2.5cm 0 },width=40mm ]{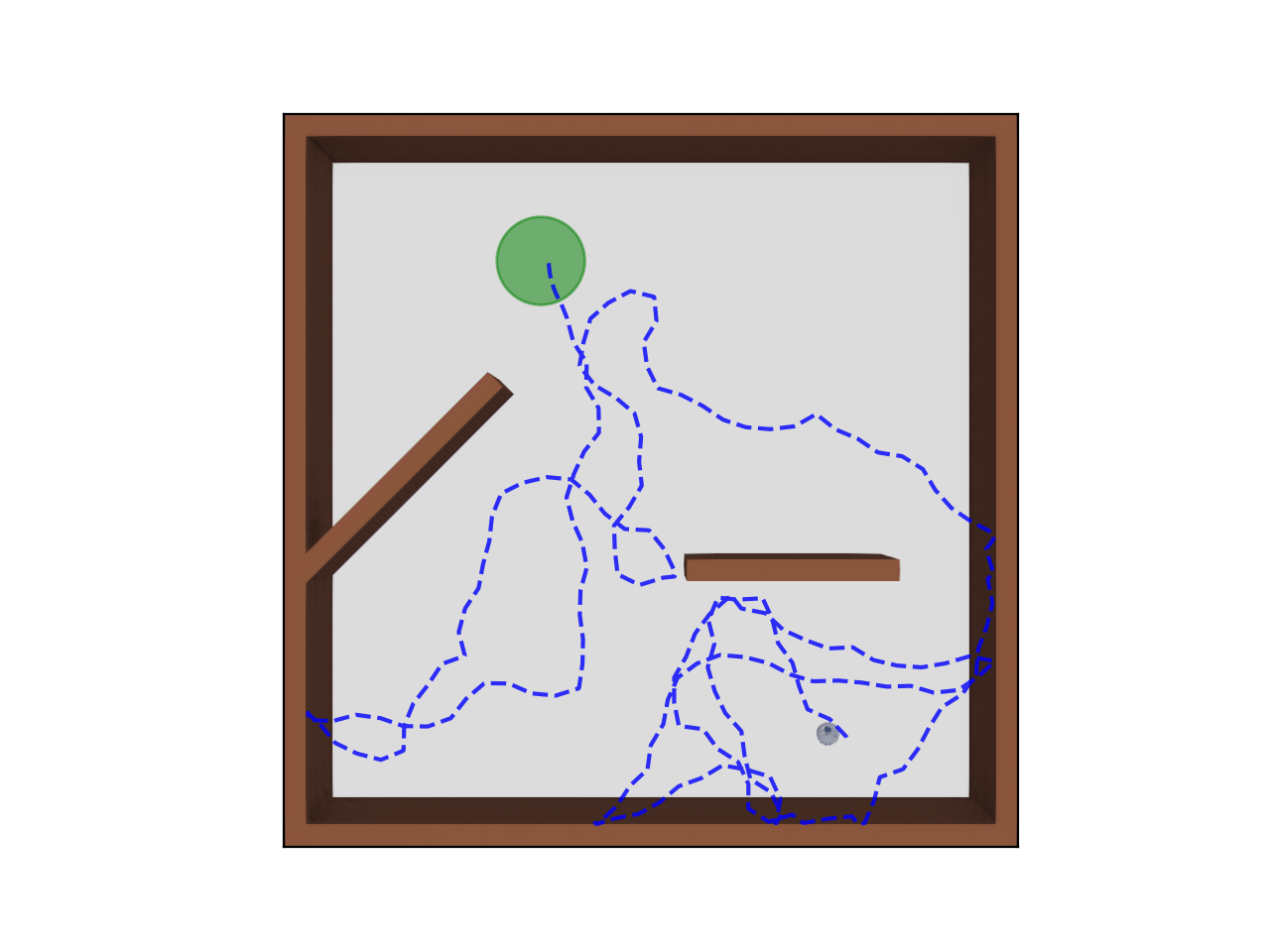}
        \includegraphics[clip,trim={2.5cm 1cm 2.5cm 0 },width=40mm ]{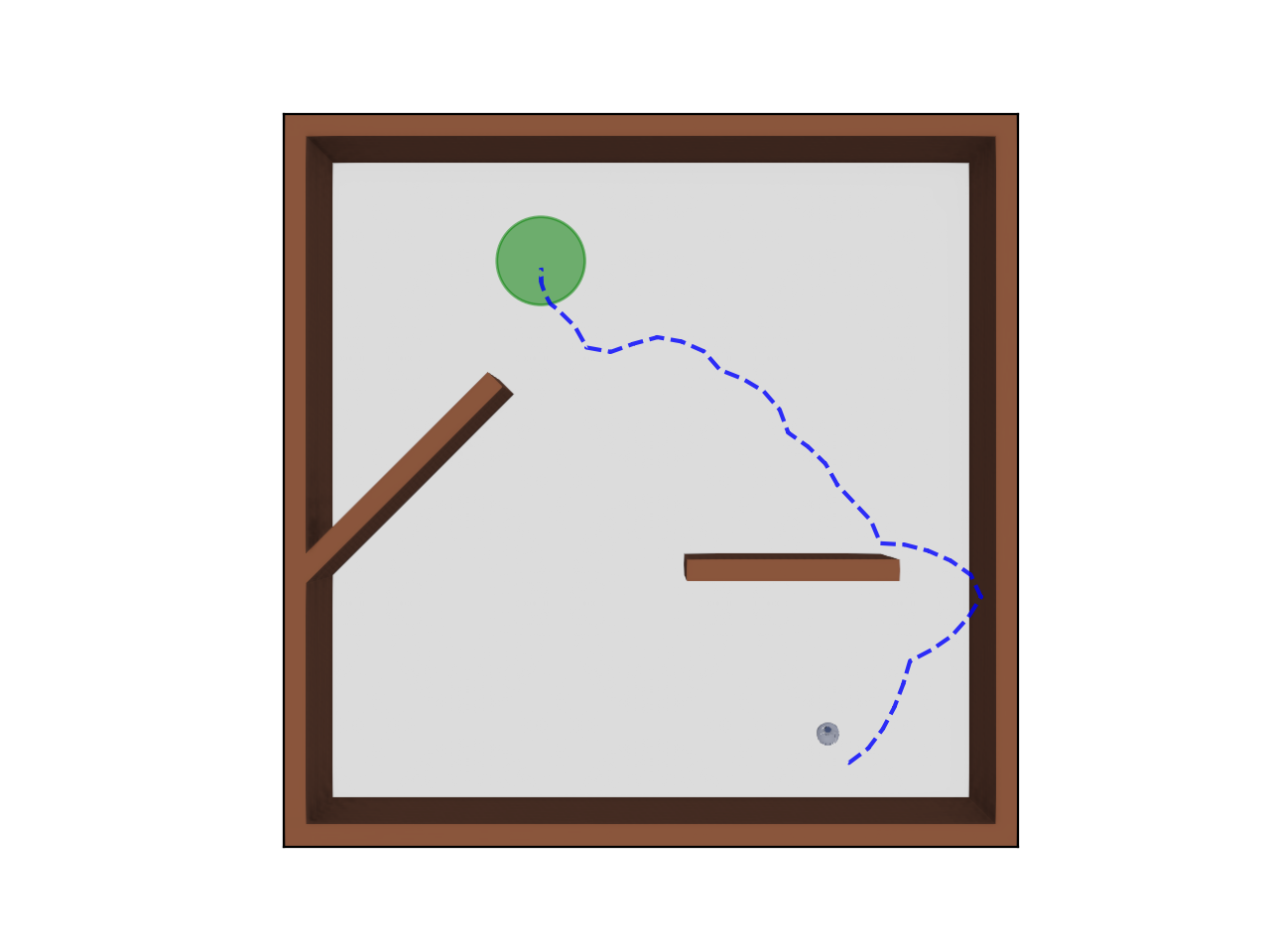}
        \includegraphics[clip,trim={2.5cm 1cm 2.5cm 0 },width=40mm ]{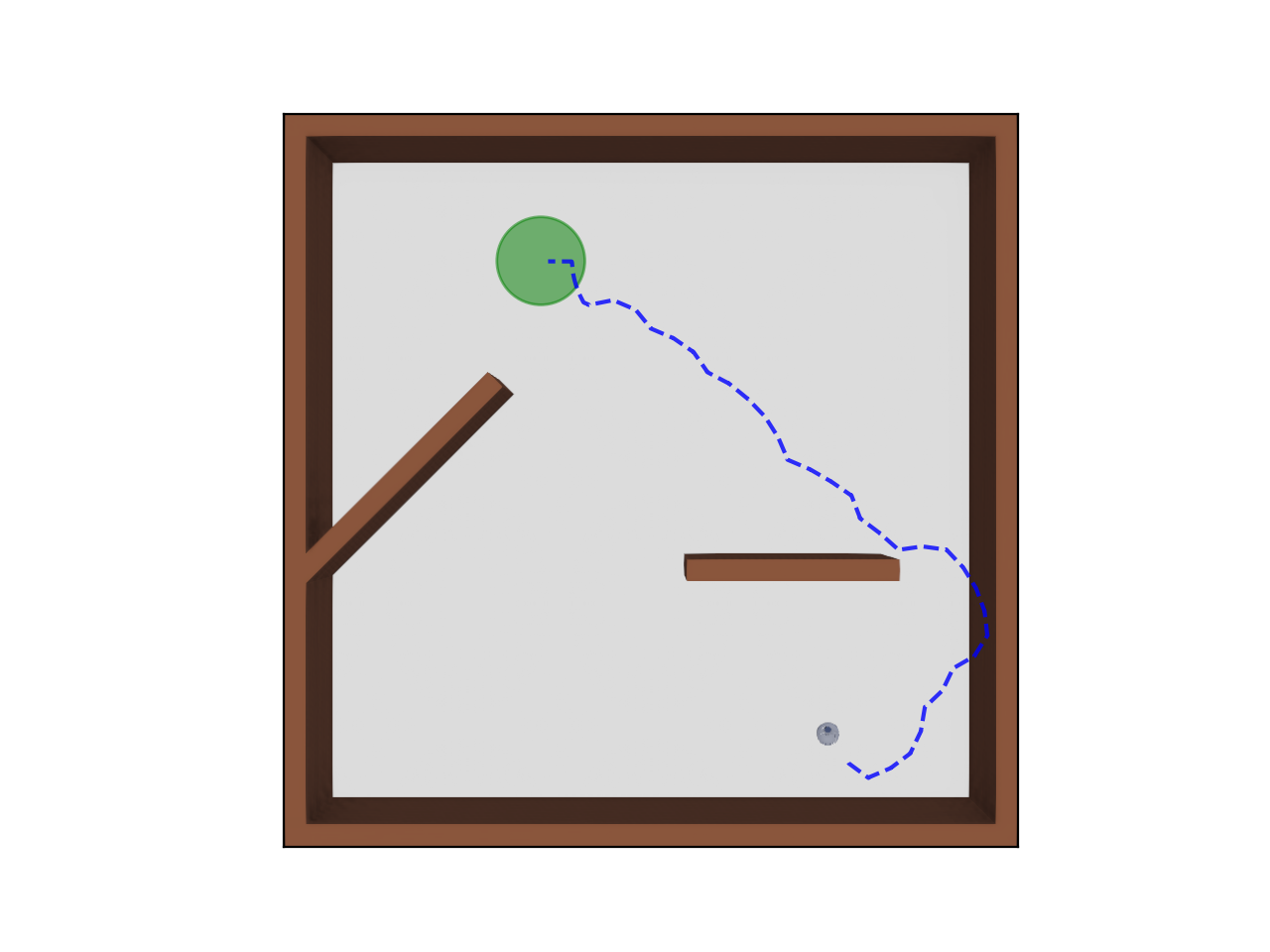}
        \includegraphics[clip,trim={2.5cm 1cm 2.5cm 0 },width=40mm ]{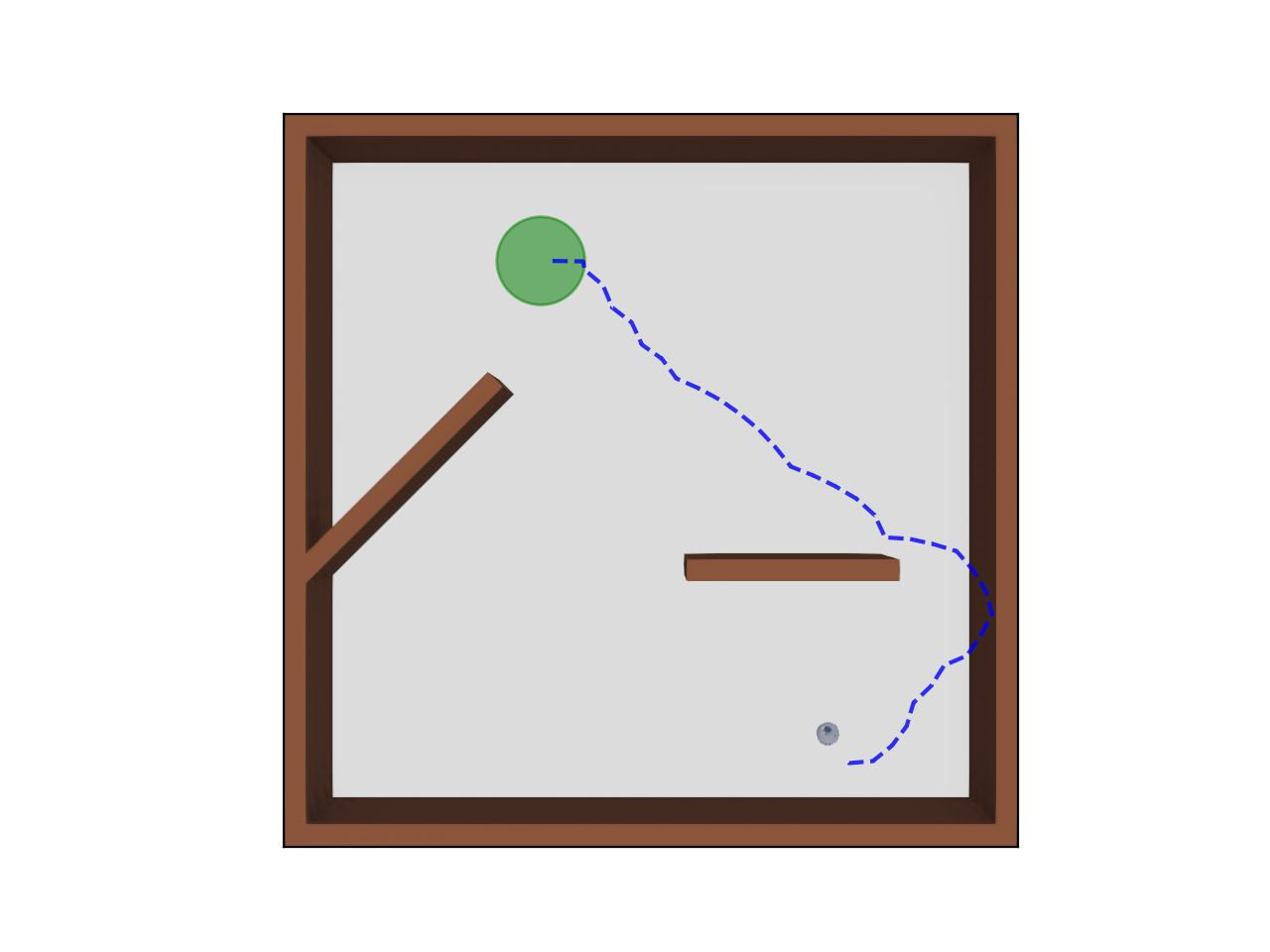}
        \caption{Maze 1}
    \end{subfigure}
    \begin{subfigure}[b]{.3\textwidth}
        \includegraphics[clip,trim={2.5cm 1cm 2.5cm 0 },width=40mm ]{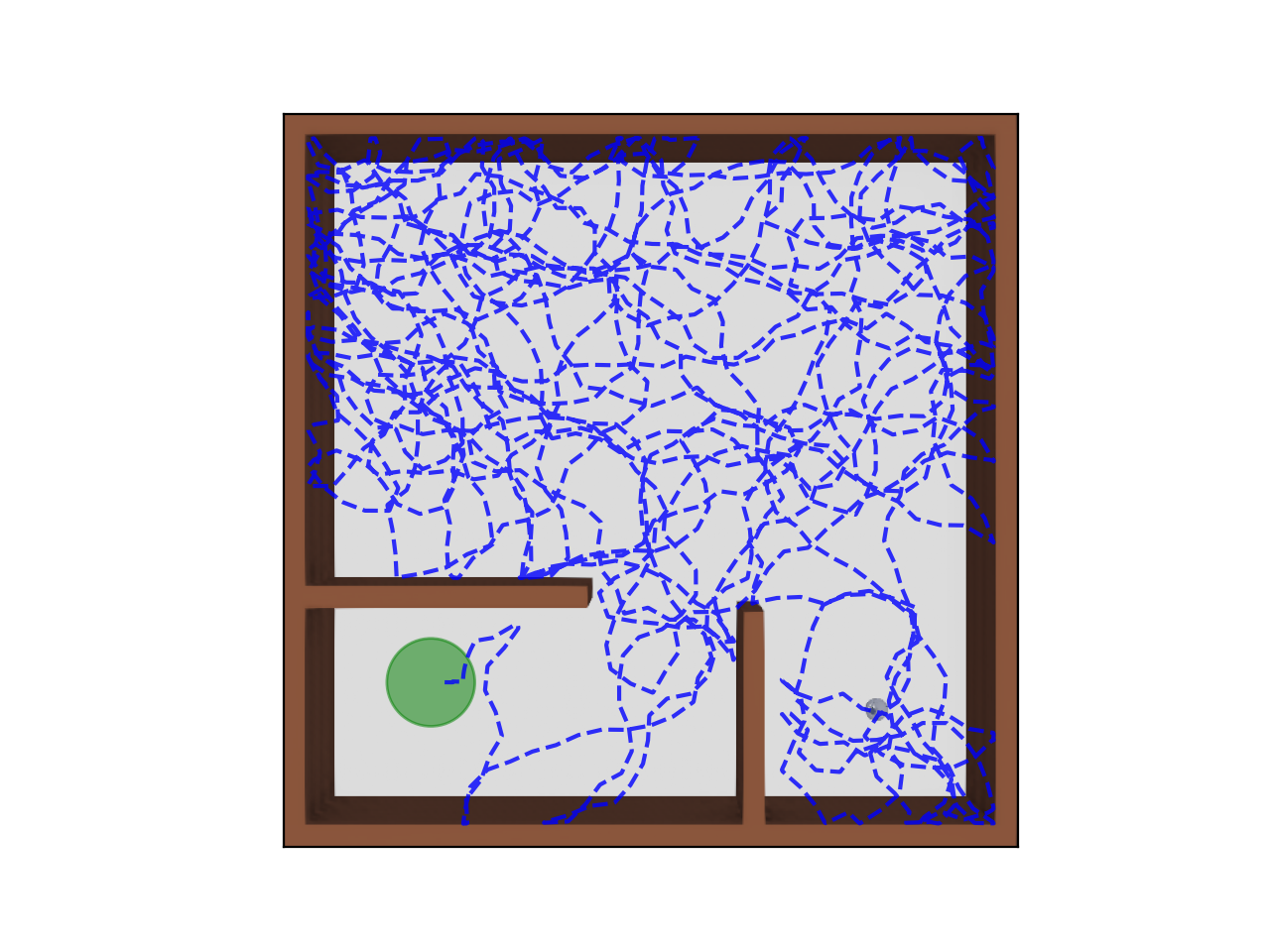}
        \includegraphics[clip,trim={2.5cm 1cm 2.5cm 0 },width=40mm ]{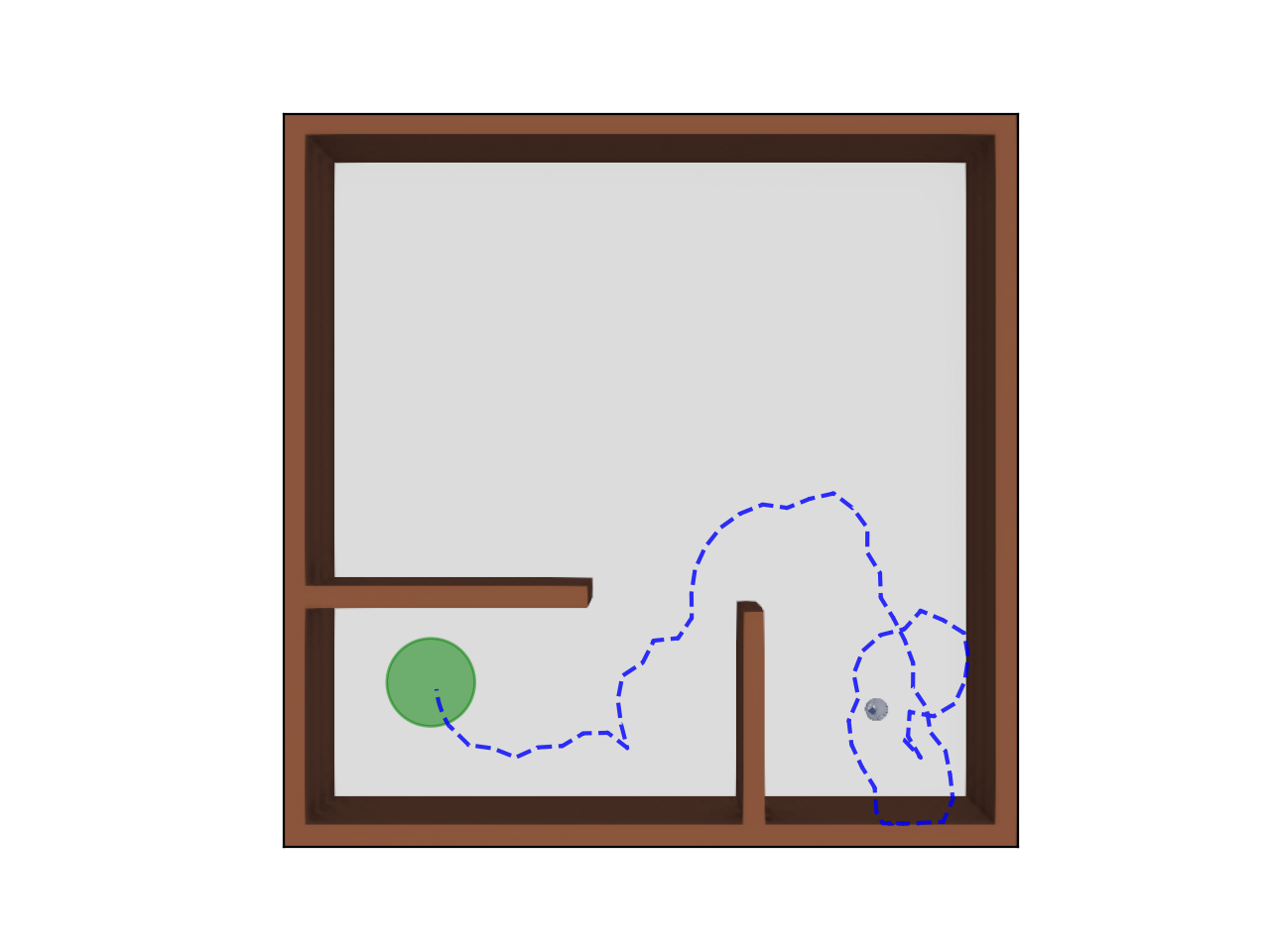}
        \includegraphics[clip,trim={2.5cm 1cm 2.5cm 0 },width=40mm ]{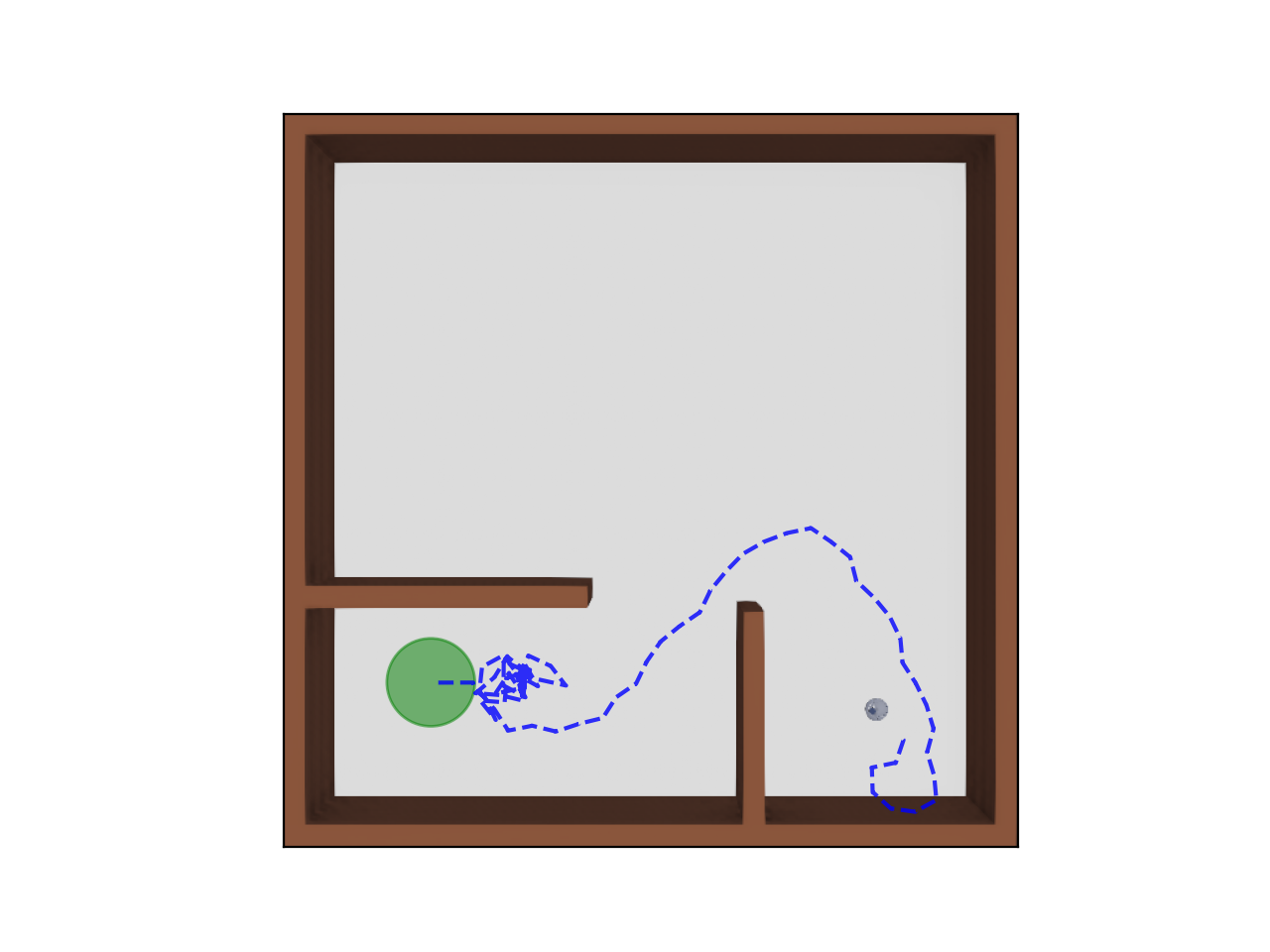}
        \includegraphics[clip,trim={2.5cm 1cm 2.5cm 0 },width=40mm ]{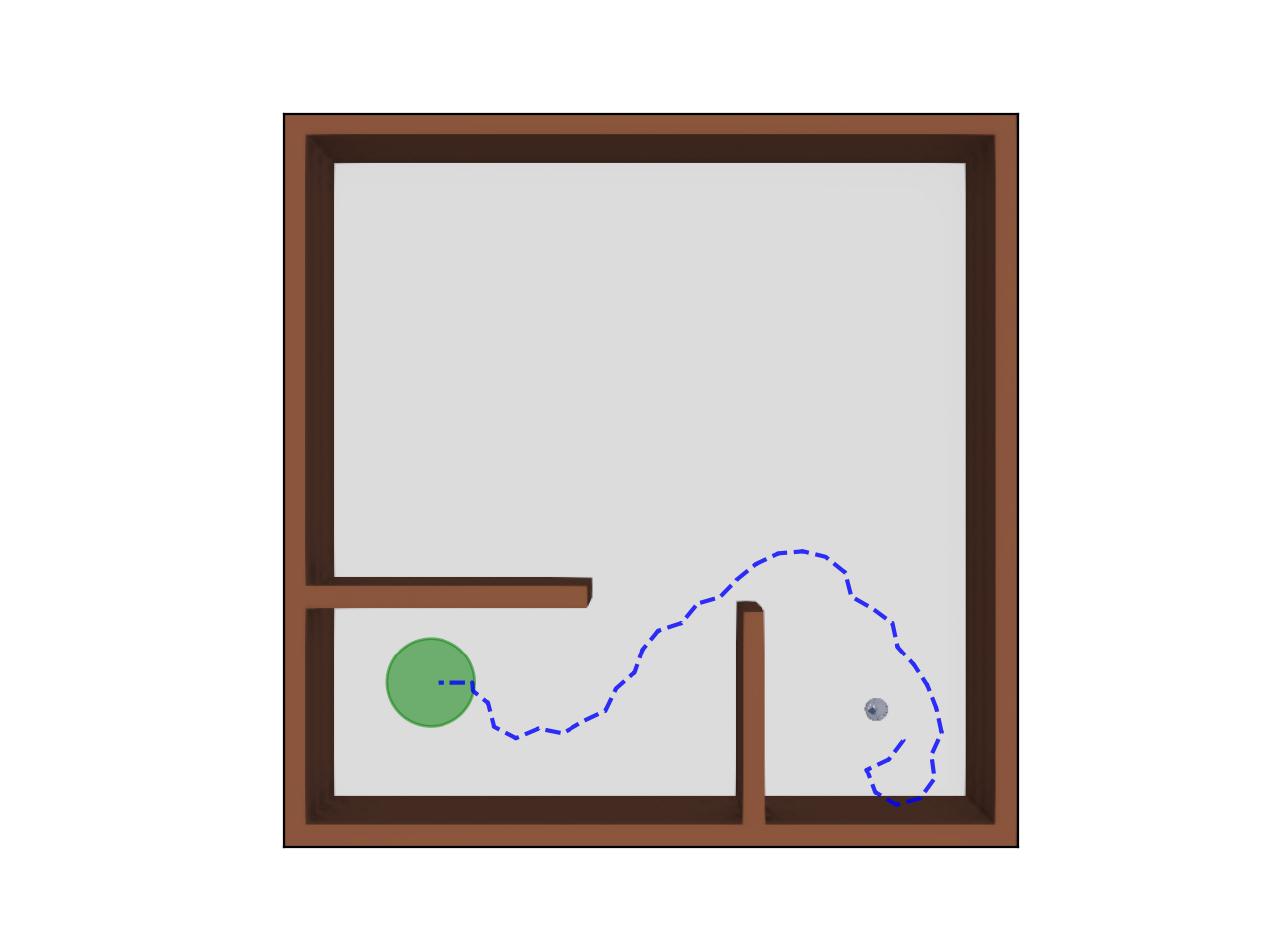}
        \caption{Maze 2}
    \end{subfigure}
    \begin{subfigure}[b]{.3\textwidth}
        \includegraphics[clip,trim={2.5cm 1cm 2.5cm 0 },width=40mm ]{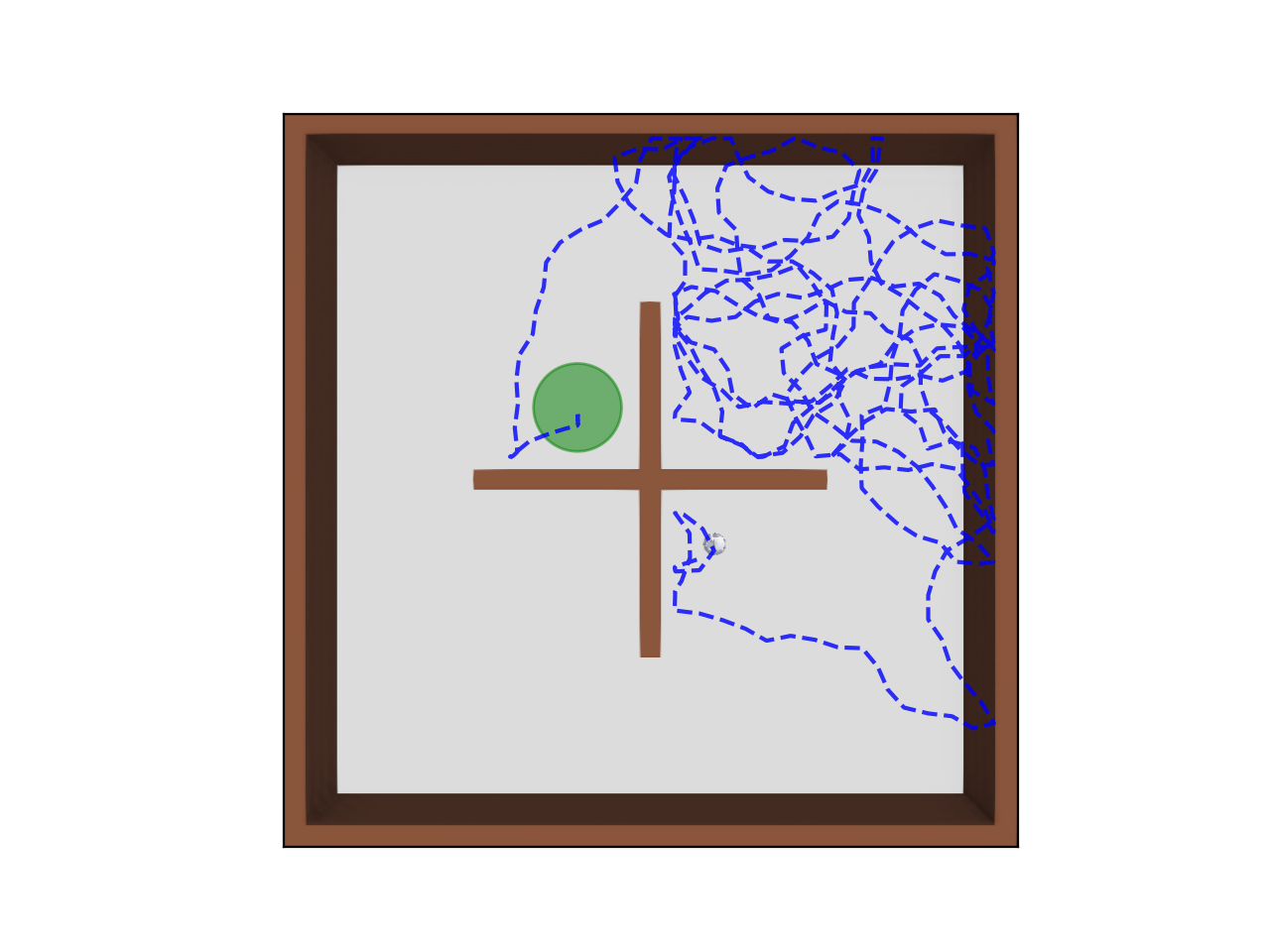}
        \includegraphics[clip,trim={2.5cm 1cm 2.5cm 0 },width=40mm ]{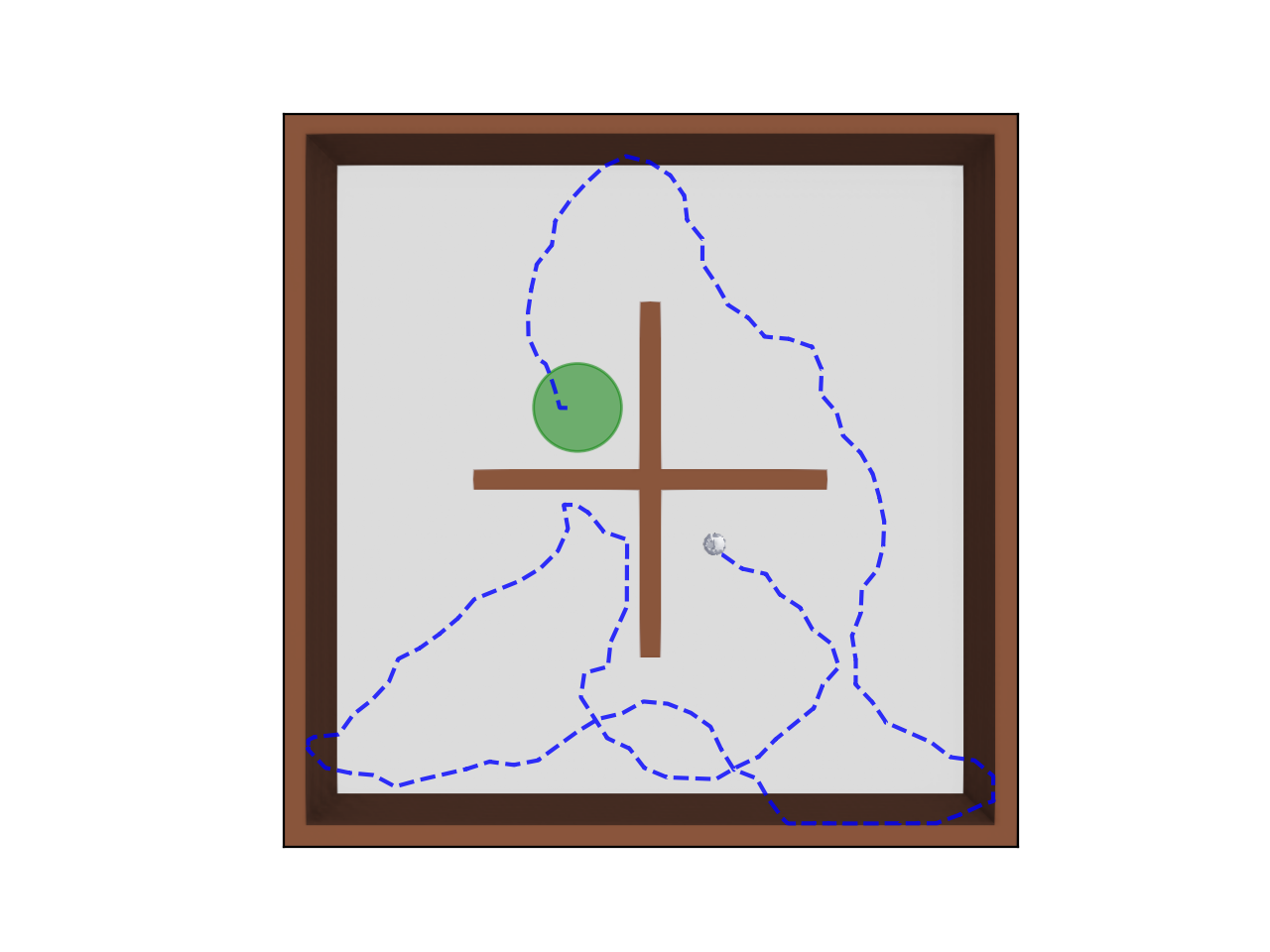}
        \includegraphics[clip,trim={2.5cm 1cm 2.5cm 0 },width=40mm ]{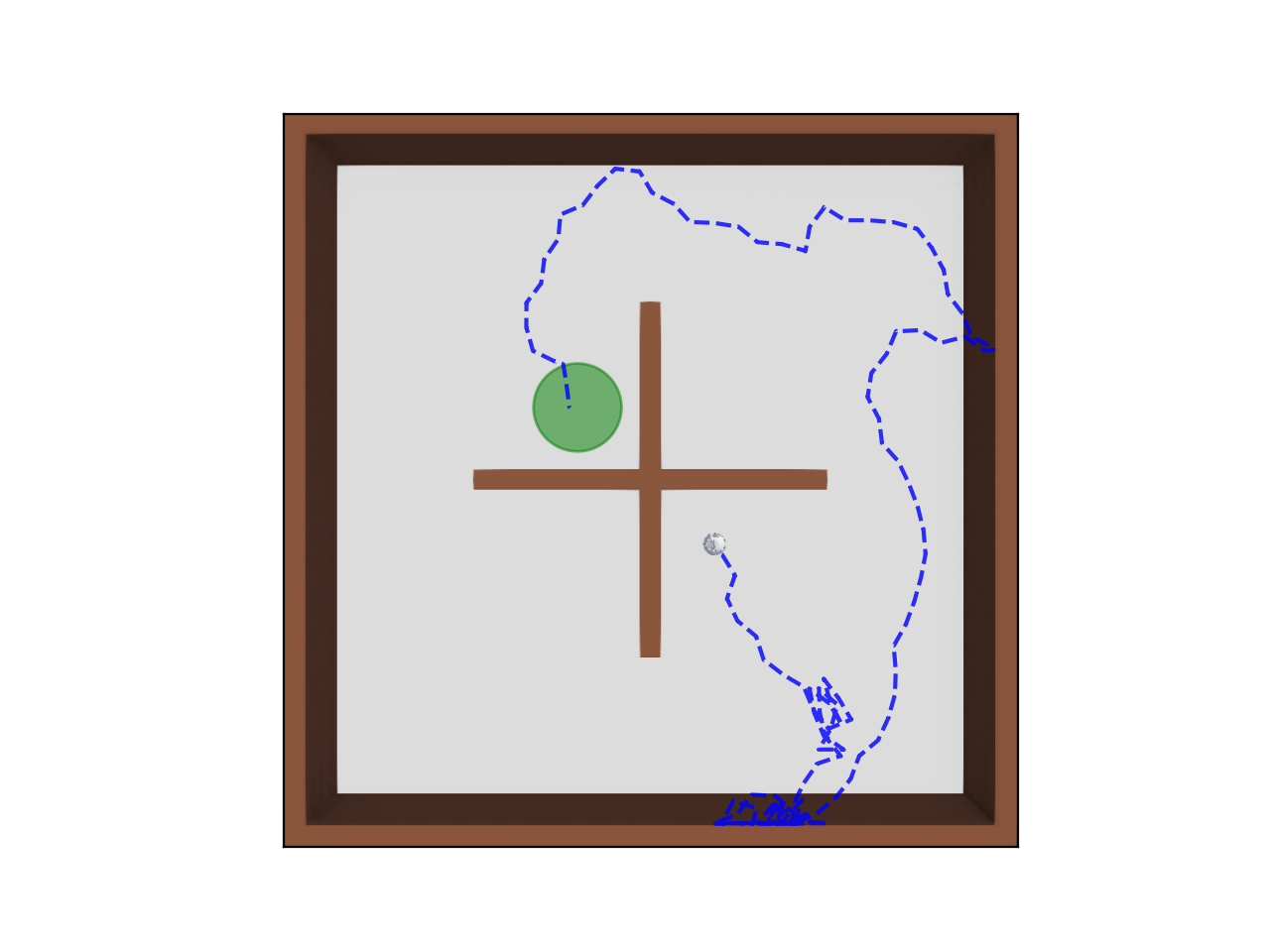}
        \includegraphics[clip,trim={2.5cm 1cm 2.5cm 0 },width=40mm ]{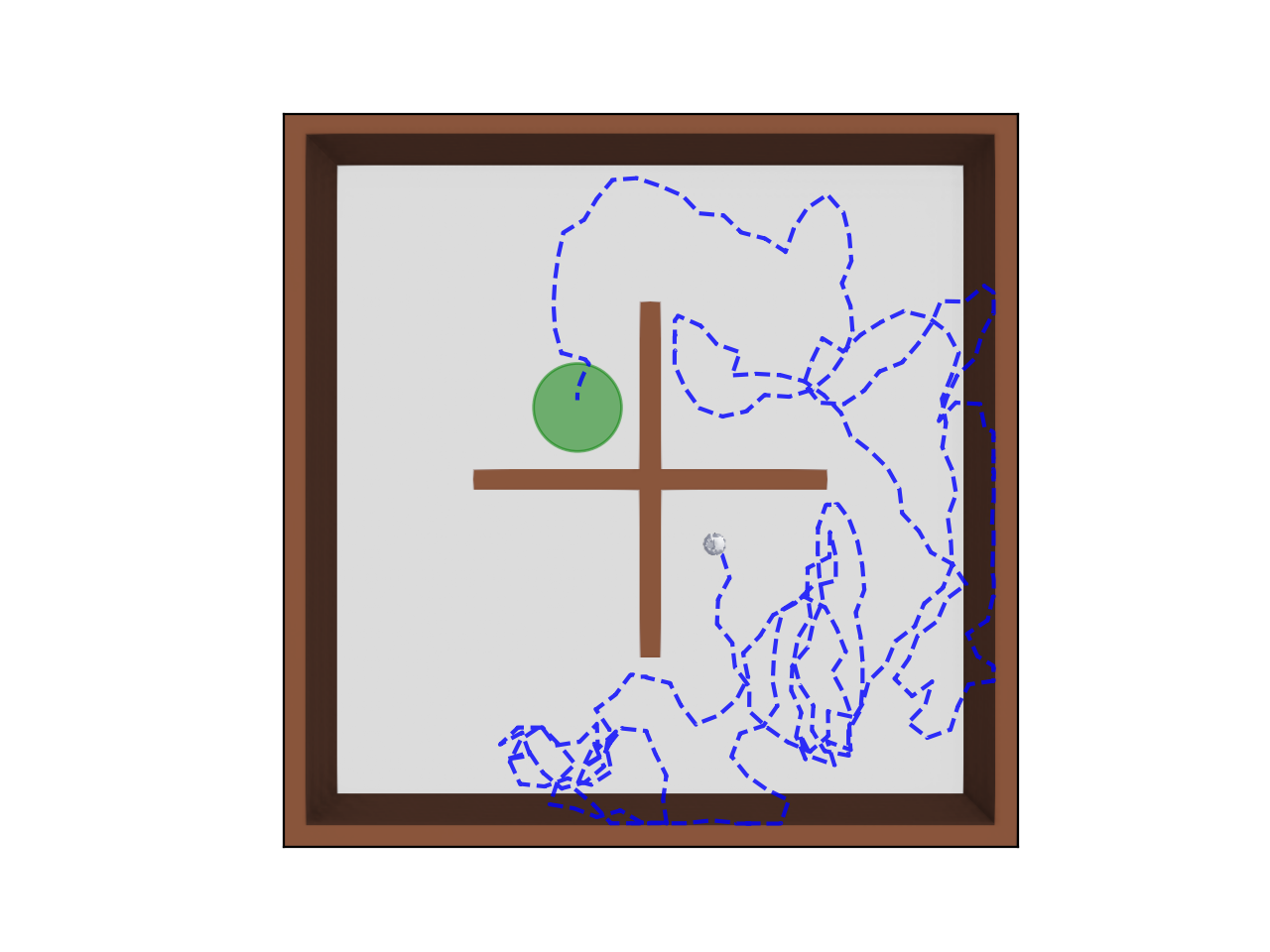}
        \caption{Maze 3}
    \end{subfigure}
  \caption{The path taken to the goal from the same starting point in the complex mazes is shown over four runs in Maze 1 (left column), Maze 2 (middle column), and Maze 3 (right column). The paths in the top panels for each column are the initial exploratory paths without prior experience of the goal location. Each subsequent trial includes the learning from previous trials. The robot is able to quickly learn a useful value map in Mazes 1 and 2, but struggles with perceptual ambiguity in Maze 3.}
  \label{fig:same_start}
\end{figure*}

\begin{figure}
    \centering
    \includegraphics[clip,trim={1.25cm 1cm 1.5cm 0 },width=.6\columnwidth]{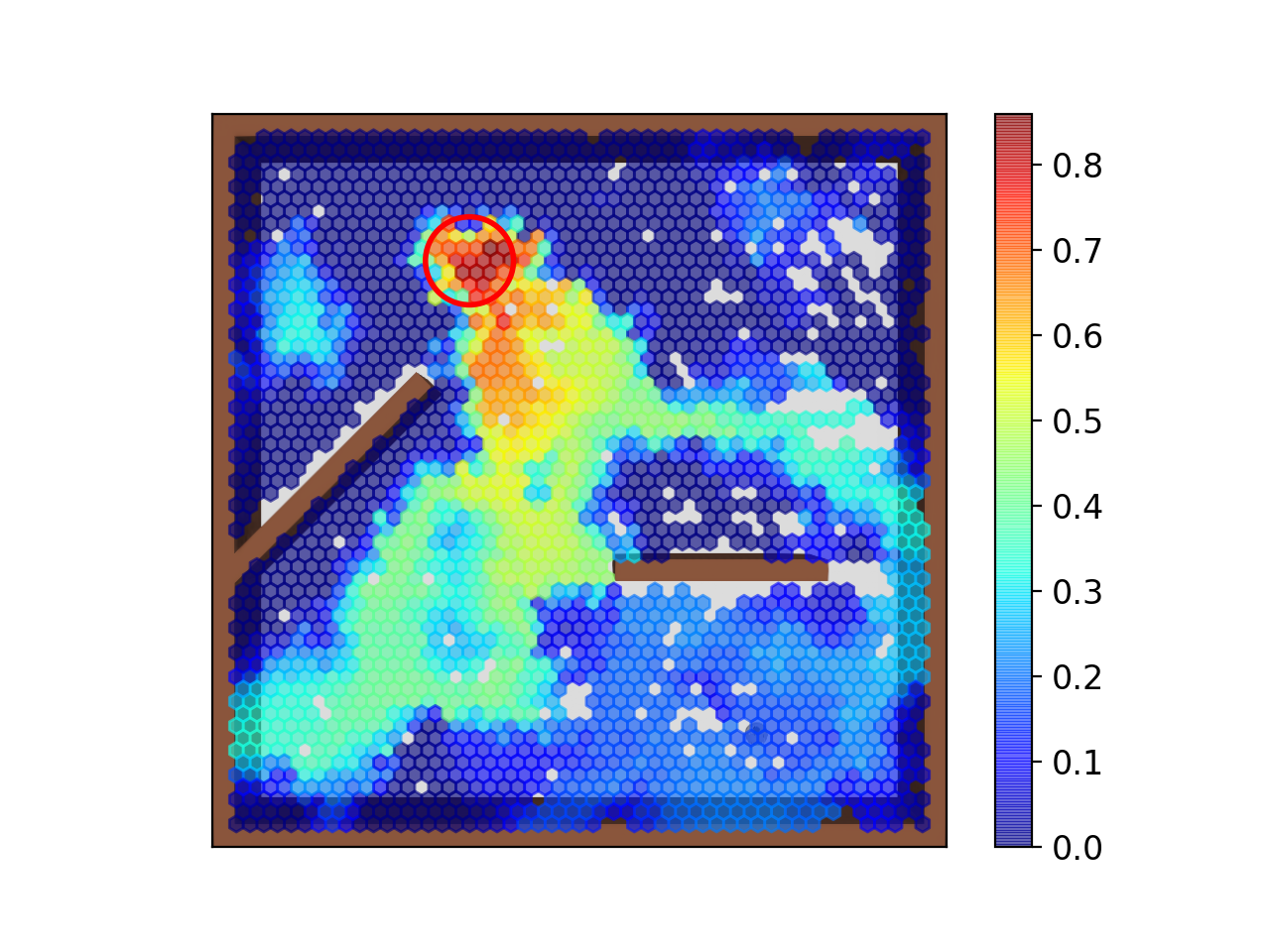}
    \includegraphics[clip,trim={1.25cm 1cm 1.5cm 0 },width=.6\columnwidth]{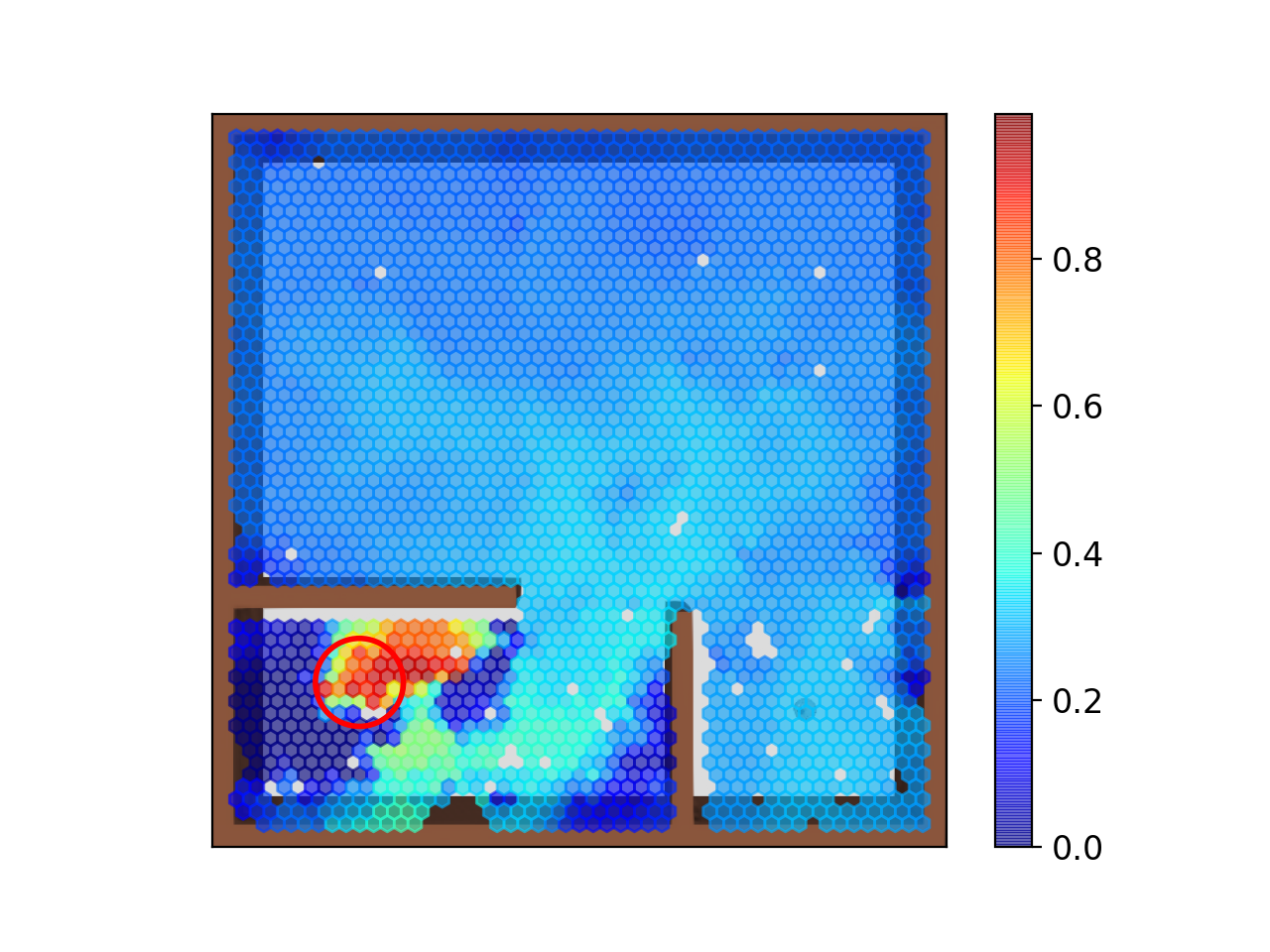}
    \includegraphics[clip,trim={1.25cm 1cm 1.5cm 0 },width=.6\columnwidth]{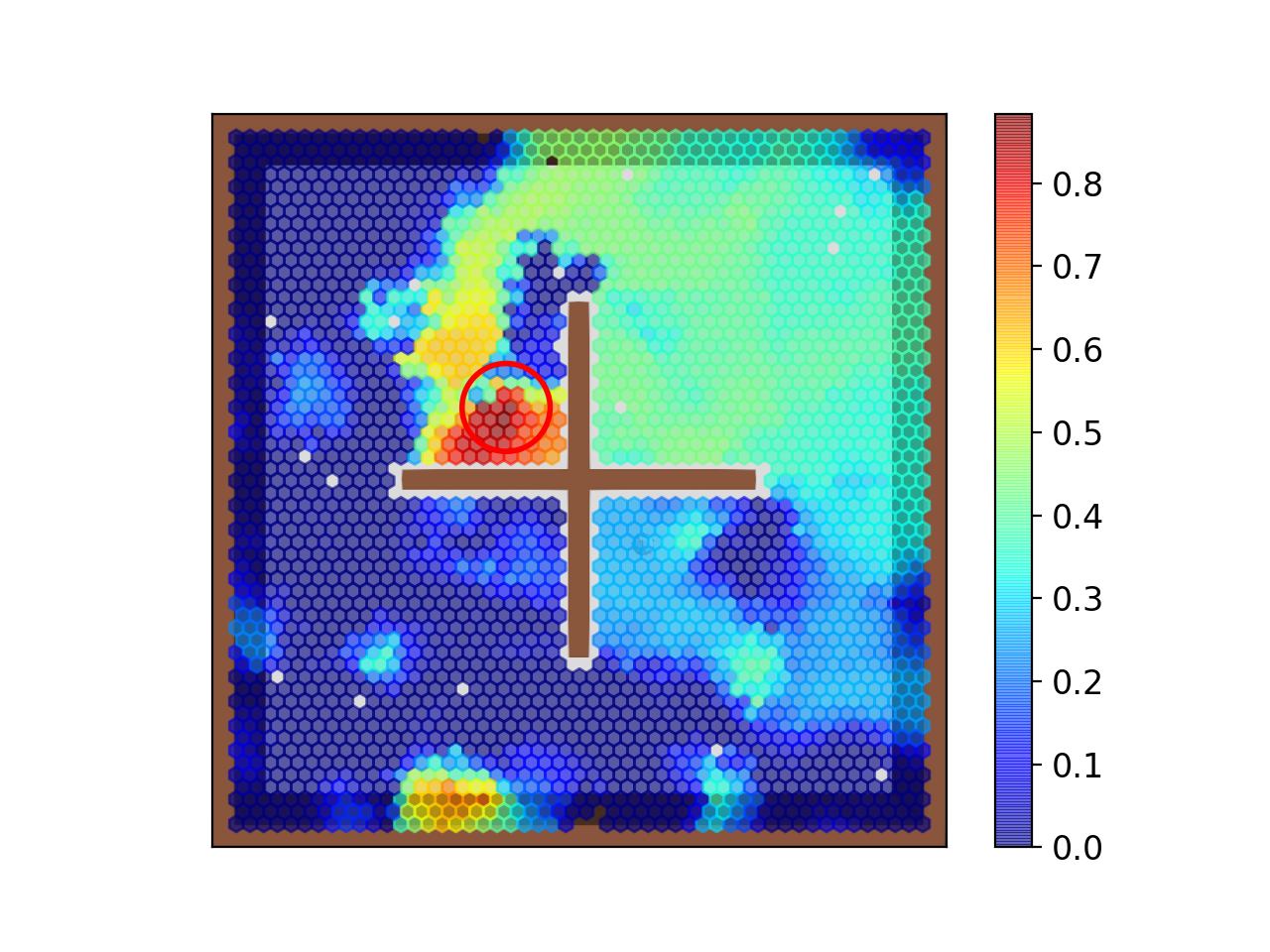}
    \caption{The implied reward maps learnt from the first runs depicted in Fig. \ref{fig:same_start} are shown here for Maze 1 (top panel), Maze 2 (middle panel), and Maze 3 (bottom panel). The reward locations are indicated by the red circles. The plots were generated using the same method as Fig. \ref{fig:reward_maps}.}
    \label{fig:non_trivial_reward_map}
\end{figure}

\begin{figure}
\centering
    \includegraphics[clip,trim={2.5cm 1cm 1.5cm 0 },width=60mm]{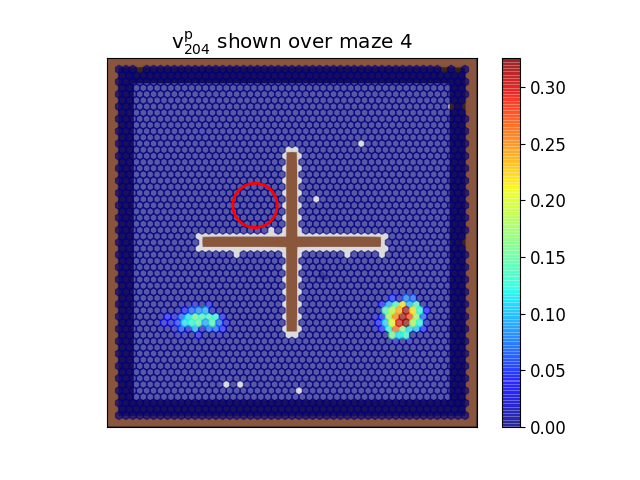}
    \includegraphics[clip,trim={2.5cm 1cm 1.5cm 0 },width=60mm]{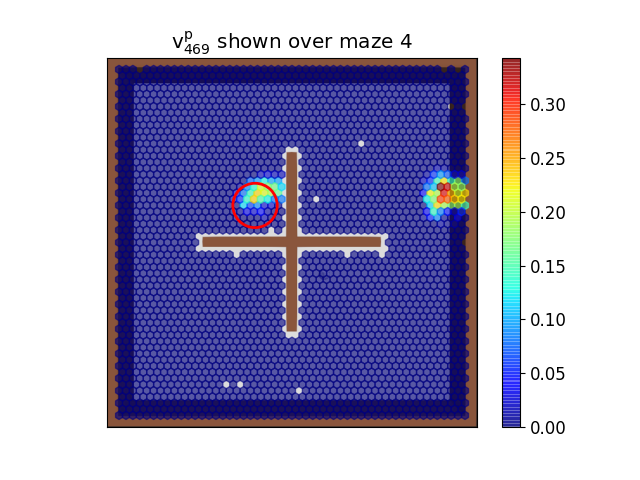}
    \caption{Place cells start to lose spatial specificity as the maze complexity grows so that locations begin to appear perceptually similar to the robot. This is evident in maze 4 where the obstacle divides the environment into four similar sub-compartments. The place cells probed here developed place fields at perceptually similar locations of three of the four sub-compartments.}
    \label{fig:repeating_place_field}
\end{figure}

\subsection{Performance in Complex Mazes} 

To test whether the model would work in more complex environments, it was evaluated in three mazes with various obstacle configurations. In each maze, the robot started from the same location on each of four trial runs. The paths taken in these runs are shown in Fig. \ref{fig:same_start}. The paths in the top row of panels are the purely exploratory ones that the robot took without prior experience of the reward location. As can be seen, the exploratory path taken in Maze 2 (middle column) was much more extensive than in the others. In Maze 1 (left column), the path samples most parts of the environment sparsely, whereas in Maze 3 (right column), exploration is dense in the upper right quadrant while other regions are unexplored. The reward maps generated after the first run in each case are shown in Fig.\ref{fig:non_trivial_reward_map}. AS expected, the inferred reward maps are highly dependent on the exploratory runs. In the subsequent exploitatory runs (roows 2 through 4 in Fig. \ref{fig:same_start}), the effects of the exploration are apparent. In Maze 1 (left column, the robots goes directly to the goal along a path that stays within the previous explored regions. It is worth noting, however, that the path represents a strong generalization, and is not just a subset of the exploratory path. In Maze 2 (middle column), the robout again finds a strongly generalized efficient path, and the quality of the path improves further on subsequent trials (since learning is still on in each trial). For Maze 3 (right column), the situation is less clear. While the robot does show some amount of generalization, it still tends to meander around -- especially if it enters a previously unexplored region. There are two reasons for this. First, the initial exploration was confined to a small part of the environment, so the robot does not have a reward map that ranges across the environment. The second problem is that the obstacle configuration divides the environment into four compartments that appear perceptually similar from the perspective of a robot whose only sensory apparatus is a rangefinder. Thus, the robot tends to suffer from \emph{perceptual aliasing}, mistaking its location for one of the other equivalent locations in some other quadrant, and heads off in the wrong direction -- sometimes entering perceptually similar regions to the goal location where it exhibits \textit{scanning} behaviour.

These limitations in the face of similar boundary conditions (or more broadly sensory input) are expected in the model. In fact, rodent place cells are also known to present multiple place fields in geometrically similar regions of environments, suggesting the same perceptual aliasing \citep{skaggs1998spatial, spiers2015place, fuhs2005influence, derdikman2009fragmentation}. Fig. \ref{fig:repeating_place_field} shows two instances of this effect in the present model,  where the sampled place cells each developed place fields in perceptually similar regions of two quadrants. In rodents however, the dorsal place cells that are putatively driven by BVC input coexist with ventral place cells with larger place fields which are believed to encode context. Such a multiscale configuration would allow the place cell network to encode the knowledge of what quadrant the robot is in and is a natural extension to this work. Moreover, it would allow for the value map to spread out farther during replay and for the computation of more efficient paths as the preplay can then be done at a coarser resolution further away from the goal -- smoothing out the finer details of the exact trajectory previously experienced. 
 % post 
In addition to a multiscale configuration of place fields, rodents are believed to combine sensory information via place cells with ideothetic information from grid cells~\citep{mcnaughton2006path,solstad2006grid,bush2014grid}. This path integration input can potentially be used to disambiguate position as well. Adding these biologically-motivated features to the current model is the subject of ongoing work and will be reported in the future.

\begin{figure*}
    \centering
    \begin{subfigure}[b]{.3\textwidth}
        \includegraphics[clip,trim={2.5cm 1cm 1.5cm 1cm },width=40mm]{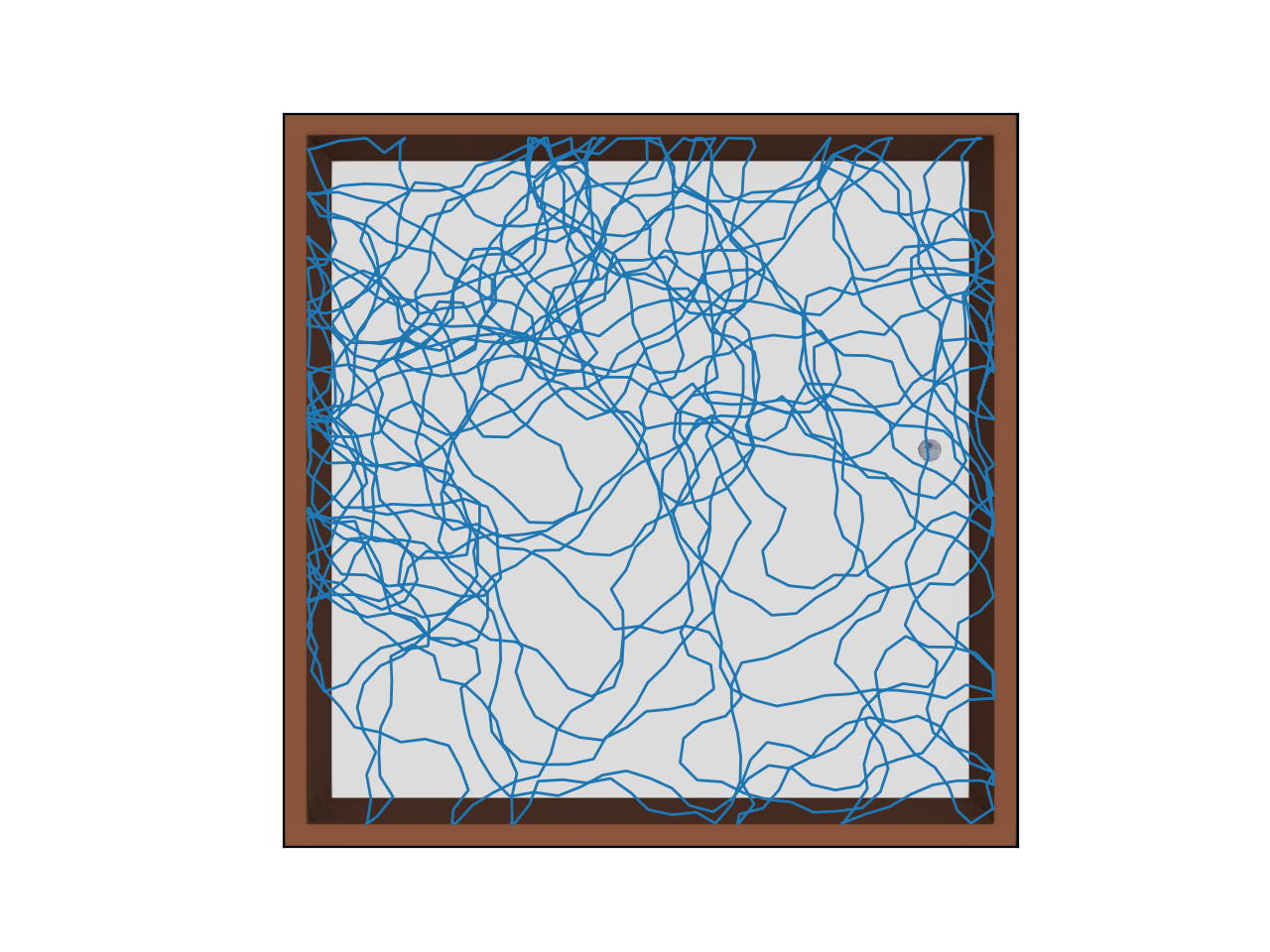}
        \includegraphics[clip,trim={2.5cm 1cm 1.5cm 1cm },width=40mm]{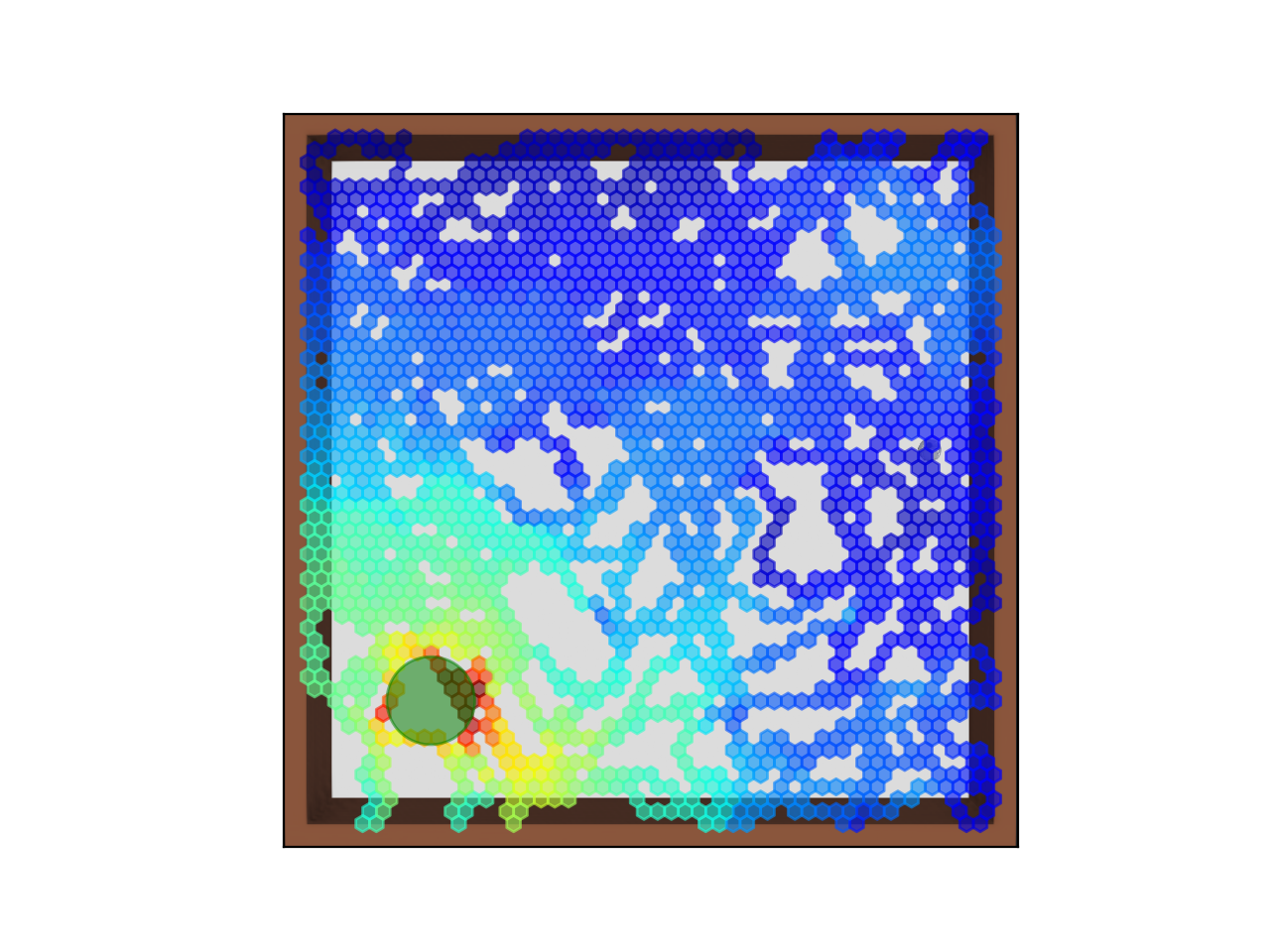}
        \includegraphics[clip,trim={2.5cm 1cm 1.5cm 1cm },width=40mm]{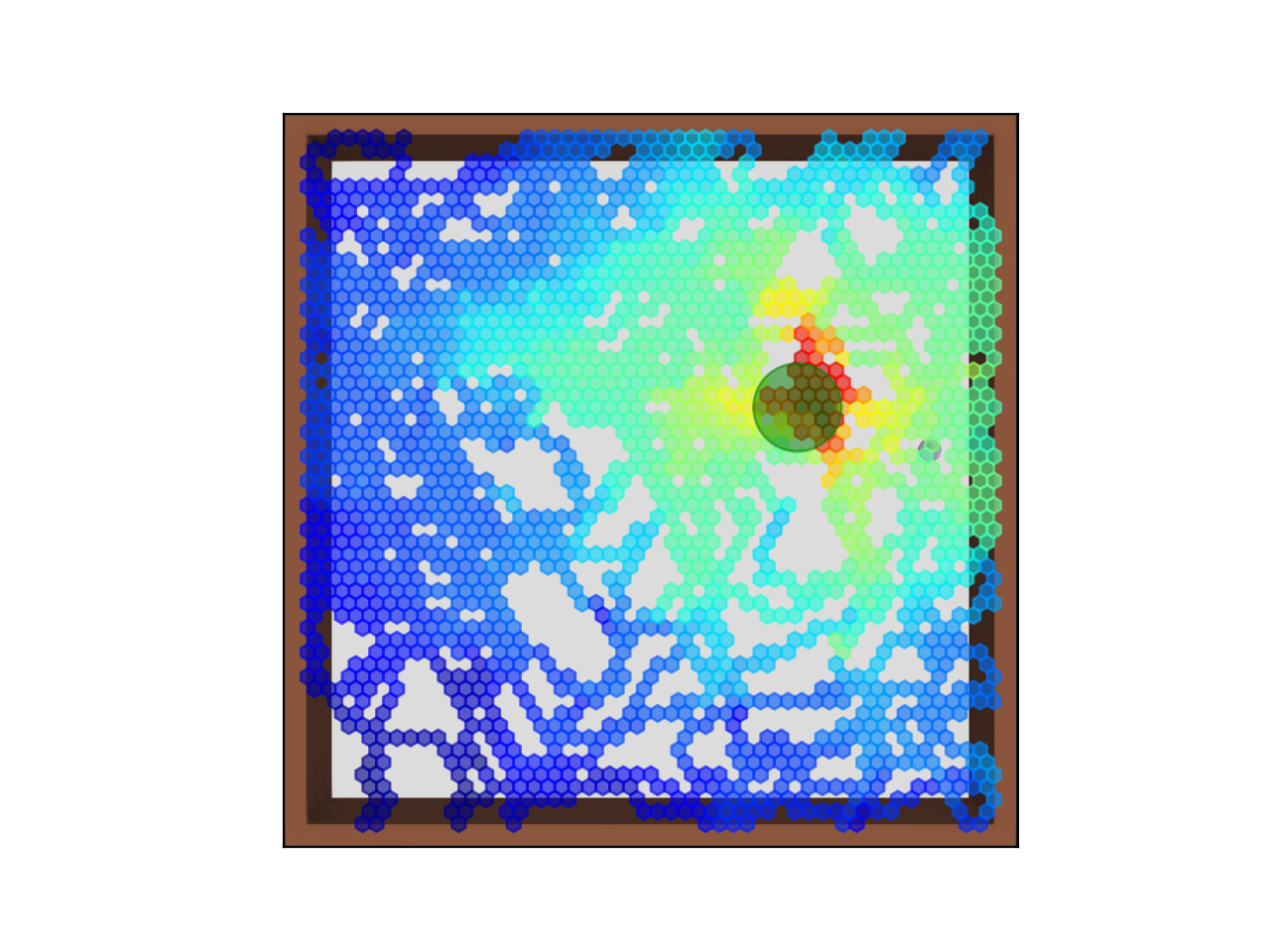}
        \includegraphics[clip,trim={2.5cm 1cm 1.5cm 1cm },width=40mm]{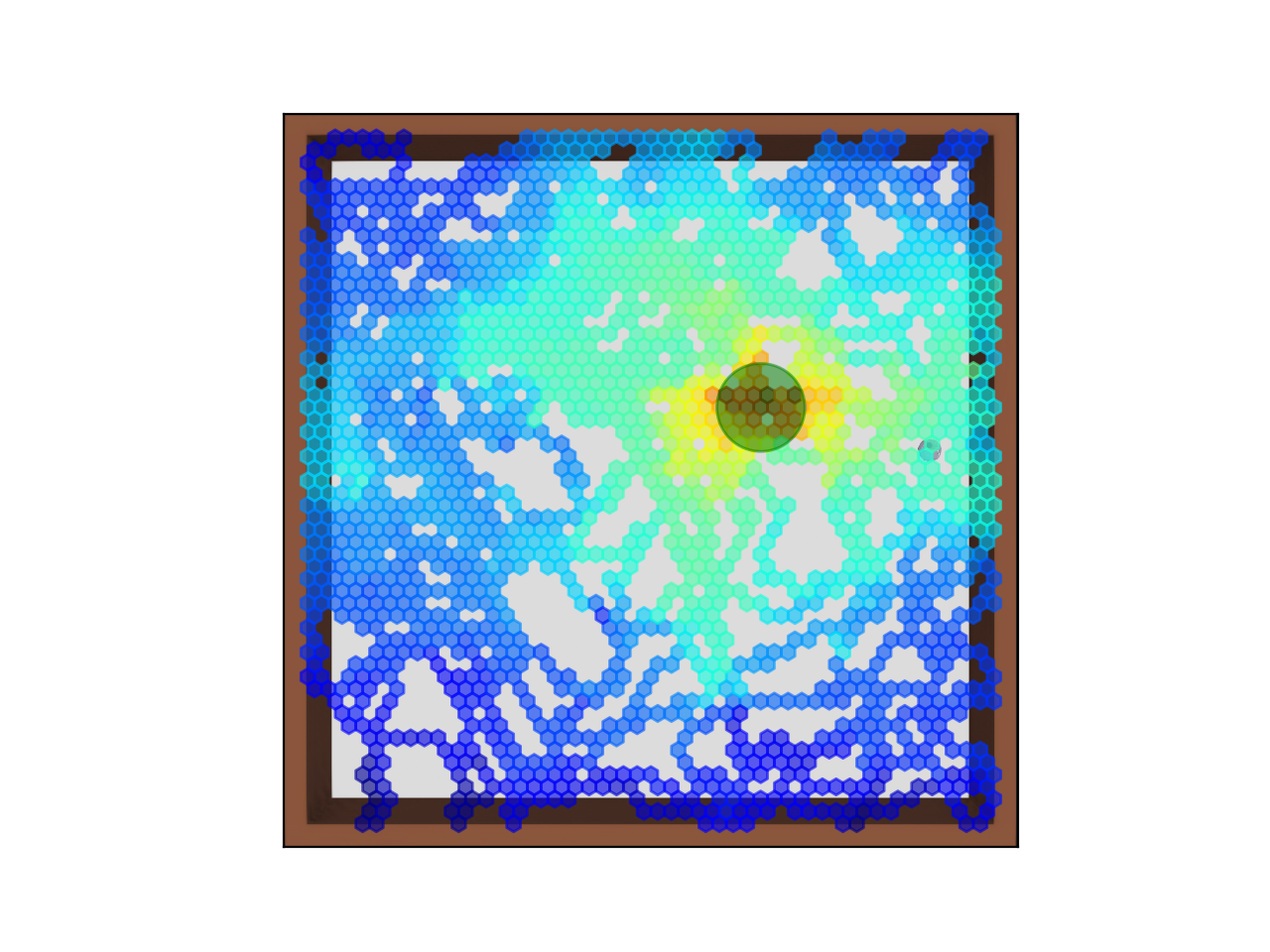}
        \caption{}
    \end{subfigure}
    \begin{subfigure}[b]{.3\textwidth}
        \includegraphics[clip,trim={2.5cm 1cm 1.5cm 1cm },width=40mm]{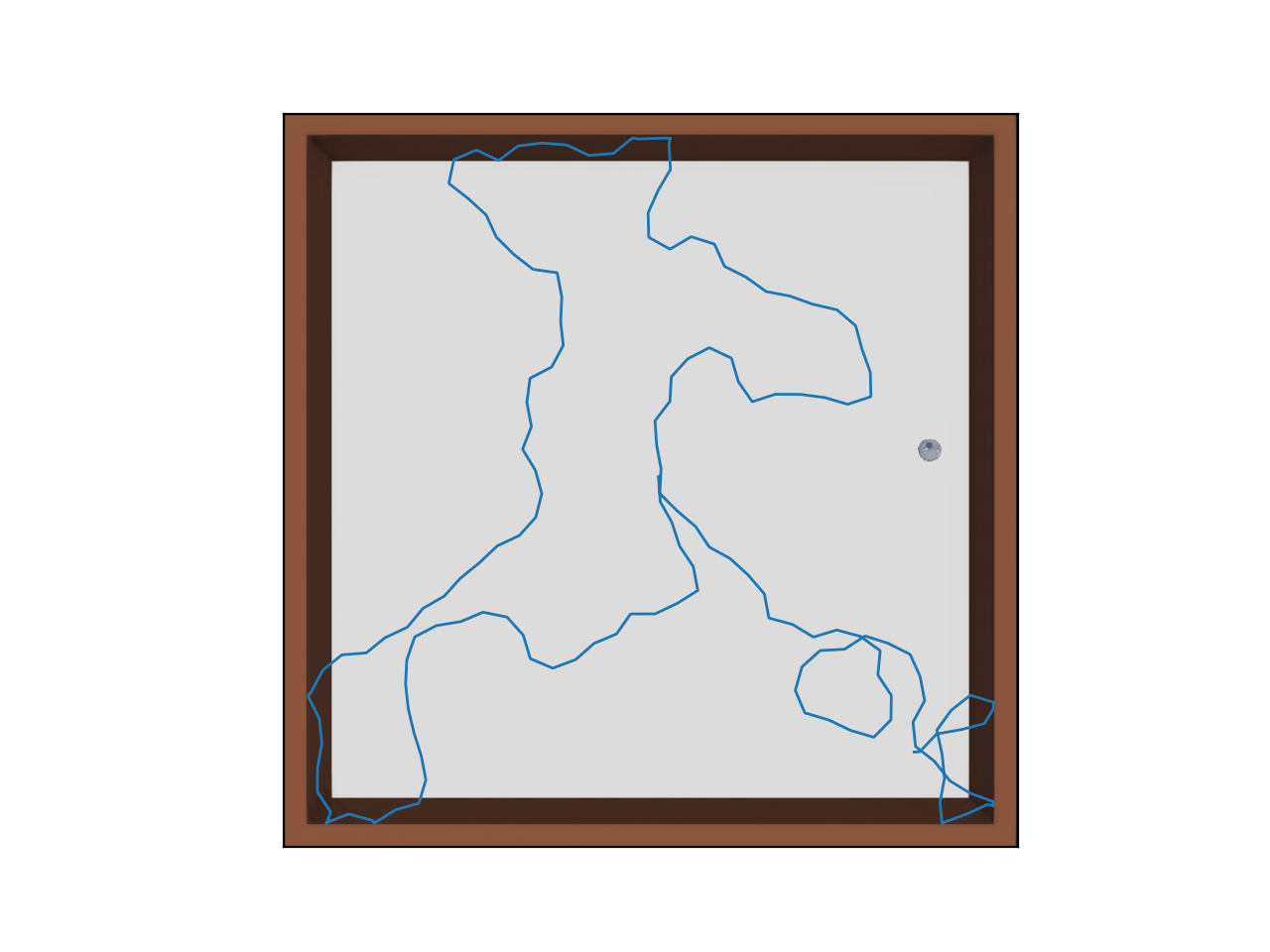}
        \includegraphics[clip,trim={2.5cm 1cm 1.5cm 1cm },width=40mm]{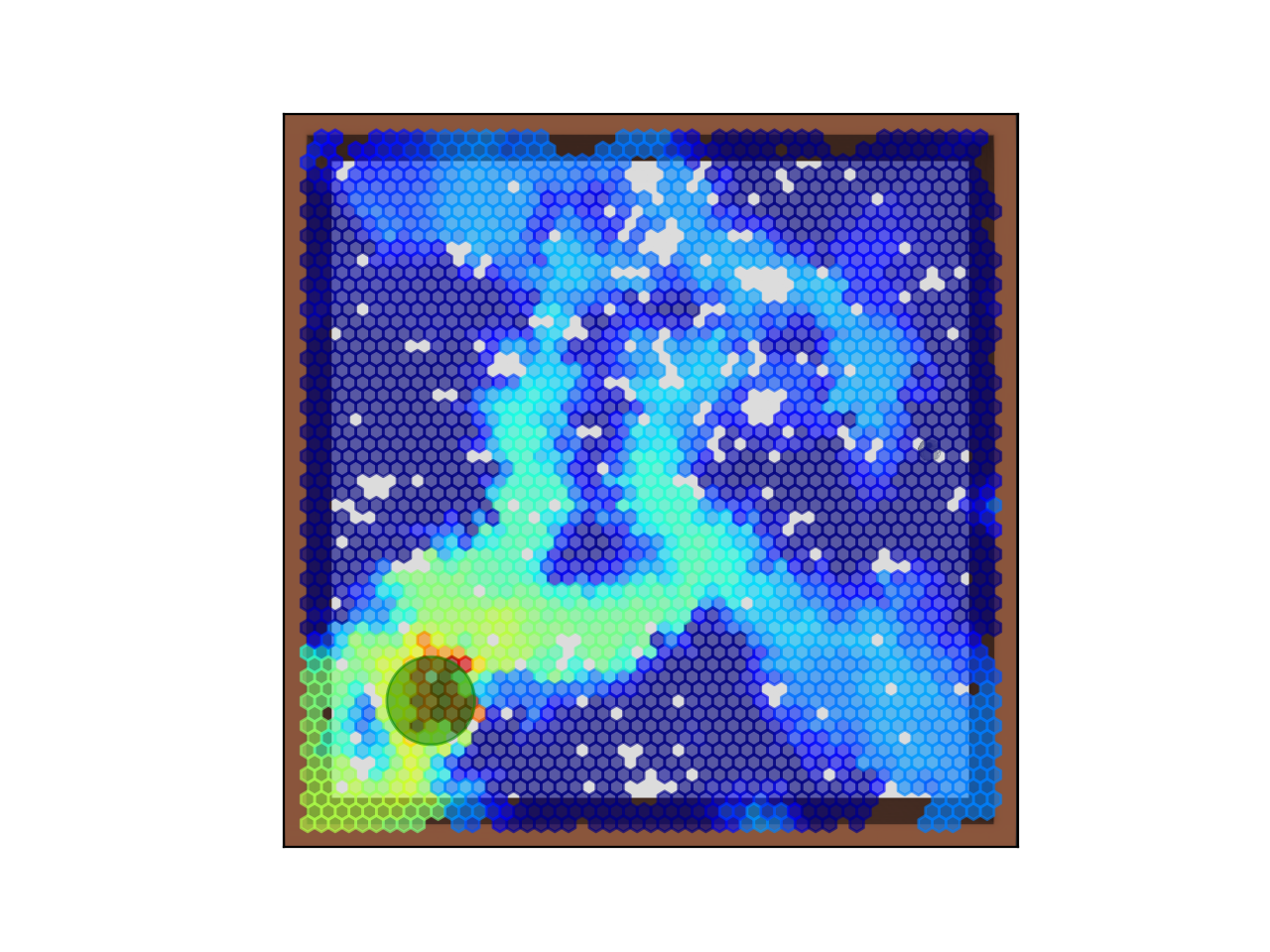}
        \includegraphics[clip,trim={2.5cm 1cm 1.5cm 1cm },width=40mm]{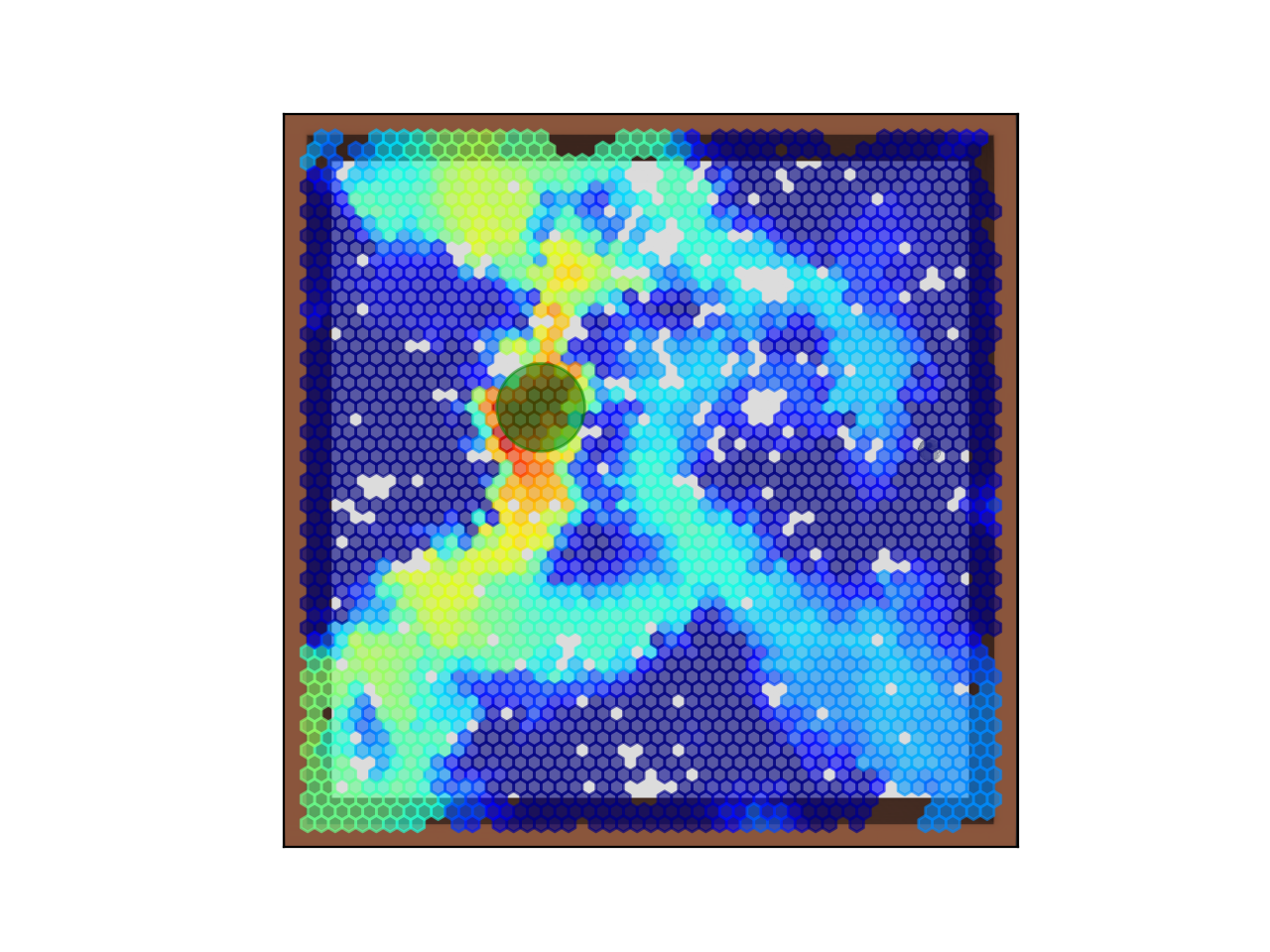}
        \includegraphics[clip,trim={2.5cm 1cm 1.5cm 1cm },width=40mm]{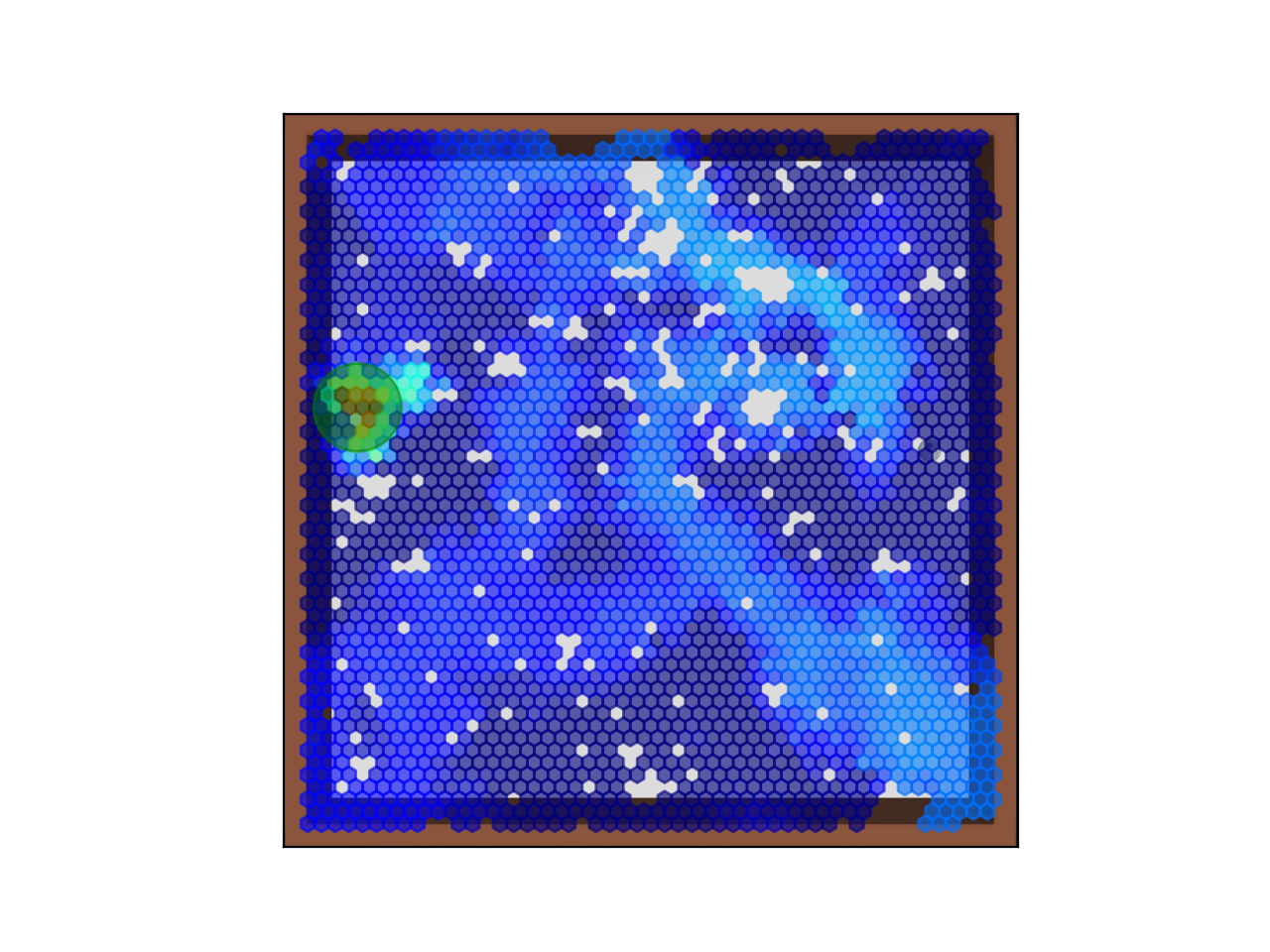}
        \caption{}
        \label{fig:generalization_limited}
    \end{subfigure}
    \begin{subfigure}[b]{.3\textwidth}
        \includegraphics[clip,trim={2.5cm 1cm 1.5cm 1cm},width=40mm]{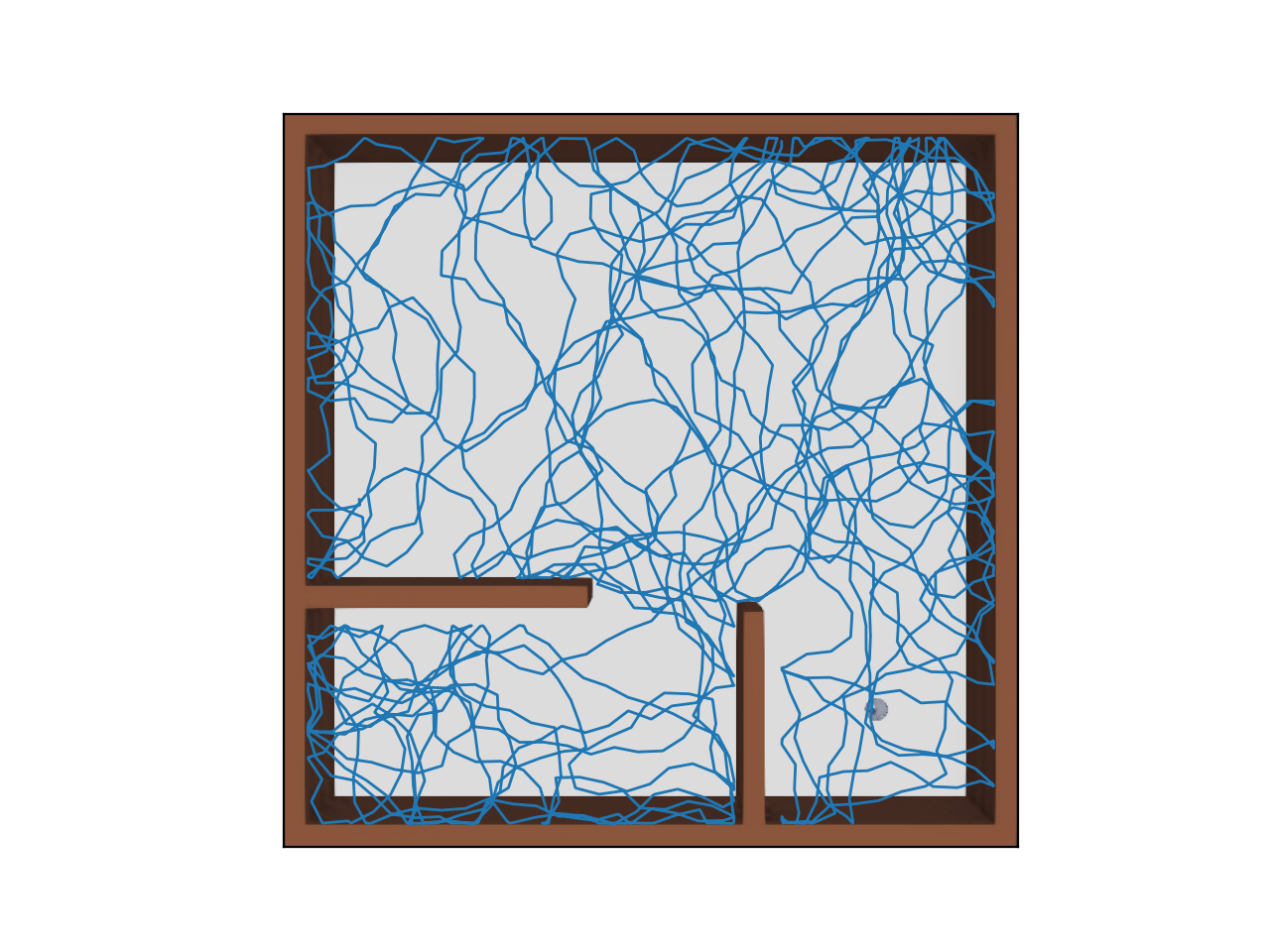}
        \includegraphics[clip,trim={2.5cm 1cm 1.5cm 1cm},width=40mm]{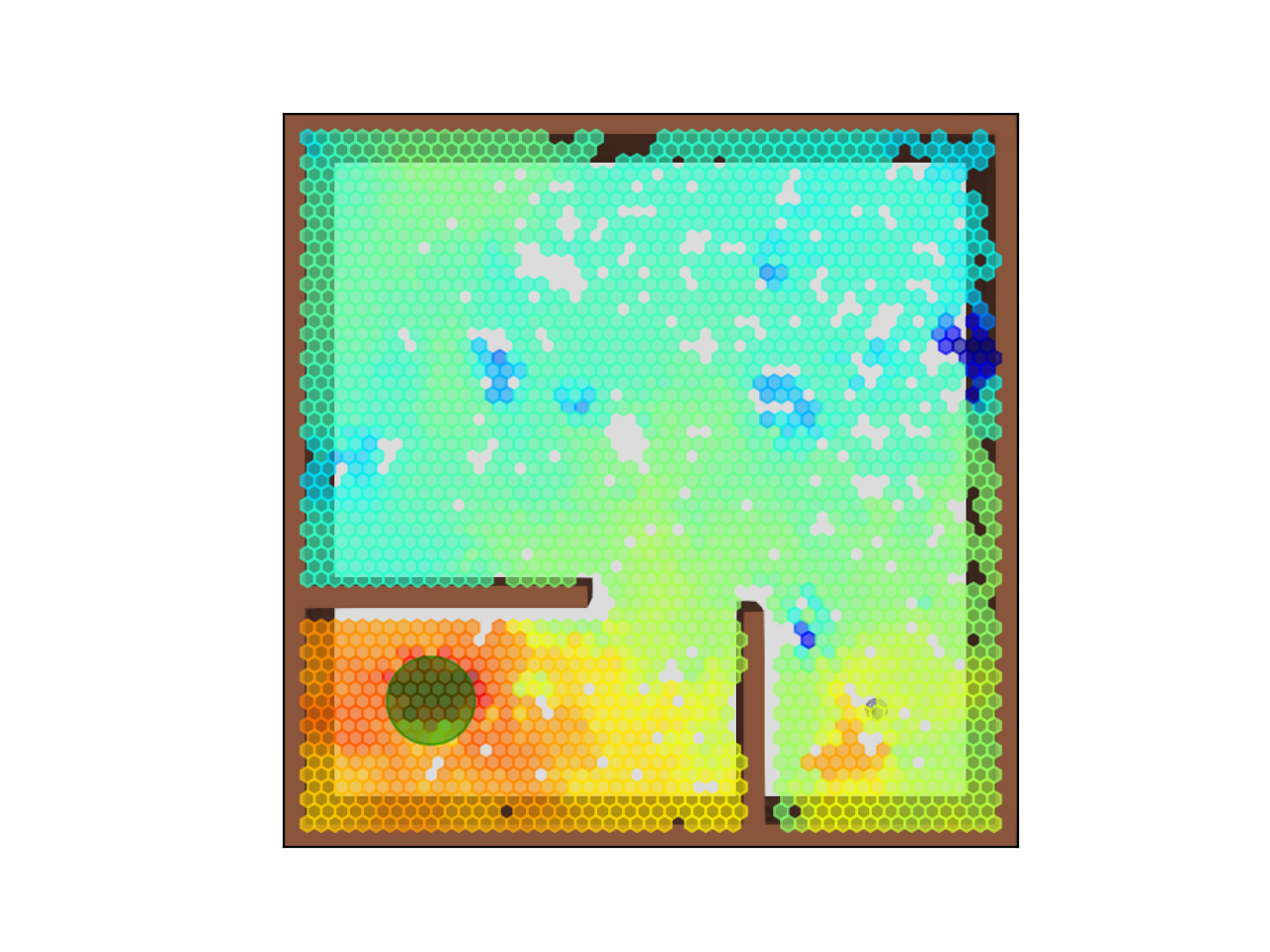}
        \includegraphics[clip,trim={2.5cm 1cm 1.5cm 1cm},width=40mm]{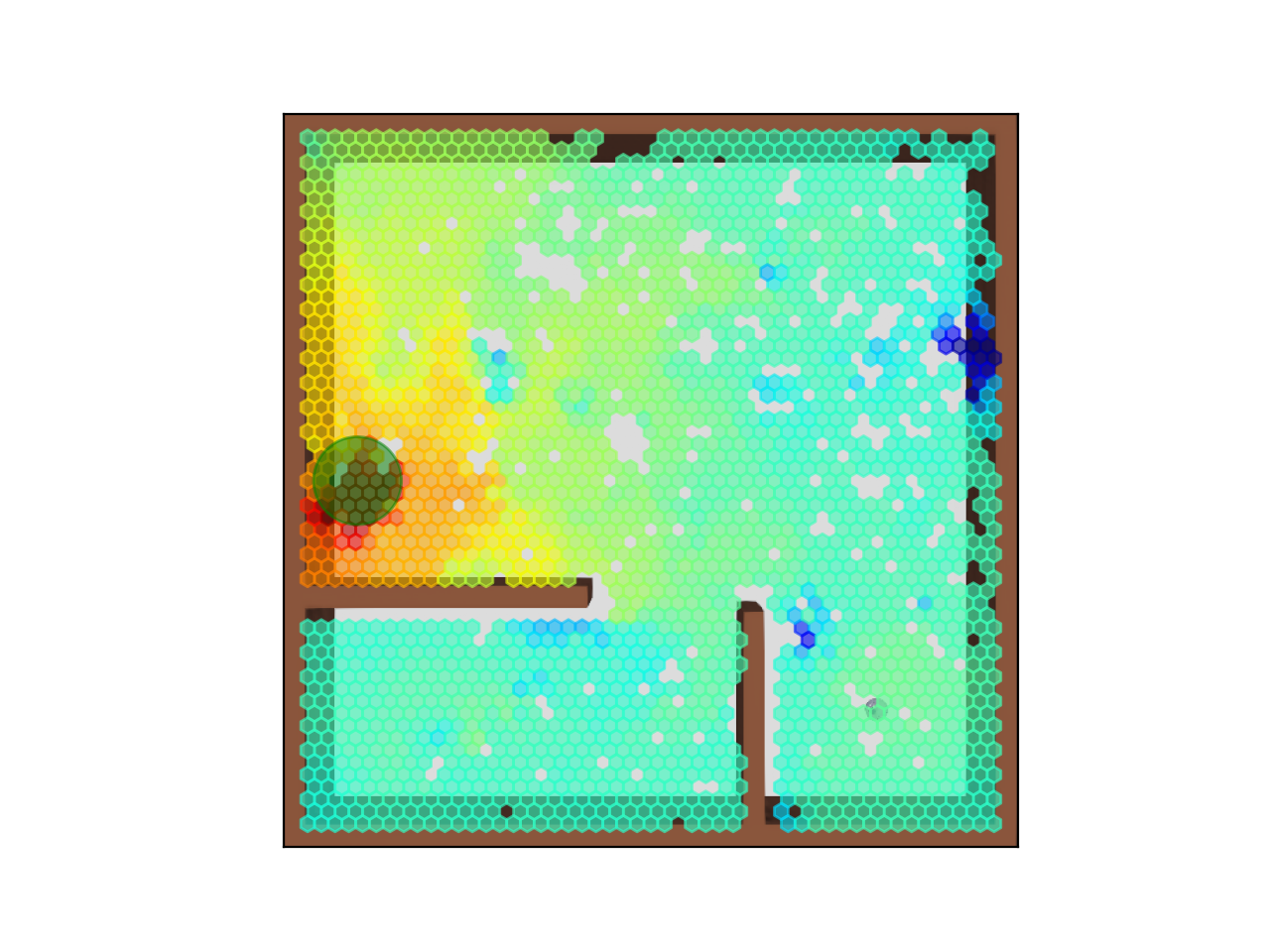}
        \includegraphics[clip,trim={2.5cm 1cm 1.5cm 1cm },width=40mm]{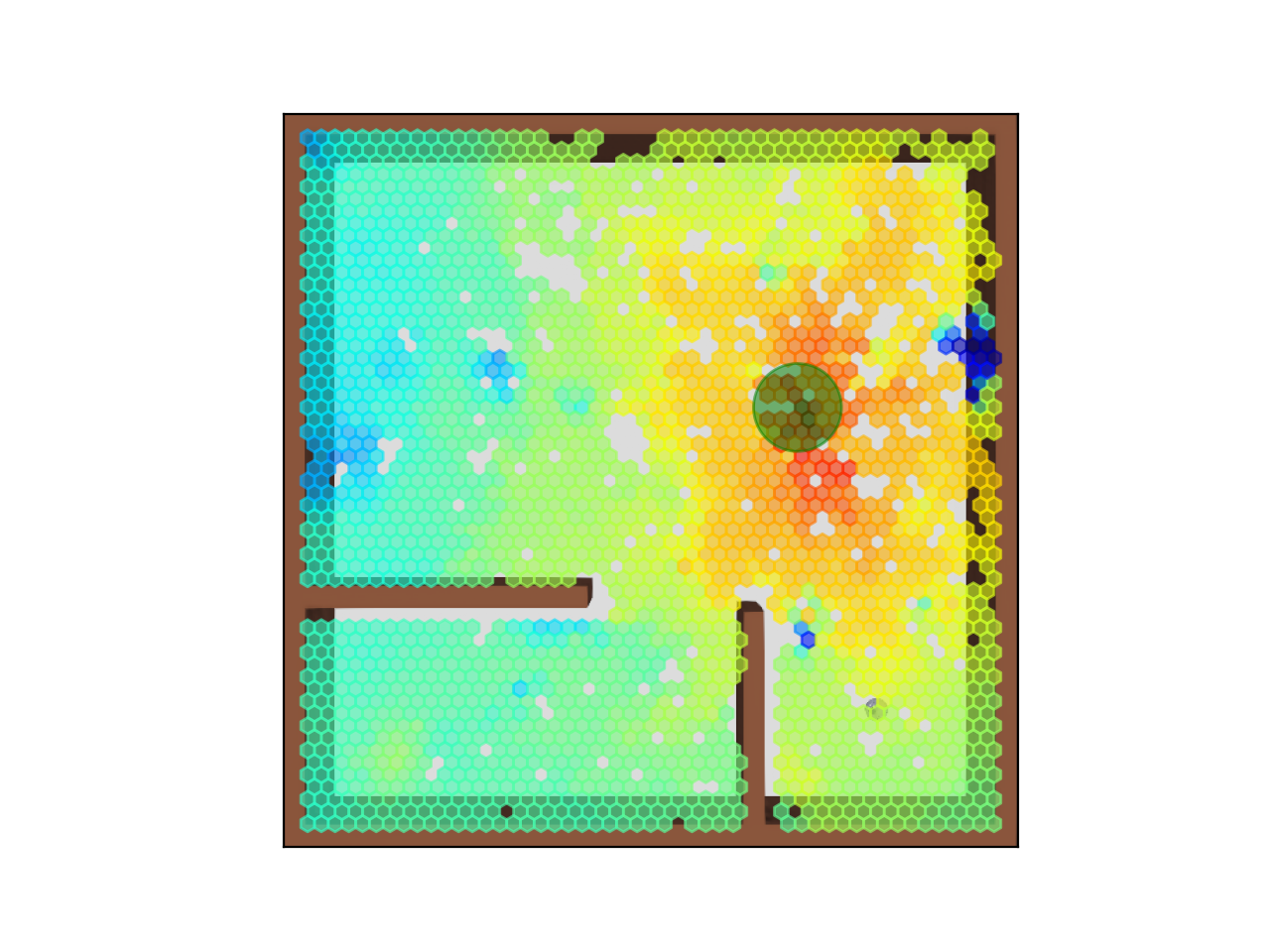}
        \caption{}
        \label{fig:generalization_limited}    
    \end{subfigure}
    
    \caption{The reward maps from a robot placed at random reward locations without any further exploration after initially exploring the open maze extensively in (a), partially in (b), and Maze 2 extensively in (c). The top row shows the exploratory path followed in each case. Rows 2 through 4 show the inferred reward maps obtained by placing the reward at various locations (depicted with the green circle) and sampling the inferred reward map for that location over the whole environment without further learning. The reward maps spread over the known environment when the reward location had been previously encountered. This spread is more limited when the reward location was previously unknown as seen in the last map in scenario (b).}
    \label{fig:generalization}
\end{figure*}

% post review
\subsection{Generalization Across Different Reward Locations}
An important issue for the model is its ability to generalize after learning, i.e., once it has explored an environment (and thus established place fields), it should be able to infer a reward map for a reward at any location in the environment {\it without further exploration}.

To test for this spatial generalization, we allowed the robot to first explore the maze in the absence of any reward location. The robot was then placed at arbitrary locations that were designated as the reward locations, and allowed to infer a reward map relative to that location purely by using the backward replay mechanism described earlier, i.e., without any further exploration. This map was then sampled by letting the robot run extensively through the maze without any further learning, generating a sampled view of the inferred reward map for visualization.

Since the inference of the reward map would clearly depend on how extensive the initial exploration was, we simulated two cases in the open maze: a) Extensive initial exploration of the maze; and b) A very limited initial exploration along a specific path. Additionally, we simulated a scenario with extensive exploration of Maze 2 (c), which is a more complex maze with obstacles.
%To verify that the model can generalize and apply the map learned from exploring to novel reward locations, the robot was made to explore the open maze once before being placed at random goal locations without any further exploration. At the reward locations, the replay process ensued to build reward maps which were visualized to determine the robot's ability to build a useful reward map from the prior exploration. This experiment was conducted twice - first with extensive exploration of the maze then with limited exploration. 
The results from this experiment are shown in Fig. \ref{fig:generalization}. The first experiment with extensive exploration in the open maze is shown in the left column, that with limited exploration in the open maze is shown in the center column, while the experiment with extensive exploration of complex Maze 2 is shown in the right column. The paths followed during the exploratory runs are shown in the first row. The remaining rows show observed reward maps with the reward placed at various arbitrary locations. It can be seen that, in each case, the reward map generalizes over the known environment without further exploration or any prior knowledge of the reward location. However, as expected, the inferred reward map works only for the regions included in the initial exploration, though not confined only to the exploratory paths because of the field generalization effect discussed earlier. This also means that when the new goal location is placed in a region not visited during prior exploration, the robot is unable to compute a useful reward map. This can be seen in the last reward map of Fig. \ref{fig:generalization_limited} where the goal was placed in an area of the map it had not previously visited at all.

\section{Conclusions}
\label{sec:conclusion}
% post
In this work, the goal was to show how replay and preplay, which have been proposed to support the efficiency of biological reinforcement learning, might be used in artificial agents with a neural place representation and reinforcement learning system inspired by the hippocampus and related regions in the rodent brain. Another goal was to develop a place cell network model that generates convex place fields from minimal preconfiguration. Including this component in the model has three benefits: 1) The place field model is not assumed ad-hoc but is built using an explicit, well-defined, and biologically plausible process; 2) The specific pattern of neural connectivity required by the navigation process is ensured by the model; and 3) The model represents an interesting hypothesis for the rapid formation of convex place fields in animals.

The proposed self-organized model is able to develop convex localized place fields from randomly initialized BVC-to-place cell connections, learn the topology of non-trivial environments, and then compute and exploit value maps of these environments for efficient goal-directed navigation. The model also displays very rapid learning with continued improvements from further experience, and can thus be considered a one-shot or few-shot reinforcement learning system. The model is, however, limited by the fidelity of its perception. Performance suffers in the presence of perceptual aliasing, though this is expected of any spatial cognition system: Even humans get lost in mazes.

The model can potentially be extended in several ways. First, as previously stated, rodents are known to combine the BVC-driven place cells which have been modelled here with context-driven place cells that have larger place fields in a hierarchical manner. Such a multiscale configuration can allow for the disambiguation of similar areas at the BVC place cell representation level based on the representation at the lower resolution contextual level. 

Second, moving beyond a multiscale configuration, the idea of place cells tuned to boundary vectors can be abstracted to place cells tuned to specific views -- be it boundary vectors, visual cues, cues of any other sensory modality, or a combination of these. Using richer sensory modalities and/or fusing complementary modalities reduces the incidence of scene ambiguity and can increase the robustness of the model.

Further work is being done to incorporate these into the model, as well as extensions that allow for the learning of multiple maps by the same place cell network without interference.

\clearpage
\bibliographystyle{elsarticle-harv}
\bibliography{references}

\end{document}